\documentclass[notitlepage,superscriptaddress,nofootinbib,amsmath,amssymb,aps,11pt]{revtex4-1}
\usepackage[]{inputenc}
\usepackage[dvips]{graphicx}
\usepackage{color}
\usepackage{bm}
\usepackage{mathpazo}
\usepackage{epstopdf}
\usepackage[bookmarks=true,colorlinks,linkcolor=blue,urlcolor=green,citecolor=red]{hyperref}
\renewcommand\a{\alpha}
\renewcommand\b{\beta}
\renewcommand\d{\delta}

\renewcommand\r{\rho}

\renewcommand\t{\tau}

\renewcommand\c{\chi}
\renewcommand\j{\psi}

\newcommand\g{\gamma}

\newcommand\m{\mu}

\newcommand\p{\pi}

\newcommand\s{\sigma}
\newcommand\f{\phi}

\renewcommand\L{\Lambda}

\newcommand\D{\Delta}

\newcommand\J{\Psi}

\newcommand{\fig}[1]{Fig.~\ref{#1}}
\newcommand{\eq}[1]{Eq.~(\ref{#1})}
\newcommand{\sect}[1]{Sec.~\ref{#1}}

\newcommand\lb{\left(}
\newcommand\rb{\right)}

\newcommand{\lan}{\langle}
\newcommand{\ran}{\rangle}

\newcommand{\non}{\nonumber\\}
\newcommand\pt{\partial}

\newcommand{\bx}{{\vec x}}
\newcommand{\br}{{\vec r}}
\newcommand{\bp}{{\vec p}}

\newcommand{\bq}{{\vec q}}

\newcommand{\bv}{{\vec v}}

\newcommand{\bB}{{\vec B}}
\newcommand{\bE}{{\vec E}}
\newcommand{\bJ}{{\vec J}}

\newcommand{\jb}{{\bar \j}}

\newcommand{\ext}{{\rm ext}}
\newcommand{\rp}{{\rm RP}}

\renewcommand{\part}{{\rm part}}
\newcommand{\gam}{ \gamma }

\usepackage{slashed}
\usepackage{mathrsfs}

\newcommand{\bA}{{\bm A}}
\newcommand{\bmu}{{\bm{\mu}}}

\newcommand{\EM}{{\rm em}}

\newcommand{\tr}{ {\rm Tr} }

\newcommand{\Lag}{ {\mathscr{L}} }
\newcommand{\M}{ {\mathcal M} }

\newcommand{\pv}{{_{\scriptscriptstyle  \rm PV}}}
\newcommand{\va}{{_{\scriptscriptstyle \rm VA} }}
\newcommand{\sa}{{_{\scriptscriptstyle \rm SA} }}
\newcommand{\ps}{ {\scriptscriptstyle \rm P} }
\newcommand{\V}{ {\scriptscriptstyle \rm V} }
\newcommand{\A}{ {\scriptscriptstyle \rm A} }

\newcommand{\swave}{{ \scriptscriptstyle 1S }}
\newcommand{\cc}{{ \scriptscriptstyle {c \bar c} }}

\newcommand{\eb}{{ \epsilon_0}}
\newcommand{\etac}{ {\eta_c} }
\newcommand{\Jp}{ {J/\psi} }

\newcommand{\scJ}{ {\scriptstyle  J} }

\newcommand{\B}{ {\mathcal B} }
\newcommand{\Ham}{{\mathcal H}}
\newcommand{\cor}{{\bar q q}}

\newcommand{\beq}{\begin{eqnarray}}
\newcommand{\eeq}{\end{eqnarray}}

\renewcommand{\vec}{\boldsymbol}
\newcommand{\be}{\begin{equation}}
\newcommand{\ee}{\end{equation}}
\newcommand{\bear}{\begin{eqnarray}}
\newcommand{\eear}{\end{eqnarray}}
\newcommand{\ba}{\begin{array}}
\newcommand{\ea}{\end{array}}

\begin{document}

\begin{flushright}
RBRC-1202
\end{flushright}

\title{Novel quantum phenomena induced by strong magnetic fields in heavy-ion collisions}
\author{\normalsize{Koichi Hattori}}
\affiliation{Physics Department and Center for Particle Physics and Field Theory, Fudan University, Shanghai 200433, China.}
\affiliation{RIKEN-BNL Research Center, Brookhaven National Laboratory, Upton, New York 11973-5000, USA}
\author{\normalsize{Xu-Guang Huang}}
\affiliation{Physics Department and Center for Particle Physics and Field Theory, Fudan University, Shanghai 200433, China.}
\affiliation{Department of Physics, Brookhaven National Laboratory, Upton, New York 11973-5000, USA}

\date{\today}

\begin{abstract}
The relativistic heavy-ion collisions create both hot quark-gluon matter and strong magnetic fields, and provide an arena to study the interplay between quantum chromodynamics (QCD) and quantum electrodynamics (QED). In recent years, it has been shown that such an interplay can generate a number of interesting quantum phenomena in hadronic and quark-gluon matter. In this short review, we first discuss some properties of the magnetic fields in heavy-ion collisions and then give an overview of the magnetic-field induced novel quantum effects. In particular, we focus on the magnetic effect on the heavy-flavor mesons, and the heavy quark transports, and also the phenomena closely related to chiral anomaly.
\end{abstract}
\maketitle

\section {Introduction}\label{sec:intro}

Understanding the phase structure and, more generally, the many-body physics of quantum chromodynamics (QCD) is one of the most outstanding challenges in the contemporary physics. At low temperature and low quark chemical potential, the QCD matter is formed by hadrons; while at high temperature and/or quark chemical potentials, we expect the QCD matter where quarks and gluons are liberated from the confinement. The ``critical temperature" of the transition from the hadronic matter to the quark-gluon matter is expected to be at the order of the QCD confinement scale, $\L_{\rm QCD}\sim 200$ MeV, at zero quark chemical potential. The unique terrestrial experiments that can achieve such an extremely high temperature are the relativistic heavy-ion collisions which have been carried out in the Relativistic Heavy Ion Collider (RHIC) at Brookhaven National Laboratory (BNL) and in the Large Hadron Collider (LHC) at CERN. The experimental data produced in RHIC and LHC have shown many features that support the formation of the hot quark-gluon plasma (QGP) and have also revealed a number of interesting properties of the QGP.

In noncentral heavy-ion collisions, strong magnetic fields can also be created~
\cite{Rafelski:1975rf,Voskresensky:1980nk,Schramm:1991ex,Schramm:1991vu}. This is because that the two colliding nuclei generate two electric currents in opposite directions and thus produce a magnetic field perpendicular to the reaction plane (the plane defined by the impact parameter and the beam direction). The typical strength of the magnetic fields can be roughly estimated by using the Biot-Savart formula, $eB\sim \g \a_{\rm EM} Z/ R_A^2$, where $\g$ is the Lorentz gamma factor associated with the moving nuclei, $\a_{\rm EM}$ is the fine structure constant, $Z$ is the charge number of the nucleus, and $R_A$ is the radius of the nucleus. Thus one finds that the magnetic field in RHIC Au + Au collisions at $\sqrt{s}=200$ GeV can be as large as $10^{19}$ Gauss and in LHC Pb + Pb collisions at $\sqrt{s}=2.76$ TeV can reach the order of $10^{20}$ Gauss. More detailed numerical investigations indeed confirmed the generation of strong magnetic fields in heavy-ion collisions and also revealed many extreme aspects of the magnetic fields~\cite{Kharzeev:2007jp,Skokov:2009qp,Voronyuk:2011jd,Bzdak:2011yy,Ou:2011fm,Deng:2012pc,Bloczynski:2012en,Bloczynski:2013mca,Zhong:2014cda,Zhong:2014sua,Li:2016tel,Holliday:2016lbx}. As the strength of the magnetic field is comparable to the QCD scale, $eB\sim \L_{\rm QCD}^2\sim 10^{18}$ Gauss, we expect that the magnetic-field induced effects may compose important ingredients in exploring the QCD physics in heavy-ion collisions. Indeed, recently, a number of novel quantum phenomena induced by strong magnetic fields in heavy-ion collisions have been found and attracted a lot of attentions in the high-energy and nuclear physics community. In this article, we will give a short review of the recent progress on the study of these novel quantum phenomena. We will focus on the following subjects: chiral magnetic effect~\cite{Kharzeev:2007jp,Fukushima:2008xe}, chiral separation effect~\cite{Son:2004tq,Metlitski:2005pr}, chiral magnetic wave~\cite{Kharzeev:2010gd,Gorbar:2011ya,Burnier:2011bf}, magnetic-field effects
on the heavy-flavor mesons~\cite{Marasinghe:2011bt,Yang:2011cz,Machado:2013rta,Machado:2013yaa,Alford:2013jva,
Liu:2014ixa,Cho:2014exa,Cho:2014loa,Gubler:2015qok,Bonati:2014ksa,Bonati:2015dka,Bonati:2016kxj,
Rougemont:2014efa,Dudal:2014jfa,Guo:2015nsa,Suzuki:2016kcs,Yoshida:2016xgm}
and the heavy-quark transport \cite{Fukushima:2015wck,Finazzo:2016mhm,Das:2016cwd}.
We will also discuss the experimental implications of these phenomena.

In addition to the above mentioned subjects, strong magnetic fields may also drive a number of other phenomena which will not be discussed in this article. These phenomena include the magnetic catalysis of chiral symmetry breaking~\cite{Gusynin:1994xp,Gusynin:1994re,Gusynin:1995nb,Shovkovy:2012zn}, the inverse magnetic catalysis or magnetic inhibition at finite temperature and density~\cite{
Preis:2010cq,Bali:2011qj,Bali:2012zg,Bruckmann:2013oba,Fukushima:2012xw,Fukushima:2012kc,Kojo:2012js,Chao:2013qpa,Yu:2014sla,Feng:2014bpa,Yu:2014xoa,Cao:2014uva,Ferrer:2014qka,Braun:2014fua,Mueller:2015fka,Hattori:2015aki,Ruggieri:2016lrn},
the rotating system under a strong magnetic field \cite{Chen:2015hfc,Hattori:2016njk,Ebihara:2016fwa},
the possible $\r$ meson condensation in strong magnetic field~\cite{Chernodub:2010qx,Chernodub:2011mc,Hidaka:2012mz,Liu:2014uwa,Liu:2015pna}, the neutral pion condensation in vaccum~\cite{Cao:2015cka}, the magnetic-field induced anisotropic viscosities in hydrodynamic equations~\cite{Braginskii:1965,Lifshitz:1981,
Huang:2009ue,Huang:2011dc,Tuchin:2011jw,Finazzo:2014cna},
and the magnetic-field induced particle production~\cite{Tuchin:2010vs,Tuchin:2010gx,Tuchin:2012mf,Tuchin:2013ie,Tuchin:2014nda,Tuchin:2014pka,Basar:2012bp, Fukushima:2012fg, Hattori:2013cra, Ayala:2016lvs, Zakharov:2016mmc}.
Some of these intriguing effects are reviewed in Refs.~\cite{Kharzeev:2013jha,Andersen:2014xxa,Miransky:2015ava}.

In \sect{sec:fields}, we briefly review some features of the magnetic fields in heavy-ion collisions. In \sect{sec:anoma}, we discuss the chiral magnetic effect, chiral separation effect, and chiral magnetic wave, and the current status of the experimental search of these anomalous transport phenomena. In \sect{sec:HQ}, we discuss how the magnetic fields
modify the properties of the heavy-flavor mesons and the heavy-quark transport in a hot medium.
Phenomenological implications are also discussed. We summarize in \sect{sec:summ}.

\section {Magnetic fields in heavy-ion collisions}\label{sec:fields}
In relativistic heavy-ion collisions, due to the fast motion of the ions which carry positive charges, large magnetic fields can be generated in the reaction zones.
In this section, we briefly review some general properties of such-generated magnetic fields
by both numerical and analytic models.
More detailed information of the numerical study can be found in the original references and in Ref.~\cite{Huang:2015oca}.

\subsection {Numerical approach}\label{sec:fields:num}
\begin{figure}[!t]
\begin{center}
\includegraphics[width=7.0cm]{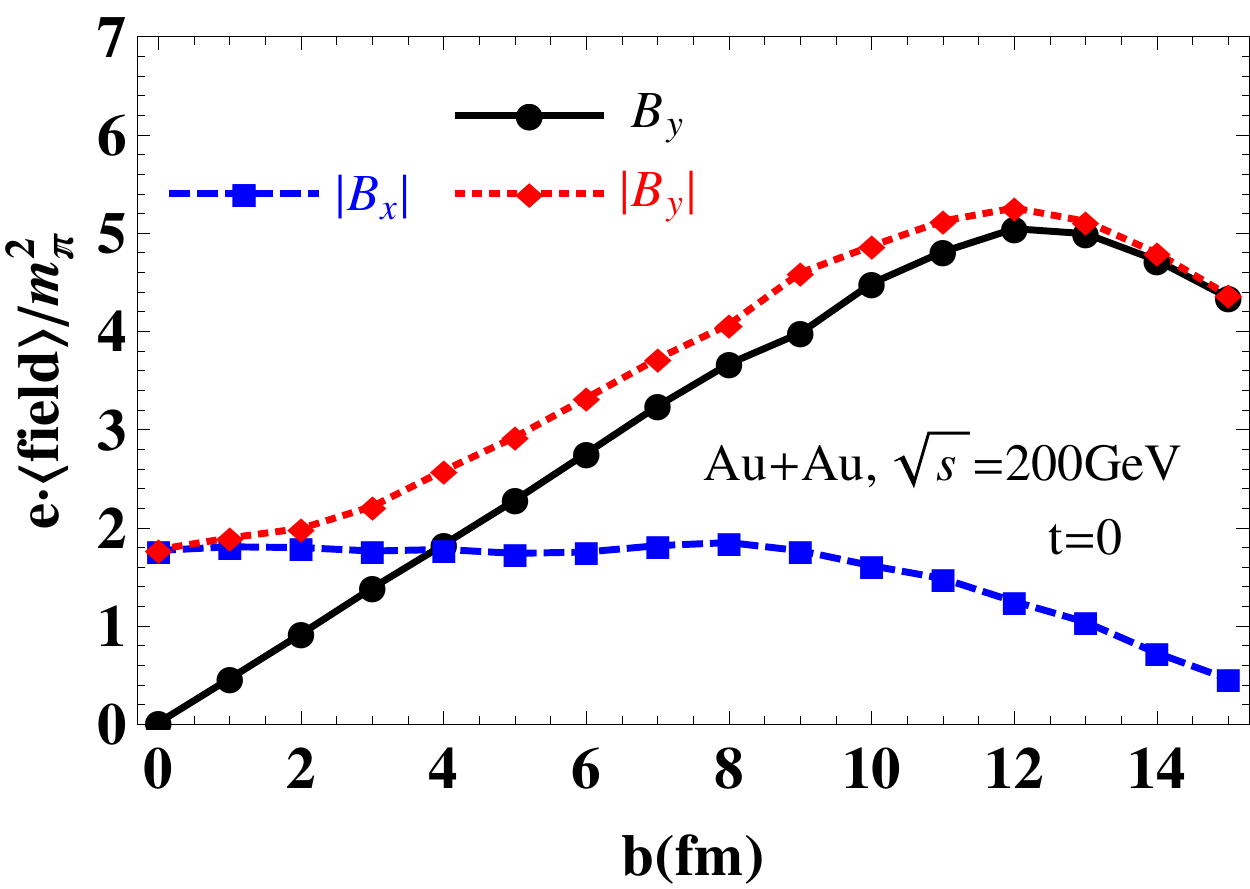}
\includegraphics[width=7.0cm]{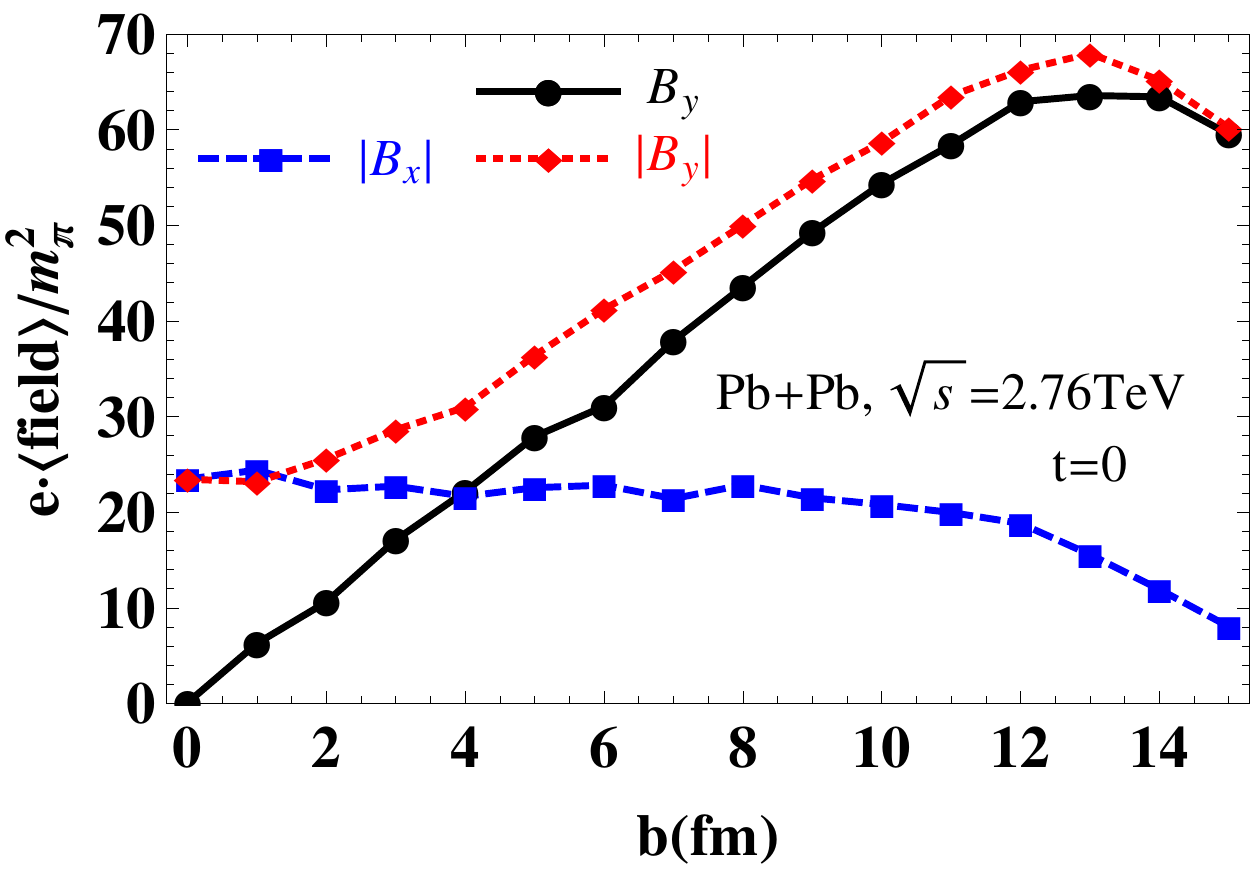}
\caption{The magnetic fields at $t=0$ and $\br={\bf 0}$
as functions of the impact parameter $b$ where $\lan \cdots\ran$ denotes average over events. (Reproduced from Ref.~\cite{Deng:2012pc}.)}
\label{bdepe}
\end{center}
\end{figure}

A number of numerical simulations have revealed that the magnitudes of the magnetic fields in heavy-ion collisions can be very large~\cite{Kharzeev:2007jp,Skokov:2009qp,Voronyuk:2011jd,Bzdak:2011yy,Ou:2011fm,Deng:2012pc,Bloczynski:2012en,Bloczynski:2013mca,Zhong:2014cda,Zhong:2014sua,Li:2016tel,Holliday:2016lbx}. In \fig{bdepe}, the different spatial components of the magnetic fields at $\br={\bf 0}$ (i.e., the center of the overlapping region) and $t=0$ (i.e., the time when the two colliding nuclei completely overlap) are presented as functions of the impact parameter $b$. The fields are calculated by using the Lienard-Wiechert potential on the event-by-event basis where the events are generated by the HIJING event generater~\cite{Deng:2012pc}. Please find the discussions on the quantum correction to the otherwise classical Leinard-Wiechert potential in Refs.~\cite{Huang:2015oca,Holliday:2016lbx}. The most important information from \fig{bdepe} are three folds.

First, it shows that heavy-ion collision experiments at RHIC and LHC generate very strong magnetic fields. The event-averaged magnetic fields at Au + Au collisions at $\sqrt{s}=200$ GeV can reach about $10^{19}$ Gauss and at Pb + Pb collisions at $\sqrt{s}=2.76$ TeV can reach about $10^{20}$ Gauss. They are much stronger than the magnetic fields of neutron stars including the magnetars which may have surface magnetic fields of the order of $10^{14}-10^{15}$ Gauss~\cite{Olausen:2013bpa,Turolla:2015mwa}. They are also much larger than the masses squared of electron, $m_e^2$, and light quarks, $m_u^2, m_d^2$, and thus are capable of inducing significant quantum phenomena.

Second, although we do see that only the $y$-component ($y$ axis is set to be perpendicular to the reaction plane) of the magnetic fields remains after the event average owing to the left-right symmetry of the noncentral collisions, we can find that the event-by-event fluctuation of the nuclear distribution results in nonzero $x$-component of the magnetic fields as well. These are reflected in the averaged absolute values of the fields in \fig{bdepe} and is most evident for central collisions~\cite{Bzdak:2011yy,Deng:2012pc}.

Third, the fields in LHC Pb + Pb at $2.76$ TeV are roughly $13.8=2.76/0.2$ times larger than that in RHIC Au + Au collisions at $200$ GeV. In fact, as investigated in Refs.~\cite{Bzdak:2011yy,Deng:2012pc}, to high precision, the magnitudes of the magnetic fields linearly depend on the collision energy $\sqrt{s}$. This can be understood from the fact that the fields are proportional to the Lorentz gamma factor of the beam motion which is equal to $\sqrt{s}/(2m_N)$ where $m_N$ is the mass of a nucleon.

The simulations depicted in \fig{bdepe} are only for fields at $t=0$. As the hot medium generated in heavy-ion collisions is expanding fast, then a natural question is that: how do the magnetic fields evolve in time in the expanding medium? This is still an unsolved problem. We summarize the recent progress briefly here and in next subsection.

\begin{figure*}[!t]
\begin{center}
\includegraphics[height=4.5cm]{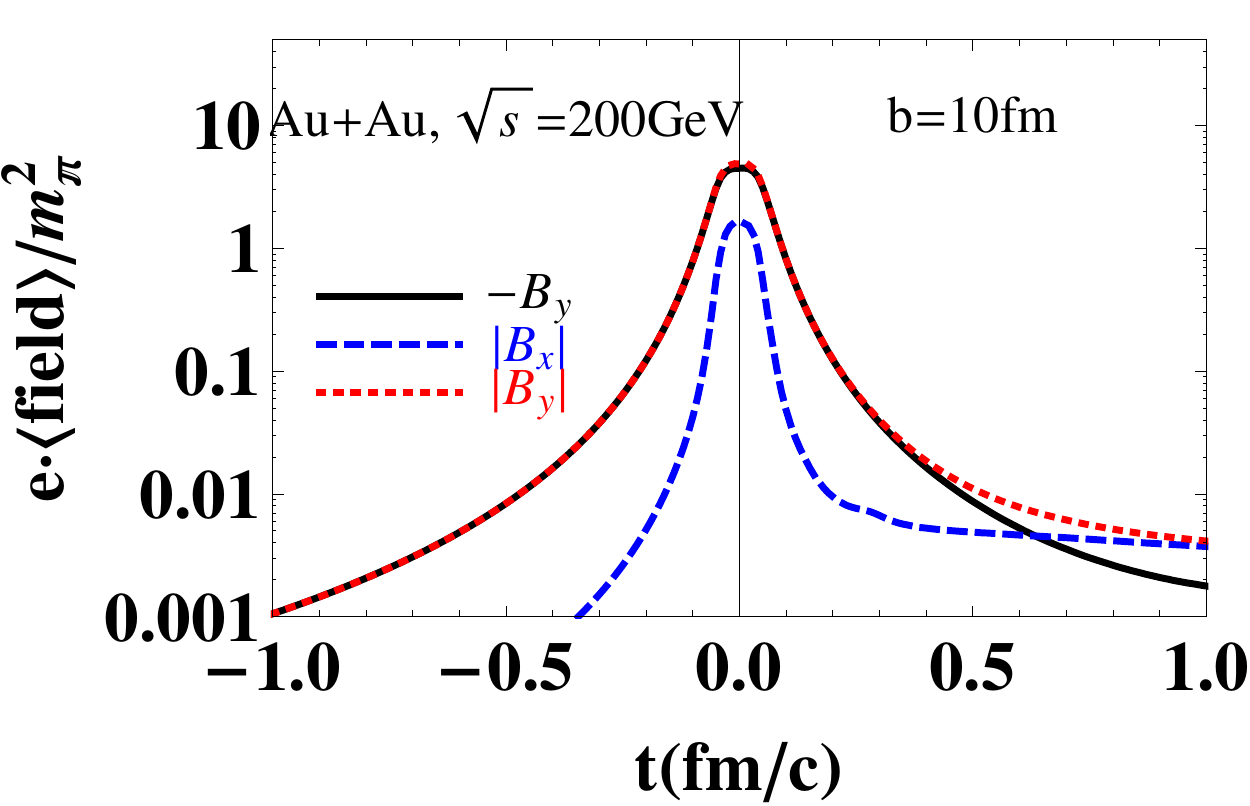}
\includegraphics[height=4.5cm]{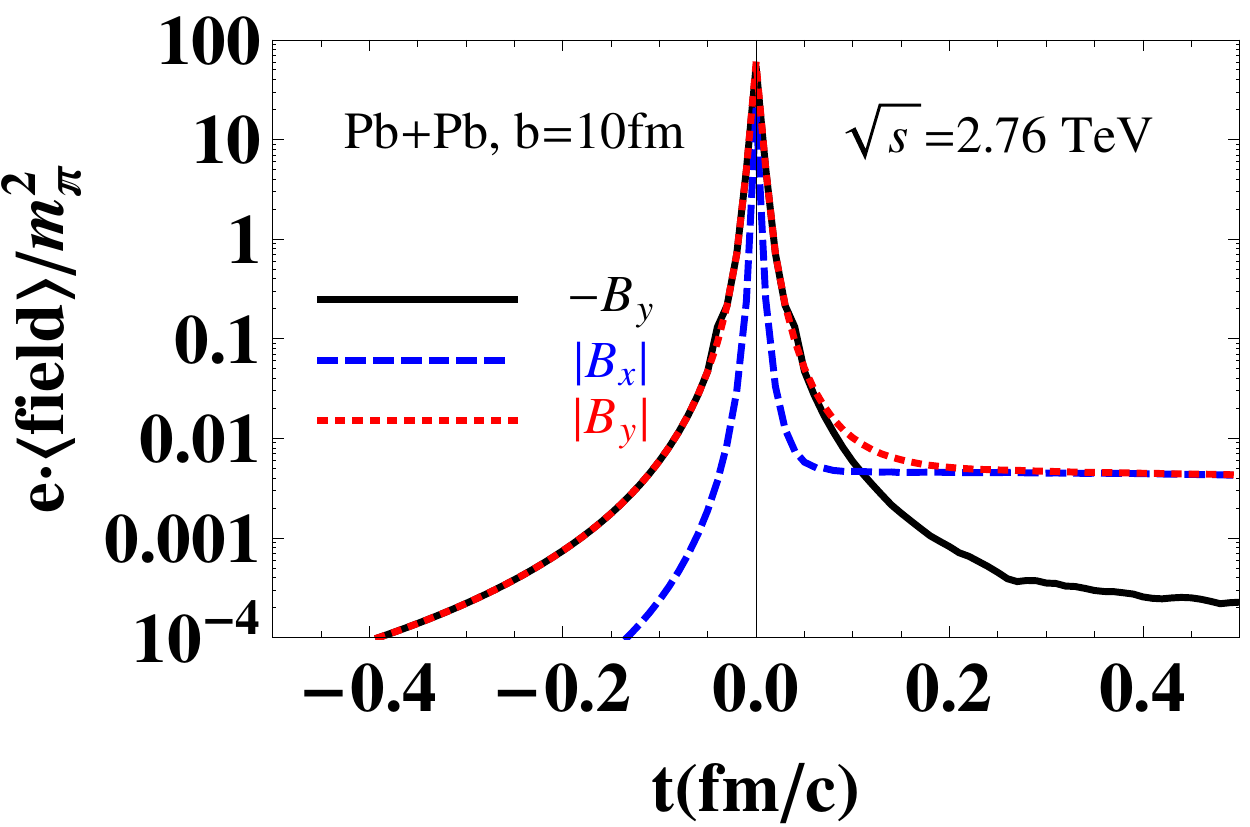}
\caption{The time evolution of the magnetic fields at $\br=0$
with impact parameter $b=10$ for Au + Au collision at $\sqrt{s}=200$ GeV and
Pb + Pb collision at $\sqrt{s}=2.76$ TeV. The effect due to the finite conductivity of quark-gluon matter is not considered. (Reproduced from Ref.~\cite{Deng:2012pc}.)}
\label{tdepe}
\end{center}
\end{figure*}
As we explained, the strong initial magnetic fields are consequences of the smallness of the system and the relativistically fast motion of the colliding nuclei. The main contributors to the initial magnetic fields are the spectator protons (which do not participate into the collision) in the colliding nuclei and they leave the collision zone very fast in a time scale characterized by the Lorentz-suppressed radius of the nucleus, i.e., $\t_B=R_A/\g  \approx 2m_N R_A/\sqrt{s}$. Thus in relativistic heavy-ion collisions, one can expect that the magnetic fields generated by the spectators will decay very fast after a short time scale $\t_B$. This fact is indeed verified by numerical simulations. In \fig{tdepe}, one can find the numerical results of the time evolutions of the magnetic fields at $\br={\bf0}$ in collisions with $b=10$ fm for Au + Au collision at $\sqrt{s}=200$ GeV and for Pb + Pb collision at $\sqrt{s}=2.76$ TeV~\cite{Deng:2012pc}. We can see that the magnetic fields decay very fast after the collision reflecting the fact that the spectators are leaving the collision region very fast. Once the spectators are all far away from the collision region, the remnant charge carriers which participate into the collisions and thus moves much slower than the spectators along the beam direction become the dominant sources for the magnetic fields and they essentially slow down the decay of the magnetic fields.

\subsection {Analytical approach}\label{sec:fields:ana}
To understand the basic picture of the creation of the magnetic fields,
we compared the numerical results shown in the previous section
with a simple analytical model proposed in Ref.~\cite{Liu:2014ixa}.
With the relativistic beam energy, the momenta of the colliding nuclei are so large
that the nuclear stopping power is not really strong enough to stop them:
the incident nuclei pass each other and recede from the collision point
with almost the same rapidity as the incident beam rapidity.
Therefore, in this simple model, we assume that a magnetic field is created by protons
freely streaming with the beam rapidity, and thus do not distinguish the spectators and participants.
From the comparison between the numerical and analytic modeling,
we will also discuss effects of the nuclear stopping power.
\begin{figure}[t!]
\begin{minipage}{0.56\hsize}
 \begin{center}
   \includegraphics[clip,width=0.9\columnwidth]{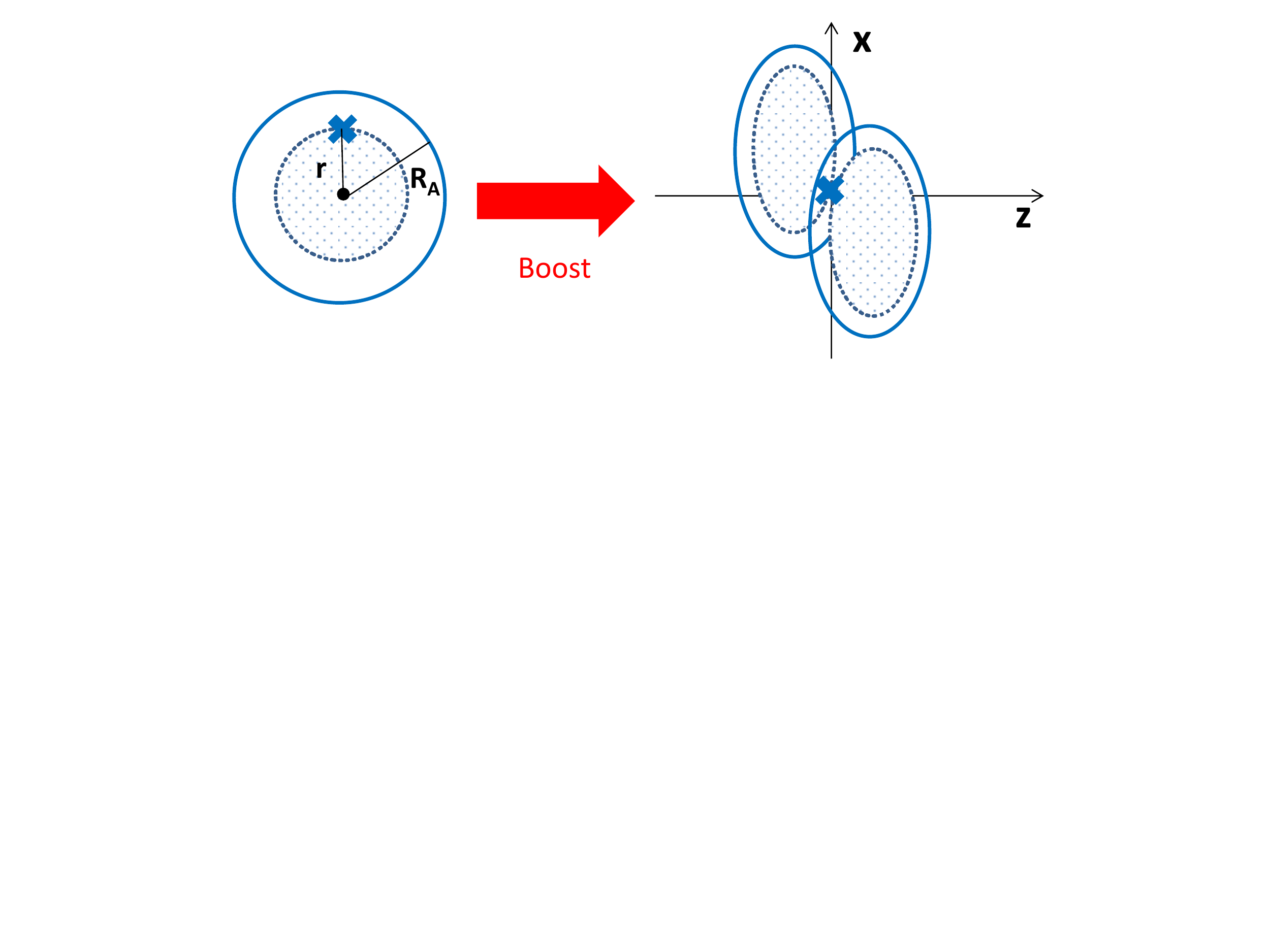}
    \end{center}
 \end{minipage}
\begin{minipage}{0.43\hsize}
 \begin{center}
   \includegraphics[clip,width=0.9\columnwidth]{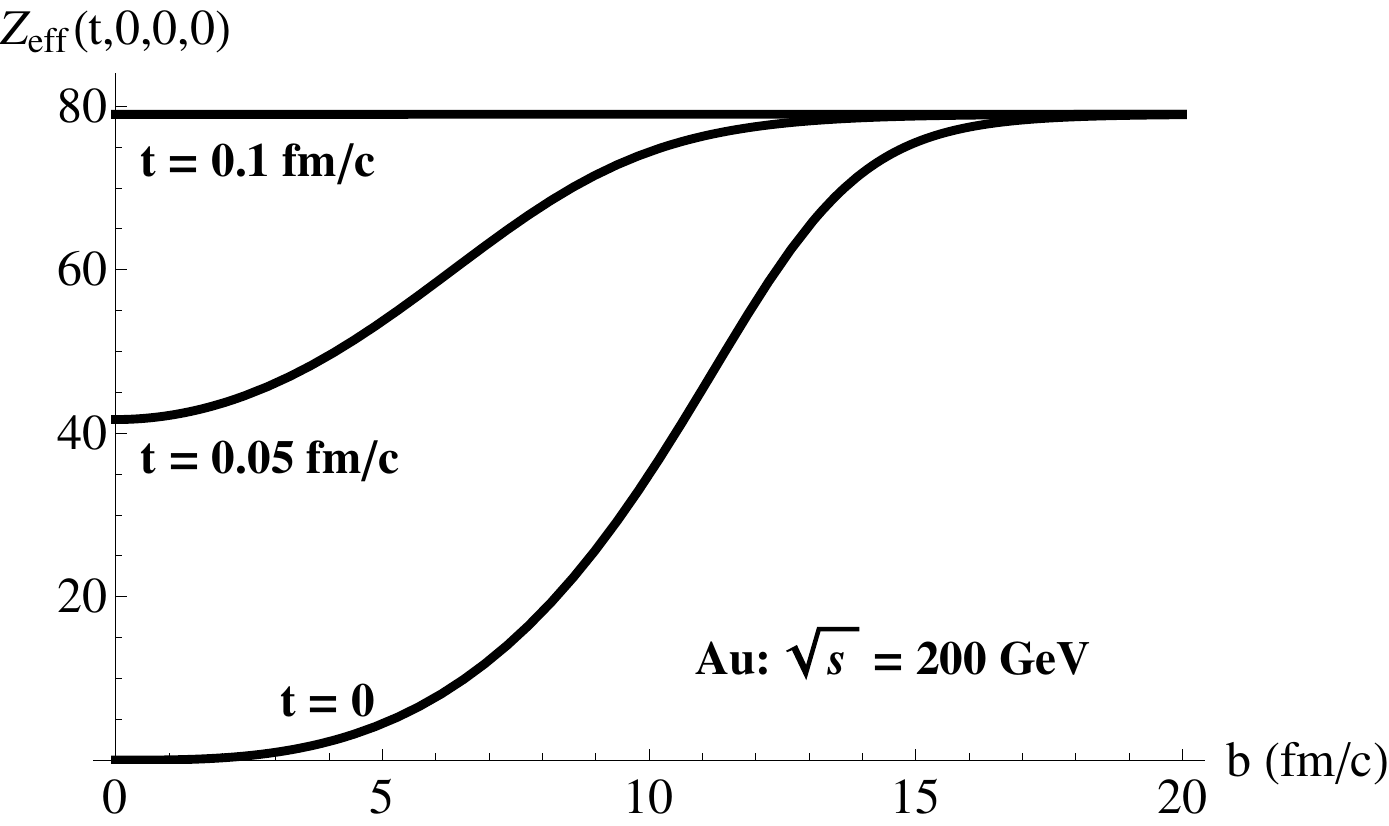}
    \end{center}
 \end{minipage}
\caption{Effective charges contributing to the creation of the magnetic field
in the rest and the center-of-mass (COM) frame of the collision (Left).
Impact parameter dependence of the effective charge of a nucleus at the origin of the COM coordinate (Right).}
\label{fig:Zeff}
\end{figure}

As we have specified the classical motion of charged particles,
the magnitude of the magnetic field can be estimated by using the Lienard-Wiechert potential
\begin{eqnarray}
e \bB^{\pm} =
\frac{ \alpha_{{}_{\rm EM}}  Z_{\rm eff}^\pm }{ (r^{\pm})^ 3 }
\sinh(\pm Y_{\rm beam})  (\tilde \bx^\pm \times {\bm e}_z)
\label{eq:LW}
\, ,
\end{eqnarray}
where $ Y_{\rm beam}$ is the absolute value of the beam rapidity.
The positive and negative signs correspond to the nuclei
moving in the positive and negative $z $-direction, respectively. The coordinate system is illustrated in \fig{fig:Zeff} (Left).
We have defined the coordinate vector  $ \tilde \bx  $
from the the center of the nucleus $(\pm b/2,0,v^\pm t) $ to the observer at $\bx $,
and its length in the rest frame
$ r^\pm = \sqrt{ (x \pm b/2 )^2 + y^2 + \gamma^2 (z-v^\pm t)^2 } $
with the impact parameter $ b$, the velocity of the nucleus $v^\pm = \pm \tanh Y_{\rm beam}  $,
and the gamma factor $ \gam = 1/\sqrt{1- (v^\pm)^2}$.

The Lienard-Wiechert potential in Eq.~(\ref{eq:LW}) is obtained
by boosting the electro-static potential in the rest frame of the nuclei (see Fig.~\ref{fig:Zeff}).
As well known, the net electro-static potential can be calculated
from the effective point-like charge sitting at the center of the sphere.
Therefore, the effective charge $ Z_{\rm eff}$ appearing in Eq.~(\ref{eq:LW}) should
be the total amount of the charges inside the sphere, i.e.,
\begin{eqnarray}
Z_{\rm eff}^\pm (t,\bx) = 4\pi \int_0^{r^{\pm}} \!\! dr^\prime \,  r^{\prime\, 2}
\rho_{\scriptscriptstyle } (r^\prime)
\label{eq:Zeff}
\, ,
\end{eqnarray}
where $ \rho_{\scriptscriptstyle } (r^\prime) $ is the charge distribution inside the nuclei.
Note that the upper boundary of the integral is given by the distance $ r^\pm$ from the center to the observer.
When the observer is in the exterior of the nucleus $ r^\pm \geq R_A$,
one simply has the total charge of the nucleus $Z_{\rm eff}(r^\pm>R_A) = Z $.
The impact parameter and time dependences of the effective charge is shown in
the right panel of Fig.~\ref{fig:Zeff}.
We find significant dependence on both the impact parameter and the time collapsed after the collision,
which will be important to reproduce the results from the numerical simulation shown in the previous section.

\begin{figure}[t!]
\begin{minipage}{0.48\hsize}
 \begin{center}
   \includegraphics[clip,width=\columnwidth]{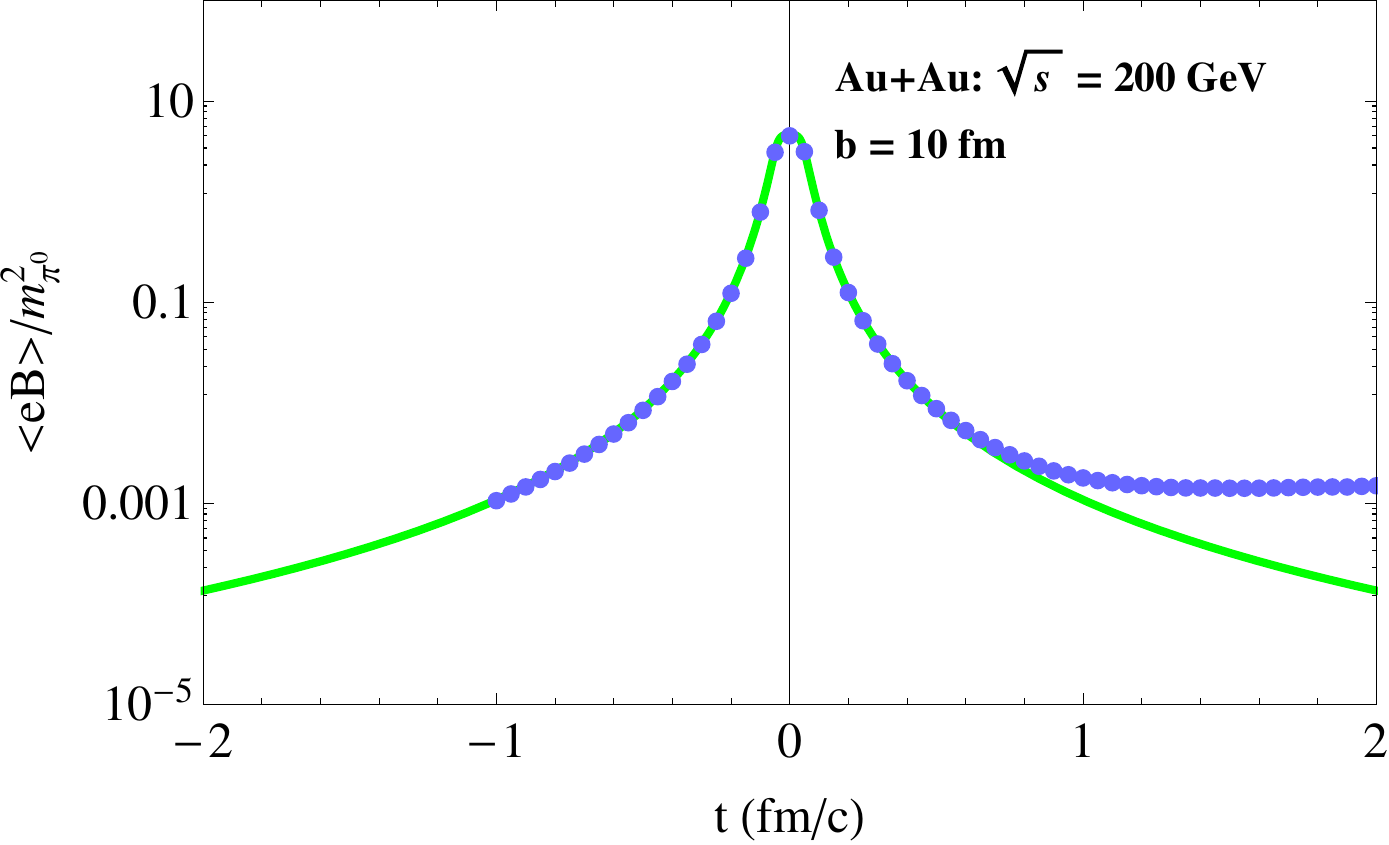}
    \end{center}
 \end{minipage}
\begin{minipage}{0.48\hsize}
 \begin{center}
   \includegraphics[clip,width=\columnwidth]{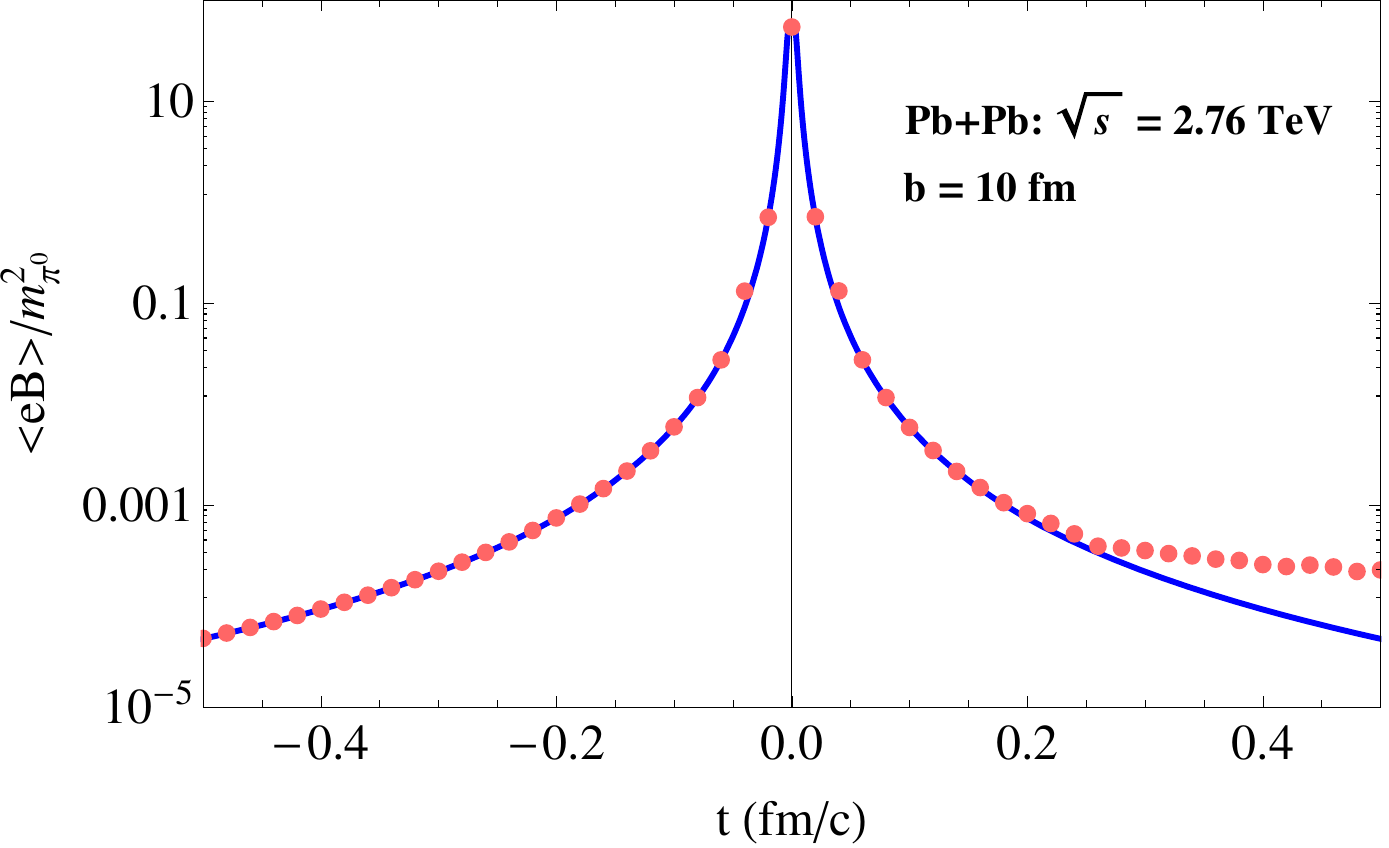}
    \end{center}
 \end{minipage}
 \caption{The analytic model (solid lines) reproduces well the time dependence of the magnetic field at  $\bx = 0 $
 from the numerical simulation (dots) shown in Fig.~\ref{tdepe}. The mismatches seen at late time are due to the lack of considering nuclear stopping power in the analytical modeling. }
 \label{fig:Bt}
\end{figure}
\begin{figure}[t]
\begin{minipage}{0.48\hsize}
 \begin{center}
   \includegraphics[clip,width=0.9\columnwidth]{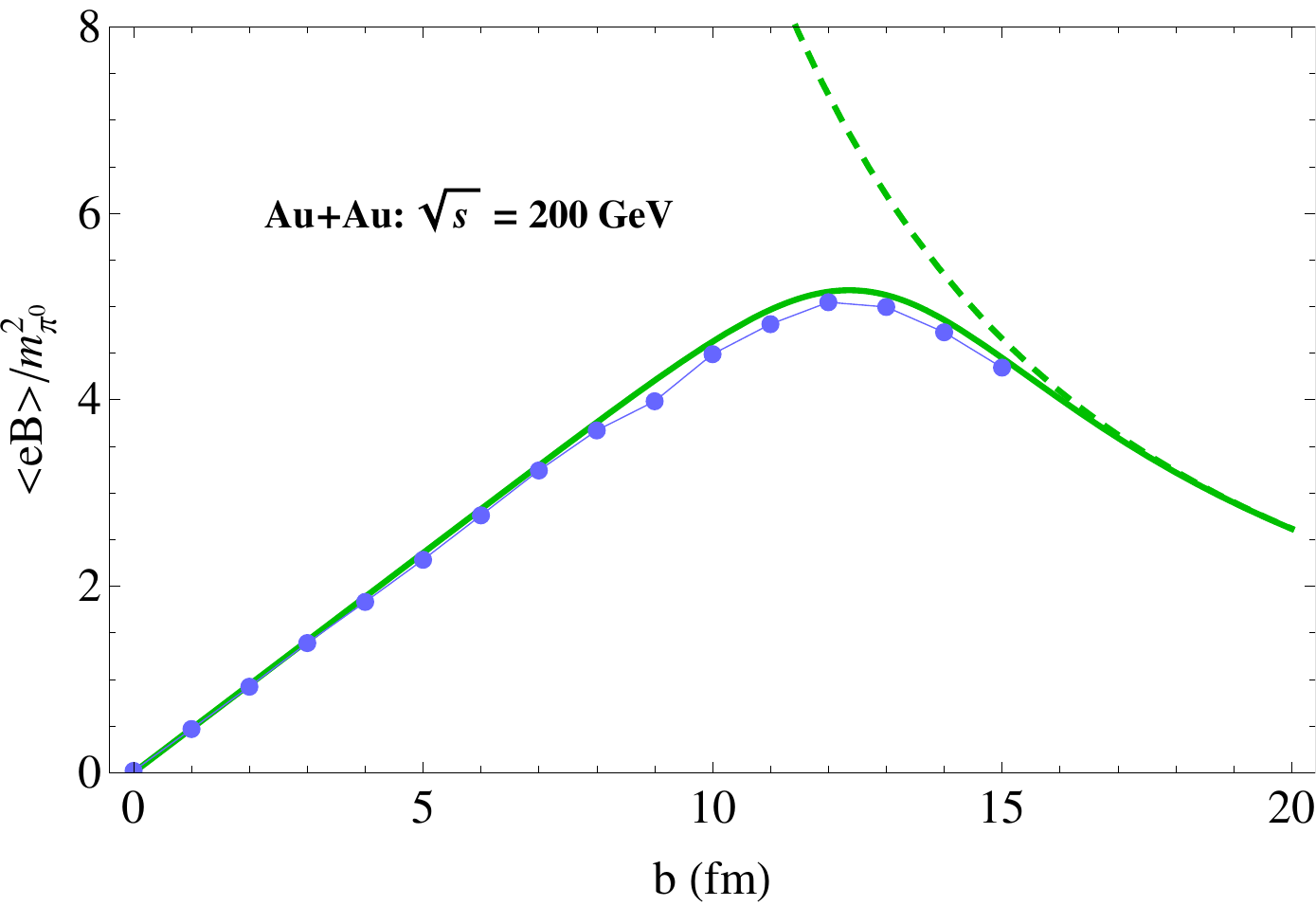}
    \end{center}
 \end{minipage}
\begin{minipage}{0.48\hsize}
 \begin{center}
   \includegraphics[clip,width=0.9\columnwidth]{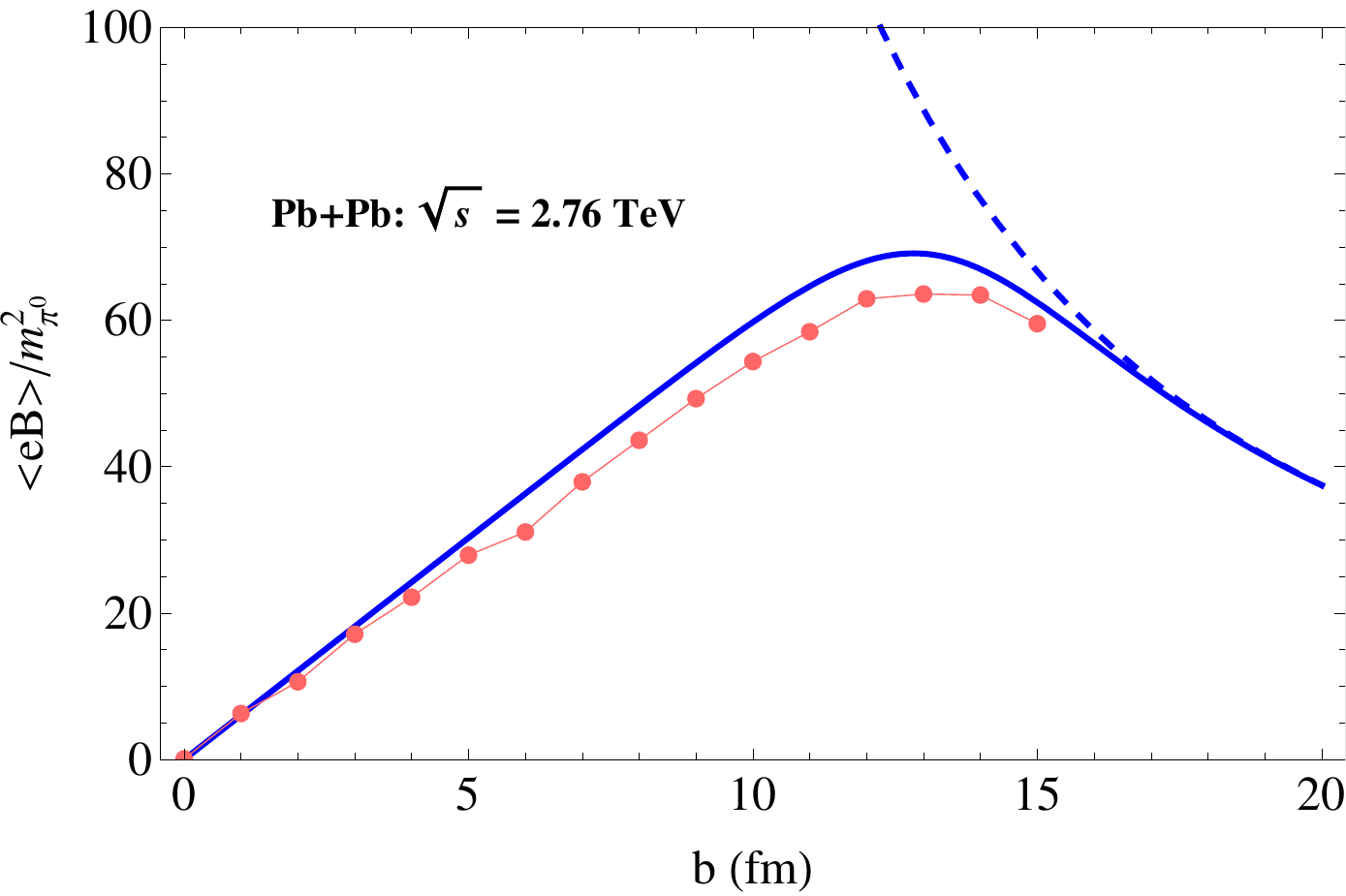}
    \end{center}
 \end{minipage}
\caption{The analytic model (solid lines) reproduces well the impact parameter dependence of
the event-averaged magnetic field at $ t=0$ and $\bx = 0 $ from the numerical simulation (dots) shown in Fig.~\ref{bdepe}.
The pointlike nucleus model (dashed lines) fails to reproduce the numerical results
when the nuclei have an overlap with each other, i.e., $b \lesssim 2R_A $.}
 \label{fig:Bb}
\end{figure}

\begin{figure}[t!]
\begin{minipage}{0.48\hsize}
 \begin{center}
   \includegraphics[clip,width=0.88\columnwidth]{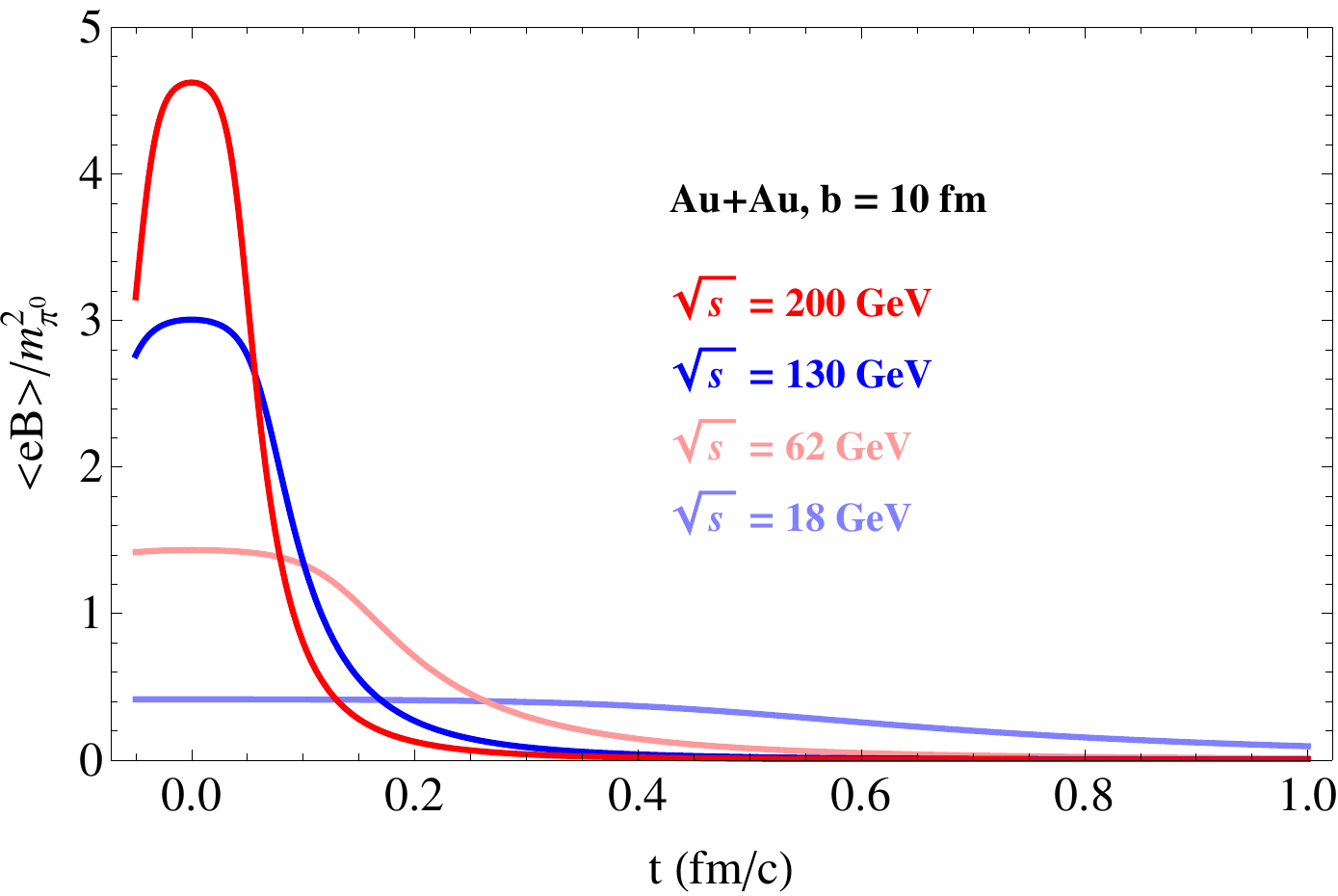}
    \end{center}
 \end{minipage}
\begin{minipage}{0.48\hsize}
 \begin{center}
   \includegraphics[clip,width=1\columnwidth]{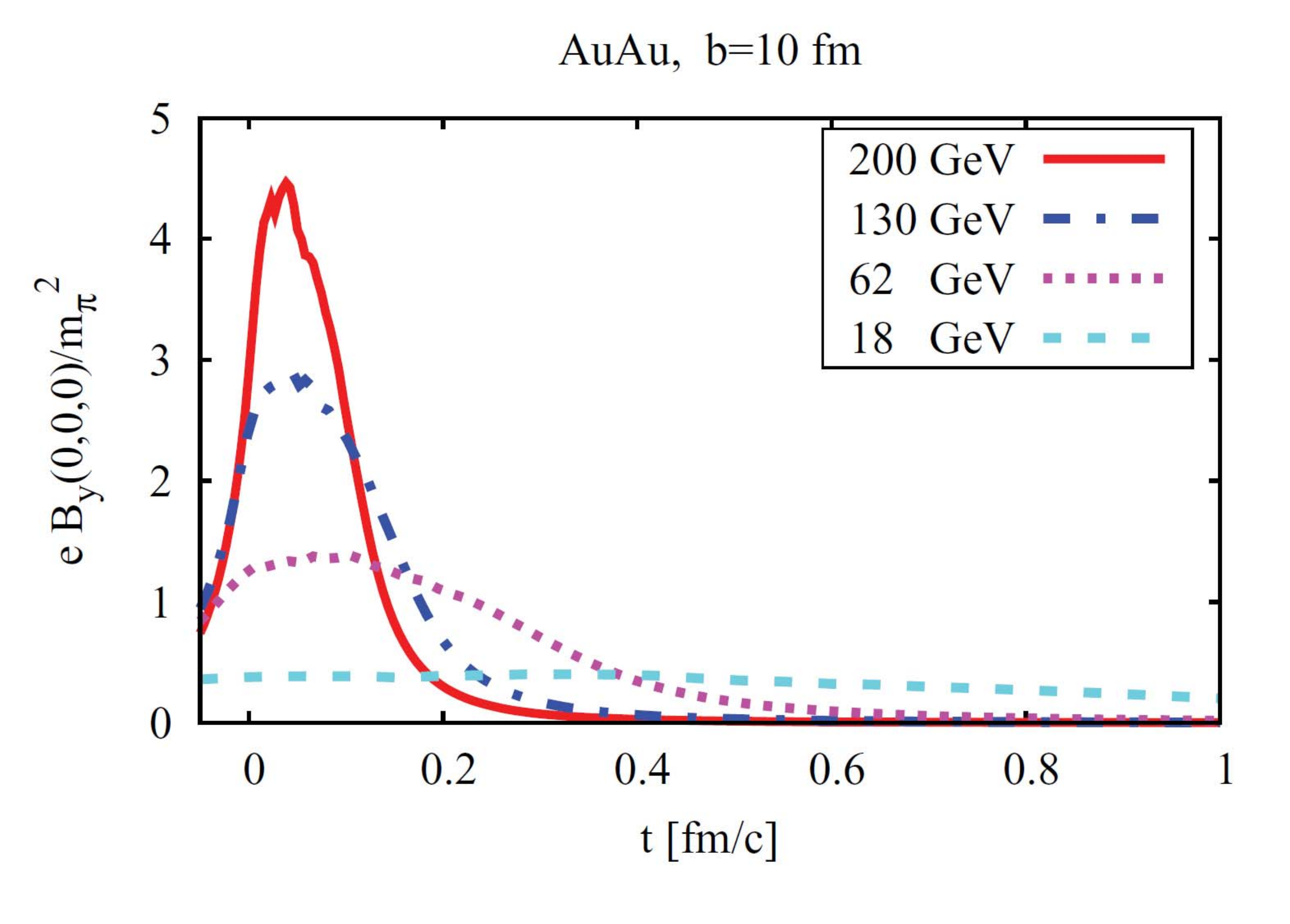}
    \end{center}
 \end{minipage}
     \caption{Beam-energy dependence of the magnetic field from the analytic model (left)
and the numerical simulation (right) taken from Ref.~\cite{Voronyuk:2011jd}.}
 \label{fig:Be}
\end{figure}
Plugging the effective charge (\ref{eq:Zeff}) to the Lienard-Wiechert potential (\ref{eq:LW}),
we obtain the magnitude of the magnetic field shown in Figs.~\ref{fig:Bt}--\ref{fig:Be}.
Here, we have taken the Woods-Saxon form for
the charge distribution $ \rho_{\scriptscriptstyle } (r) =  N_Z /\{ 1 + \exp[ (r - R_A)/a] \}$
with $N_Z = Z \{  4\pi \int_0^{R_A} \! dr^\prime  r^{\prime\, 2} \rho_{\scriptscriptstyle } (r^\prime) \}^{-1}$
where $ Z = 79$, $R_A = 6.38 $ fm, and $ a = 0.535$ fm for Au;
and $Z = 82 $, $R_A = 6.62 $ fm, and $a = 0.546$ fm for Pb. We do not consider the event-by-event fluctuation of the charge distribution in our analytical modeling so that we in Figs.~\ref{fig:Bt}--\ref{fig:Be} compare the analytical modeling with the numerical simulations of event-averaged magnetic fields.

Figure~\ref{fig:Bt} shows the time dependence of the magnetic field
obtained form the analytic model (solid lines) and the event-averaged numerical simulations (dots).
The latter has been already shown in Fig.~\ref{tdepe}.
We find a good agreement between the results from these two models
in the time range before and after the collision except for the late time.
The origin of the deviation in the late time is, as can be guessed from the asymmetric time dependence
in the numerical result with respect to the origin of time,
attributed to the effect of the nuclear stopping power.
Because of the deceleration by the nuclear stopping power,
participant nucleons (also called remnants) will stay in the collision region longer than in the free streaming case,
so that they will maintain the magnetic field for a longer time.

Figure~\ref{fig:Bb} shows the impact parameter dependence of the magnetic field.
Again, the sold lines and dots show the results from the analytic model
and the event-averaged numerical simulations, respectively,
and we find a good agreement between them.
The slight deviation may be originated from the energy loss
due to the nuclear stopping power in the numerical simulation.
In addition to these results, we show the magnetic field created by
point-like nuclei carrying the total charge $Z $, i.e., $Z_{\rm eff} (t,\bx) = Z $ in Eq.~(\ref{eq:LW}).
We find that this simplest model completely fails to reproduce
the numerical results when there is an overlap between the nuclei ($ b \lesssim 2R_A $),
indicating the importance of taking into account the effective charge (\ref{eq:Zeff}).

Finally, Fig.~\ref{fig:Be} shows the beam energy dependence of the magnetic field
from the analytic model (left panel) and the numerical simulation (right panel) \cite{Voronyuk:2011jd}.
While we find a good agreement, the nuclear stopping power will play an important role
in the low energy collisions such as those in the beam energy scan program in RHIC,
and in FAIR and J-PARC.

In the above discussions, we omit one important issue which may substantially modify the time evolution of the magnetic fields, that is, the finite electric conductivity $\s$ of the produced quark-gluon matter. If $\s$ is large, the Faraday induction effect may be strong as well and the lifetime of the magnetic fields can be significantly prolonged. As a matter of fact, the equilibrium QGP is a very good conductor according to the theoretical and lattice QCD studies~\cite{Arnold:2003zc,Gupta:2003zh,Aarts:2007wj,Ding:2010ga,Francis:2011bt,Ding:2014dua,Brandt:2012jc,Amato:2013naa,Aarts:2014nba,Ding:2016hua}; see also Ref.~\cite{Ding:2015ona} for review.

Now let us understand how the large $\s$ induces the strong Faraday induction and prolongs the lifetime of the magnetic fields; more discussions can be find in review~\cite{Huang:2015oca}. Let us write down the Maxwell's equations in the background of a charged flow characterized by velocity field $\bv$,
\begin{eqnarray}
\label{maxwell}
&\displaystyle\nabla\times\bE=-\frac{\pt\bB}{\pt t},&\\
\label{maxwell2}
&\displaystyle\nabla\times\bB=\frac{\pt\bE}{\pt t}+\bJ,&
\end{eqnarray}
where $\bJ$ is the electric current which according to the Ohm's law is determined by
\begin{eqnarray}
\label{ohm}
\bJ=\s\lb\bE+\bv\times\bB\rb.
\end{eqnarray}
Substituting $\bJ$ into \eq{maxwell2} and eliminating $\bE$ by using \eq{maxwell}, one can obtain an equation that governs the time evolution of the magnetic field
\begin{eqnarray}
\label{induce1}
&\displaystyle\frac{\pt\bB}{\pt t}=\nabla\times(\bv\times\bB)
+\frac{1}{\s}\lb\nabla^2\bB-\frac{\pt^2\bB}{\pt t^2}\rb,&
\end{eqnarray}
where we have regarded the QGP as locally neutral (this is a good assumption for high-energy collisions) so that $\nabla\cdot\bE=\r=0$.

The Faraday prolongation of the lifetime of the magnetic fields is most easily seen in two limits. The first limit is the no-flow limit, i.e., the limit when $\bv$ can be neglected. In this case, if the electric conductivity is so large that $\s\gg 1/t_c$ with $t_c$ the characteristic time scale over which the field strongly varies, one can neglect the second-order time derivative and render \eq{induce1} a diffusion equation~\cite{Tuchin:2010vs}:
\begin{eqnarray}
\displaystyle\frac{\pt\bB}{\pt t}=\frac{1}{\s}\nabla^2\bB.
\end{eqnarray}
Clearly, in this case, the decay of the magnetic field is due to diffusion, and the diffusion time is given by
\begin{eqnarray}
t_D=L^2\s,
\end{eqnarray}
where $L$ characterizes the length scale of the QGP system. For sufficiently large $\s$, the diffusion time $t_D$ could be larger than $\t_B$~\cite{Tuchin:2010vs}.

The second limit is the ideal magnetohydrodynamic limit, i.e., when $\s$ is very large and at the same time the flow velocity $\bv$ is also large so that we can neglect the term divided by $\s$ in \eq{induce1} and obtain
\begin{eqnarray}
\label{induce1new}
&\displaystyle\frac{\pt\bB}{\pt t}=\nabla\times(\bv\times\bB).&
\end{eqnarray}
This equation is known to lead to the so-called {\it frozen-in theorem}, i.e., the magnetic lines are frozen in the ideally conducting
plasma. Thus in this limit, the decay of the mangnetic fields is totaly due to the expansion of the QGP. For high-energy heavy-ion collisions, the longitudinal expansion is well described by the Bjorken flow, $v_z=z/t$, which makes the area of cross section of the overlapping region on the reaction plane grow linearly in time. Thus the frozen-in theorem leads us to the conclusion that in the ideal magnetohydrodynamic limit, before the transverse expansion play significant role, the magnetic fields will decay inversely proportional to time, $B_y(t)\propto (t_0/t) B_y(t_0)$, which is much slower than the $1/t^3$-type decay in the insulating case. The transverse expansion of the system can happen because of the transverse gradient of the fluid pressure. When this transverse expansion becomes fast, the decay of the fields will become fast as well, see discussions in Refs.~\cite{Deng:2012pc,Huang:2015oca}.

So far, we discussed two special limiting cases from which we can analytically extract the effect of Faraday induction. However, the realistic environment produced by heavy-ion collisions is not the case as described by the two limits. More over, the system is essentially non-equilibrium. Thus, the precise time evolution of the magnetic fields is really a hard task to achieve, and some recent progresses can be found in Refs.~\cite{Deng:2012pc,Tuchin:2010vs,Tuchin:2013ie,Tuchin:2013apa,Tuchin:2014iua,Tuchin:2014hza,Tuchin:2015oka,Gursoy:2014aka,Zakharov:2014dia,McLerran:2013hla,Li:2016tel}.

In the above discussions, we focus only on collisions of the same, spherical nuclei, i.e., the Au + Au collisions and Pb + Pb collisions. Heavy-ion program at RHIC also performed collisions of different nuclei, like Cu + Au collisions, as well as collisions of prolate nuclei like the U + U collisions. Of cause, these collisions can also generate strong magnetic fields; see for example, the studies in Refs~\cite{Hirono:2012rt,Deng:2014uja,Voronyuk:2014rna,Toneev:2016tgj,Bloczynski:2013mca,Chatterjee:2014sea} and recent review ~\cite{Huang:2015oca}. It is worth mentioning that, owing to the charge number asymmetry between Cu and Au nuclei, the Cu + Au collisions can generate a strong electric field pointing from Au nucleus to Cu nucleus~\cite{Hirono:2012rt,Deng:2014uja,Rybicki:2015kma,Voronyuk:2014rna,Toneev:2016tgj}. Such $\bE$ field may lead to a splitting between the directed flow $v_1$ of positively and negatively charged hadrons which has been recently test by STAR Collaboration at RHIC~\cite{Adamczyk:2016eux}.

\section {Magnetic-field induced anomalous transports in heavy-ion collisions}\label{sec:anoma}
The strong magnetic fields can induce a variety of novel quantum effects in the quark-gluon plasma which we will discuss in the following sections. In this section, we will focus on the ones that are deeply related to the chiral anomaly in QCD and QED, namely, the chiral magnetic effect (CME) and its relatives, the chiral separation effect, the chiral magnetic wave, etc. All these effects represent currents or collective modes that transport vector charges or axial charges in QGP which appear only at quantum level so that they do not have classical counterparts. The recent reviews of these anomalous transports can be found in Refs.~\cite{Kharzeev:2009fn,Kharzeev:2013ffa,Fukushima:2012vr,Kharzeev:2015kna,Liao:2014ava,Huang:2015oca,Kharzeev:2015znc}.

\subsection {The chiral magnetic effect}\label{sec:cme}
The CME represents the generation of vector current induced by an external magnetic field in chirality-imbalanced (P-odd) medium~\cite{Kharzeev:2007jp,Fukushima:2008xe}. The CME composes a very general class of anomalous transport phenomena spreading in a wide contexts in physics, ranging from astrophysics, condensed matter physics, nuclear physics, to particle physics. We apologize for not being able to mention all the relevant works. We refer the readers to other review articles as mentioned above for more information. The CME can be expressed as
\begin{eqnarray}
\label{cme}
\bJ_V &=&
\s_{\rm CME}
\bB,\\
\label{cmecond}\s_{\rm CME}&=&\frac{e^2}{2\p^2}\m_A,
\end{eqnarray}
where $e$ is the charge of the fermion, $\bJ_V$ is the vector current defined by $J^\m_V=e\lan\jb \g^\m\j\ran$, and $\m_A$ is the chiral chemical potential which parameterizes how large the chirality imbalance of the medium is. Note that, the above expression for CME current is for each species of massless fermions. For QGP at high temperature where the masses of up and down quarks can be neglected, the total CME conductivity is given by $\s_{\rm CME}=N_C\m_A\sum_f q_f^2/(2\p^2)$ with $q_f$ the charge of quark of flavor $f$ and $N_c=3$ the number of color.

Although the CME stems from the profound chiral anomaly in gauge theory, its physical origin can be understood in an intuitive way by adopting the Landau quantization picture. We know that the electrically charged fermions in a magnetic field occupy a tower of Landau levels (i.e., the energy spectrum of charged fermions in a magnetic field is discrete) in such a way that the lowest Landau level contains fermions with only one spin polarization (i.e., the spin of positively (negatively) charged fermions is polarized to be along (opposite to) the direction of the magnetic field) while higher Landau levels are occupied by equal numbers of up-polarized and down-polarized fermions. Now, let us consider the situation that the number of right-handed (RH) and the number of left-handed (LH) fermions are unequal, say, for example, $N_R>N_L$. Thus there are more RH fermions than LH fermions confined in the lowest Landau level which means that a current consist of positively charged fermions will flow along the magnetic field as RH fermions lock their momenta with their spins, while a current consist of negatively charged fermions will flow opposite to the magnetic field; the net effect is the appearance of an electric current along the magnetic field. This is the CME current. Note that the higher Landau levels do not contribute to the CME current as higher Landau levels are occupied symmetrically by spin-up and spin-down fermions and their contribution to the CME current cancel.

Now let us quantify the above intuitive argument to see how we can arrive at \eq{cme}. To simplify our life, we consider a system of massless Dirac fermions with only positive charge $e$ in an constant magnetic field $\bB$ pointing along $z$ direction which is supposed to be so strong that all the fermions are confined in the lowest Landau level. The number densities of RH and LH fermions are given by
\begin{eqnarray}
\label{density}
n_{R/L}\equiv\frac{d^3 N_{R/L}}{dxdydz}=\frac{eB}{2\p}\frac{p_F^{R/L}}{2\p},
\end{eqnarray}
where the prefactor $eB/(2\p)$ is the density of states of the lowest Landau level in the transverse directions and $p_F$ denotes the Fermi momentum (which is the momentum that separates the empty states and occupied states). Moreover, chiral fermions on the lowest Landau level move at the speed of light in a way correlated to their spins which tell us that the RH fermions generate a current along $z$ direction, $J_R=en_R$, while LH fermions generate a current along $-z$ direction, $J_L=-en_L$. Therefore the net vector current reads
\begin{eqnarray}
J_V=J_R+J_L=\frac{e^2B}{4\p^2}(p_F^{R}-p_F^L),
\end{eqnarray}
which, once we equate $p_F^{R/L}$ with $\m_{R/L}$ and $\m_A$ with $(\m_R-\m_L)/2$, leads us to \eq{cme} and \eq{cmecond}. Here we use the fact that for massless fermions, the chemical potential is equal to the Fermi momentum. 

We emphasize that although the above argument is for non-interacting system and relies on Landau quantization picture, the CME conductivity $\s_{\rm CME}=e^2\m_A/(2\p^2)$ is actually fixed by the chiral anomaly equation and universal in the sense that it does not involve concrete microscopic scattering among fermions. This is in parallel to the fact that the chiral anomaly equation itself is universal so that it does not receive perturbative correction from scattering between fermions.

Apparently, the occurrence of CME requires an environmental parity violation as being characterized by $\m_A$. A nonzero $\m_A$ may be induced by different mechanisms in different physical systems. For example, the electro-weak plasma can have a nonzero $\m_A$ as the weak interaction does not respect parity symmetry~\cite{Vilenkin:1980fu}. Recently, it is realized in condensed matter experiments the so-called Weyl or Dirac semimetals whose band structures permit level crossing around which the low energy excitations are Weyl or Dirac fermions. The CME was observed in these semimetals by applying parallel electric and magnetic fields which induce a difference between the chemical potentials of RH and LH fermions via the chiral anomaly~\cite{Li:2014bha,Xiong:2015nna,Shekhar:2015rqa,Huang:2015weyl}; see also discussions in Refs.~\cite{Miransky:2015ava,Huang:2015oca,Kharzeev:2015znc,Zubkov:2016tcp}. In the hot quark-gluon matter produced in heavy-ion collisions, $\m_A$ may appear due to the topological transition between the degenerate vacua of the gluonic sector of QCD characterized by different winding numbers~\cite{Kharzeev:2004ey,Kharzeev:2007jp,Iatrakis:2015fma}. The rate of such topological transition is exponentially suppressed at zero temperature as two neighbor vacua are separated by a finite energy barrier of the order of QCD confinement scale $\L_{\rm QCD}\sim 200$ MeV. However, at high temperature, the topological transition can be induced by another, classical, thermal excitation called sphaleron~\cite{Manton:1983nd,Klinkhamer:1984di} which, instead of tunneling through the barrier, can take the vacuum over the barrier at a large rate. At equilibrium, the probability of happening a topological transition that drives the winding number to increase is equal to the probability of happening an opposite topological transition that decreases the winding number. Thus, what really matters is the fluctuation of the topological transition above global equilibrium which results in local domains in the quark-gluon matter within which the gluonic configuration has a nonzero winding number. The quarks coupled with such gluons via the chiral anomaly can acquire nonzero chirality and thus a nonzero $\m_A$. Furthermore, $\m_A$ in heavy-ion collisions may also be induced by other mechanisms. For example, the parallel chromoelectric and chromomagnetic fields, $\bE^a\cdot\bB^a\neq 0$, in the initial glasma can also induce a nonzero $\mu_A$~\cite{Kharzeev:2001ev,Lappi:2006fp,Hirono:2014oda}.

\subsection {Experimental signals of the chiral magnetic effect}\label{sec:corrcme}
\begin{figure}[!htb]
\begin{center}
\includegraphics[width=7cm]{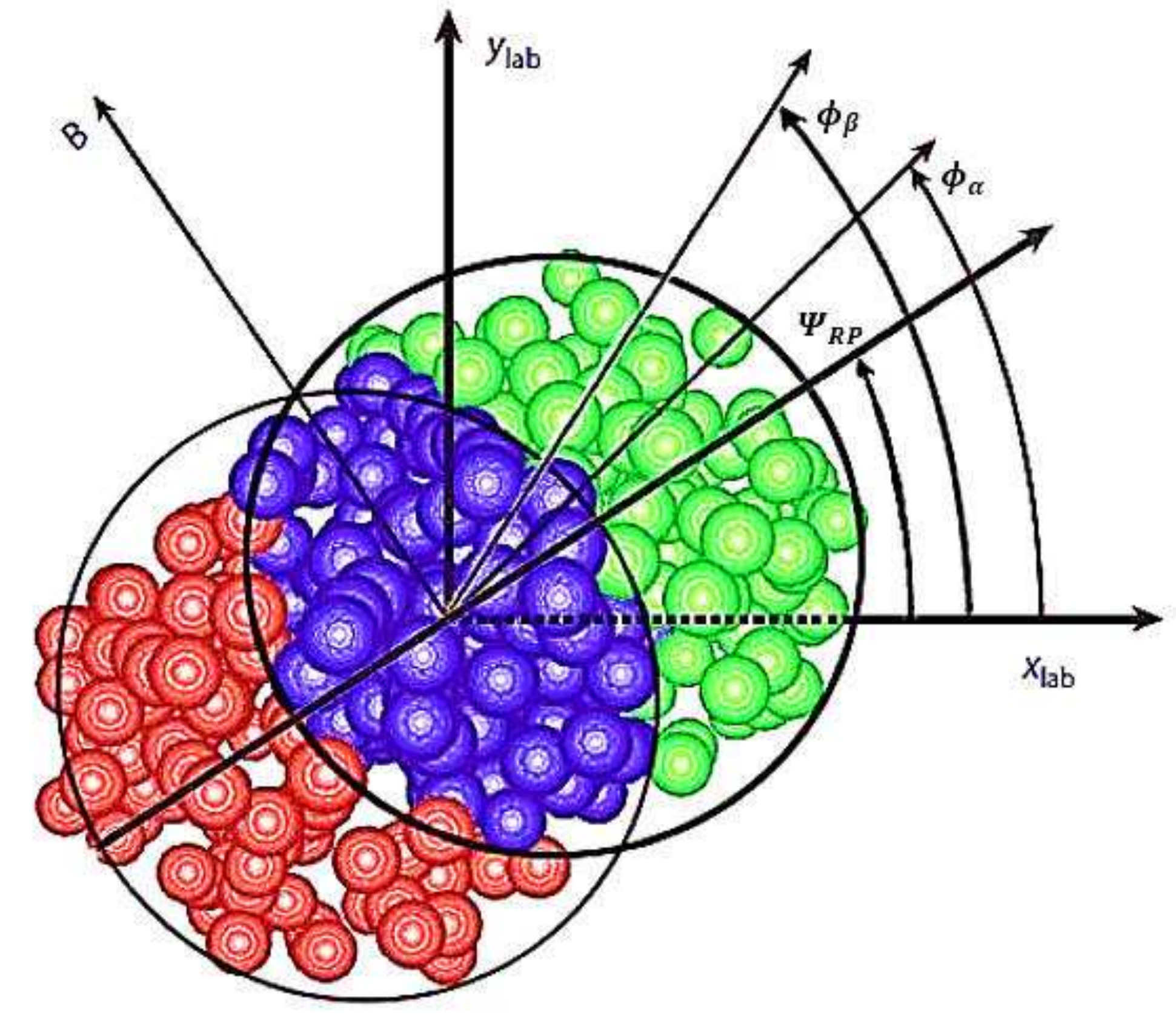}
\caption{Illustration of a typical noncentral collision.}
\label{illcme}
\end{center}
\end{figure}

If the CME occurs in the quark-gluon matter produced in the heavy-ion collisions, what will be its experimental signals? We have already seen that the magnetic fields in heavy-ion collisions are, on average over events, perpendicular to the reaction plane; thus the CME is expected to induce an electric current which on average is also perpendicular to the reaction plane. Such an electric current will drive a charge dipole in the fireball which, after the hydrodynamic evolution of the fireball, will be converted to a charge dipole in momentum space. However, as $\m_A$ can flip the sign from event to event, the charge dipole could also flip its direction from event to event, and thus the dipolar distribution in different events cancel each other after event average and cannot be measured directly. Instead, one should look for an observable which measures the dipolar fluctuation over events. Such an observable can be chosen as the following charge-dependent two-particle azimuthal correlation~\cite{Voloshin:2004vk,Kharzeev:2004ey}:
\begin{eqnarray}
\label{gamma}
\g_{\a\b}&\equiv&\lan\cos(\f_\a+\f_\b-2\J_\rp)\ran,
\end{eqnarray}
where the indices $\a$ and $\b$ denote the charge of the hadrons, $\f_\a$ and $\f_\b$ are the azimuthal angles of hadrons of charge $\a$ and $\b$, respectively, $\J_\rp$ is the reaction plane angle, and $\lan\cdots\ran$ denotes average over events; see \fig{illcme} for illustration. It is easy to understand that a charge separation with respect to the reaction plane will give a positive $\g_{+-}$ (and $\g_{-+}$) and a negative $\g_{++}$ or $\g_{--}$. We will refer to $\g_{+-}=\g_{-+}$ as $\g_{\rm OS}$ and $\g_{++}\approx \g_{--}$ as $\g_{\rm SS}$ where ``OS" stands for ``opposite-sign" and ``SS" stands for ``same-sign". In real experiments, the measurements are done with three-particle correlations where the third hadron (of arbitrary charge) is used to reconstruct the reaction plane.

\begin{figure}[!htb]
\begin{center}
\includegraphics[width=7cm,height=5cm]{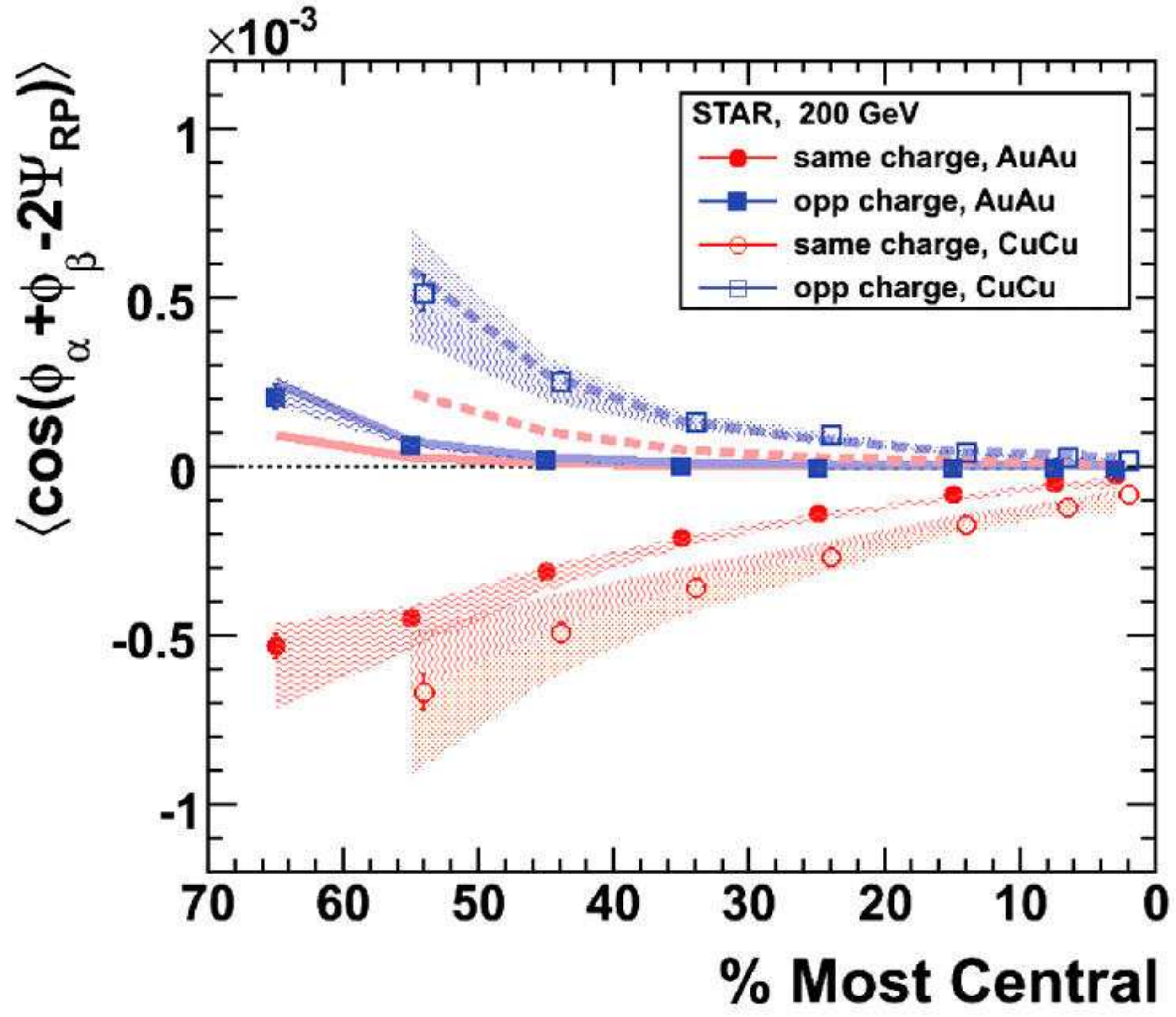}
\caption{The measured correlation $\g_{\a\b}$ by STAR Collaboration. This figure is from Ref.~\cite{Abelev:2009ac}.}
\label{starcme}
\end{center}
\end{figure}

The correlation $\g_{\a\b}$ was measured by STAR Collaboration at RHIC for Au + Au and Cu + Cu collisions at $\sqrt{s}=200$ GeV~\cite{Abelev:2009ac,Abelev:2009ad,Adamczyk:2013hsi} (see \fig{starcme}), by PHENIX Collaboration at RHIC for Au + Au collisions at $\sqrt{s}=200$ GeV~\cite{Ajitanand:2010rc},
and by ALICE Collaboration at LHC for for Pb + Pb collisions at $\sqrt{s}=2.76$ TeV~\cite{Abelev:2012pa}. At mid-central collisions, these measurements show clear positive opposite-sign correlation and negative same-sign correlation, as we expect from the picture of CME. Further more, the observed correlation increases from zero at central collisions to peripheral collisions, in consistence with what we learned from the simulation in last section that the averaged magnetic field increases with the centrality. More recently, the STAR Collaboration performed the measurement of $\g_{\a\b}$ at different beam energies~\cite{Wang:2012qs,Adamczyk:2014mzf} and observed that $\g_{\a\b}$ persists as long as the beam energy is larger than $19.6$ GeV; for further lowered beam energies, the difference between $\g_{\rm OS}$ and $\g_{\rm SS}$ steeply falls down, which may be understood by noticing that at lower energies the system is probably in a hadronic phase where the chiral symmetry is broken and the CME is strongly suppressed.

Although the experimental data supports the presence of the CME, the interpretation of the data is not unique owing to possible background effects
that are not related to the CME. First, the transverse momentum conservation (TMC) can contribute to $\g_{\a\b}$~\cite{Pratt:2010gy,Pratt:2010zn,Bzdak:2010fd} . The momentum conservation enforces the sum of the momenta projected to the transverse plane (the plane that is perpendicular to the beam direction) of all the particles to be zero at any time during the evolution of the system. Thus, the detection of a hadron of transverse momentum $\bp_\perp$ requires the presence of other produced hadrons whose transverse momenta must be summed to $-\bp_\perp$. This enforces an intrinsic back-to-back two-particle correlation:
\begin{eqnarray}
\g^{\rm TMC}_{\rm SS}&=&\g^{\rm TMC}_{\rm OS}=\lan\frac{\sum_{i\neq j}\cos(\f_i+\f_j-2\J_\rp)}{\sum_{i\neq j}}\ran\non
&\approx&-\frac{1}{N}\lan\cos(2\f_i-2\J_\rp)\ran=-\frac{v_2}{N},
\end{eqnarray}
where we used that $\sum_{i=1}^N\cos(\f_i-\J_\rp)=\sum_{i=1}^N\sin(\f_i-\J_\rp)=0$ and $N$ is the total multiplicity. From the above relation, we can estimate how large the TMC contribution is. For example, let us consider RHIC Au + Au at 200 GeV, $v_2\sim 0.1$ while $N\sim 1000$ which gives that $\g^{\rm TMC}_{\rm SS}=\g^{\rm TMC}_{\rm OS}\sim-10^{-4}$ which is roughly at the same order as the experimental result shown in \fig{starcme}. Obviously, the TMC is charge blind. Thus we can subtract the contribution from TMC by making a difference between opposite-sign and same-sign correlations:
\begin{eqnarray}
\D\g&\equiv&\g_{\rm OS}-\g_{\rm SS}.
\end{eqnarray}

However, there are other background effects contributing to $\D\g$. The most notable one may be the local charge conservation (LCC)~\cite{Pratt:2010gy,Schlichting:2010na,Schlichting:2010qia} which assumes that the pairs of opposite charges are created locally (i.e., within a small volumn) in space at the late stage of the heavy-ion collisions. Then due to the elliptic flow, opposite-sign pairs are strongly correlated while the same-sign pairs are not correlated; thus LCC gives a finite contribution to $\D\g$. The detailed analysis reveals that the LCC contribute to $\D\g$ the amount of~\cite{Huang:2015oca}
\begin{eqnarray}
\D\g^{\rm LCC}&\approx&\frac{M}{N}v_2,
\end{eqnarray}
where $M$ is the average number of hadrons in a typical local neutral cell. As $M$ is expected to be not very large, so we expect the LCC contribution is at the order of $10^{-4}$ or even larger in, e.g., RHIC Au + Au collisions at 200 GeV. This is comparable to the experimental results in \fig{starcme}.

The background effects mask the possible CME signals. To disentangle the possible CME signal and the flow-related backgrounds, one can setup experiments to either vary the backgrounds with the signal fixed or vary the signal with the backgrounds fixed (It is also important to setup delicate analysis methods to extract the elliptic flow on the event-by-event basis~\cite{Wen:2016zic,Wang:2016iov}). The former approach can be carried out by using the prolate shape of the uranium nuclei~\cite{Voloshin:2010ut}. In central U + U collisions, there would be sizable $v_2$ due to the prolate shape of the colliding nuclei while the magnetic field would be very tiny~\cite{Bloczynski:2013mca}. The STAR Collaboration collected $0-1\%$ most central events from U + U collisions at $\sqrt{s}=197$ GeV in 2012, and indeed found sizable $v_2$ while $\D\g$ is consistent with zero~\cite{Wang:2012qs}. However, it was found later that the total multiplicity is far less correlated to the number of binary collisions than expected~\cite{Adamczyk:2015obl}, so that the shape selection in experiments is very hard; see Refs.~\cite{Chatterjee:2014sea,Shou:2014cua} for recent discussions. The latter approach was proposed to be achieved by using the collisions of the isobaric nuclei, such as $^{96}_{44}$Ru and $^{96}_{40}$Zr~\cite{Voloshin:2010ut,Deng:2016knn,Skokov:2016yrj}. It is expected that Ru + Ru and Zr + Zr collisions at the same beam energy and same centrality will produce similar elliptic flow but a $10\%$ difference in the magnetic fields. Thus if the observed $\g$ correlation contains contribution from CME, the isobaric collisions could have a chance to see it by comparing the same $\g$ correlation in Ru + Ru and Zr + Zr collisions. The detailed numerical simulation was reported in Ref~\cite{Deng:2016knn} where the authors demonstrated that the two collision types at $\sqrt{s}=200$ GeV have more than $10\%$ difference in the CME signal and less than $2\%$ difference in the elliptic-flow-driven backgrounds for the centrality range of $20-60\%$ by assuming that the CME contribution to the $\g$ correlation is $1/3$. Such a difference is feasible in current experimental setups of heavy-ion collisions at RHIC. Therefore, the isobaric collisions at RHIC would be very valuable to isolate the chiral magnetic effect from the background sources.

Recently, the CMS Collaboration at LHC reported the measurement of $\g_{\a\b}$ in high multiplicity events in p + Pb and Pb + Pb collisions at $\sqrt{s}=5.02$ TeV~\cite{Khachatryan:2016got}. They found that the $\g_{\a\b}$'s in p + Pb and in Pb + Pb as functions of multiplicity behave similarly at the same multiplicities. These results give useful information about the CME search in heavy-ion collisions. First, although the magnetic field in p + Pb can be very strong, the magnetic-field direction should not be firmly correlated to the $v_2$ plane, and the lifetime of the magnetic field should be shorter at higher collision energy, thus the magnetic field is not expected to drive strong $\g_{\a\b}$ in p + Pb. The CMS measurement thus gives at least two very different possibilities. One is that, although it was not expected before, the CME still occurs in p + Pb collisions and the magnetic field, although weakly correlated to the $v_2$ plane, is still able to drive a finite signal which combined with the background effects gives the observed results. The other one is that there is no CME at $\sqrt{s}=5.02$ TeV (or very weak CME) in both p + Pb and Pb + Pb collisions, the $\g_{\a\b}$ in both p + Pb and Pb + Pb collisions are due to background effects. This is actually somehow compatible with the recent measurement of STAR collaboration in which the so-called $H$ correlation in measured at various energies through the Beam Energy Scan (BES) program~\cite{Adamczyk:2014mzf}. In the $H$ correlation, the background effects are subtracted under reasonable assumptions; see details in Ref.~\cite{Adamczyk:2014mzf}. If one extrapolates the $H$ correlation of STAR Collaboration to $\sqrt{s}=5.02$ TeV, one can find that $H$ correlation is actually small there. This supports that at higher energy the $\g_{\a\b}$ is more dominated by background contributions. Nevertheless, more measurements are definitely needed. Especially the CMS collaboration should also measure the so-called $\delta$-correlation from which we can deduce the $H$ correlation. In addition, the charge-dependent $v_2$ measurements (see next section) in p + Pb collisions will be also very important.

\subsection {Chiral separation effect and chiral magnetic wave}\label{sec:cmw}
By looking at the CME (\ref{cme}), one may wonder that whether there is a quantum-anomaly induced axial current in response to the external magnetic field and under what condition such current can appear? The answer is yes, and such a current can be caused by the chiral separation effect (CSE) which is the dual effect to the CME~\cite{Son:2004tq,Metlitski:2005pr,Newman:2005as}:
\begin{eqnarray}
\label{cse}
\bJ_A &=&
\s_{\rm CSE}
\bB,\\
\label{csecond}\s_{\rm CSE}&=&\frac{e^2}{2\p^2}\m_V,
\end{eqnarray}
where the axial current is defined by $J_A^\m=e\lan\jb\g^\m\g_5\j\ran$, and $\m_V$ is the vector chemical potential. Note that, unlike the CME, the occurrence of the CSE does not require a parity-violating environment but a charge-conjugation odd environment.

We can understand the CSE from an intuitive picture. In parallel to the discussion of CME in Sec.\ref{sec:cme}, let us consider a massless fermionic system under a strong magnetic field and let us focus on the lowest Landau level on which the spin of positively charged fermions are polarized to be parallel to the magnetic field (we call its direction the ``up" direction and its opposite direction the ``down" direction) while the spin of negatively charged fermions are polarized to be anti-parallel to the magnetic field. If a positively charged fermion on the lowest Landau level moves upward then it contributes to the RH current while if it moves downward it contributes to the LH current; thus no matter which direction it moves it contributes to an axial current along the magnetic field. Similarly, it is easy to see that the negatively charged fermions contributes an axial current in the direction opposite to the magnetic field. Now, if the system contains more positively charged fermions than negatively charged fermions, namely, if $\m_V>0$, the downward-moving axial current cannot compensate the upward-moving axial current and thus we get a net axial current along the direction of the magnetic field. This is the CSE current. The fermions occupying the higher Landau levels do not contribute to the CSE current as their spins are not polarized.

Let us formulate the above argument in a more precise way. For simplicity, we consider a system of massless Dirac fermions with only positive charge $e$ in an constant magnetic field $\bB$ in $z$ direction which is supposed to be so strong that all the fermions are confined in the lowest Landau level. The number densities of RH and LH fermions are given by \eq{density}. As the massless fermions on the lowest Landau level move at the speed of light in a way correlated to their spins which gives that the RH fermions generate a current along $z$ direction, $J_R=en_R$, while LH fermions generates a current along $-z$ direction, $J_L=-en_L$. Therefore the net axial current reads
\begin{eqnarray}
J_A=J_R-J_L=\frac{e^2B}{4\p^2}(p_F^{R}+p_F^L),
\end{eqnarray}
which, once we equate $p_F^{R/L}$ with $\m_{R/L}$ and $\m_V$ with $(\m_R+\m_L)/2$, leads us to \eq{cse} and \eq{csecond}. More rigorous derivation of CSE and various aspects regarding CSE can be found in Refs.~\cite{Son:2004tq,Metlitski:2005pr,Newman:2005as,Bergman:2008qv,Gorbar:2009bm,Basar:2010zd,Gorbar:2010kc,Hong:2010hi,Landsteiner:2011cp,Gorbar:2013upa,Yamamoto:2015fxa,Huang:2015mga}.

The CME and CSE couple together the the vector and axial densities and currents in the presence of the magnetic field and lead to collective modes called chiral magnetic waves (CMWs). This is most easily seen when we express CME and CSE in the chiral basis:
\begin{eqnarray}
\label{cmecse1}
\bJ_R &=&\frac{e^2}{4\p^2}\m_R\bB,\\
\label{cmecse2}
\bJ_L &=&-\frac{e^2}{4\p^2}\m_L\bB,
\end{eqnarray}
where the RH and LH quantities are defined as
$\bJ_{R/L}=(\bJ_V\pm\bJ_A)/2$ and $\m_{R/L} =\m_V\pm\m_A$. Substituting $\bJ_R$ and $\bJ_L$ in to the continuation equations and considering a small fluctuation in the density $J^0_{R/L}$, we obtain
\begin{eqnarray}
\label{continuation}
\pt_t\d J^0_{R/L}\pm \frac{e^2}{4\p^2\c}\bB\cdot\vec\nabla\d J^0_{R/L} &=&0,
\end{eqnarray}
where $\c=\pt J^0/\pt\m$ is the number susceptibility (we assume that the RH and LH chiralities have the same susceptibility). Therefore, we can identify two collective wave modes, one propagating along $\bB$ and one propagating opposite to the direction of $\bB$ with the same velocity $v_\c=e^2B/(4\p^2\c)$. They are the CMWs~\cite{Kharzeev:2010gd} (see also Ref.~\cite{Stephanov:2014dma} for a derivation of CMW based on the kinetic theory).

\subsection {Experimental signals for chiral magnetic wave}\label{sec:expcmw}
\begin{figure}[!htb]
\begin{center}
\includegraphics[width=10cm]{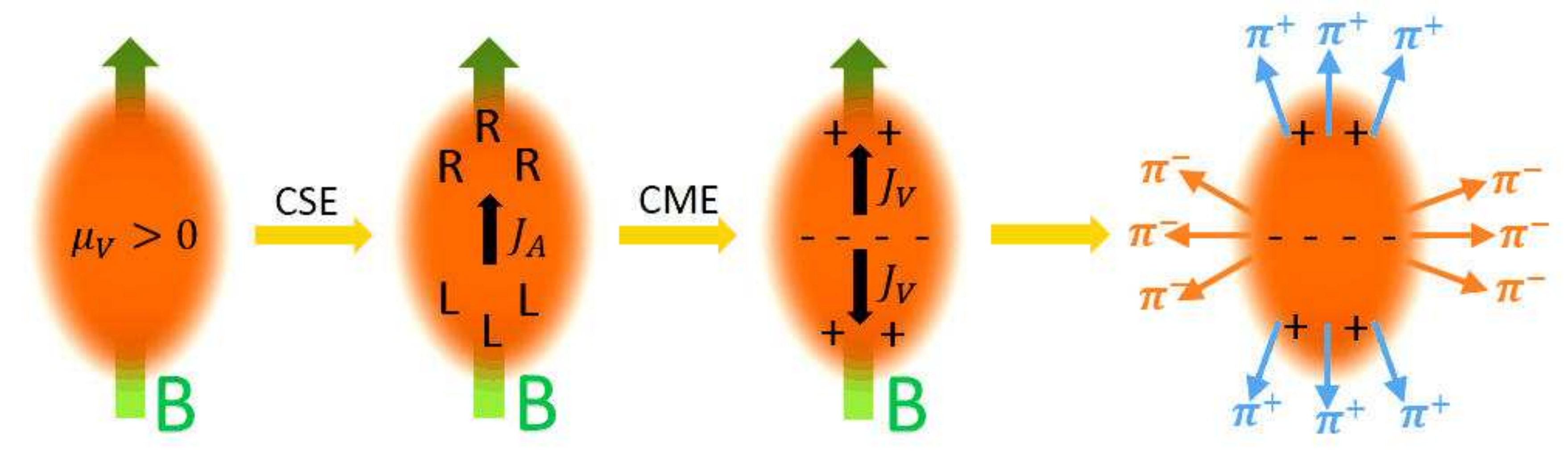}\;\;\;\;
\caption{The illustration of the CMW-induced electric quadrupole in QGP and the charge-dependent elliptic flow of pions. Figure is from Ref.~\cite{Huang:2015oca}.}
\label{figcmwv2}
\end{center}
\end{figure}

The CMWs are able to transport both chirality and electic charge and can induce an electric quadrupole in the QGP~\cite{Gorbar:2011ya,Burnier:2011bf}; see \fig{figcmwv2} for an illustration. Owing to strong in-plane pressure gradient, such an electric quadrupole can cause a splitting between the $v_2$ of $\p^+$ and $\p^-$~\cite{Burnier:2011bf}; see also Refs.~\cite{Burnier:2012ae,Taghavi:2013ena,Yee:2013cya,Hirono:2014oda} for more advanced numerical implementations. The CMW-induced charge dependent elliptic flows can be parameterized as
\begin{eqnarray}
\label{cmwpred}
v_2(\p^{\pm})=v_2\mp \frac{rA_{\rm ch}}{2},
\end{eqnarray}
where $v_2$ is the averaged elliptic flow and $A_{\rm ch}=(N_+-N_-)/(N_++N_-)$ is the net charge asymmetry parameter which is usually small (the events that can generate large $A_{\rm ch}$ is rare). The slop parameter $r$ is predicted to be proportional to $\lan(e\bB)^2\cos[2(\j_\bB-\J_\rp)]\ran$ where $\j_\bB$ is the azimuthal direction of the magnetic field~\cite{Burnier:2011bf,Bloczynski:2012en,Huang:2015mga}.

Recently, the CMW has been tested by STAR Collaboration~\cite{Wang:2012qs,Ke:2012qb,Adamczyk:2015eqo} at RHIC and by ALICE Collaboration~\cite{Adam:2015vje} at LHC. The data shows an elliptic-flow difference, $v_2(\pi^-)-v_2(\pi^+)$, linear in $A_{\rm ch}$ with a positive slope whose centrality dependence can be fitted by the CMW estimation. Furthermore, the slope parameter $r$ displays no obvious trend of the beam energy dependence for $10-60\%$ centrality at $\sqrt{s}=20-200$ GeV. This can be understood by noticing that the strength of magnetic field times its lifetime is roughly beam energy independent.
\begin{figure}[!htb]
\begin{center}
\includegraphics[height=4cm]{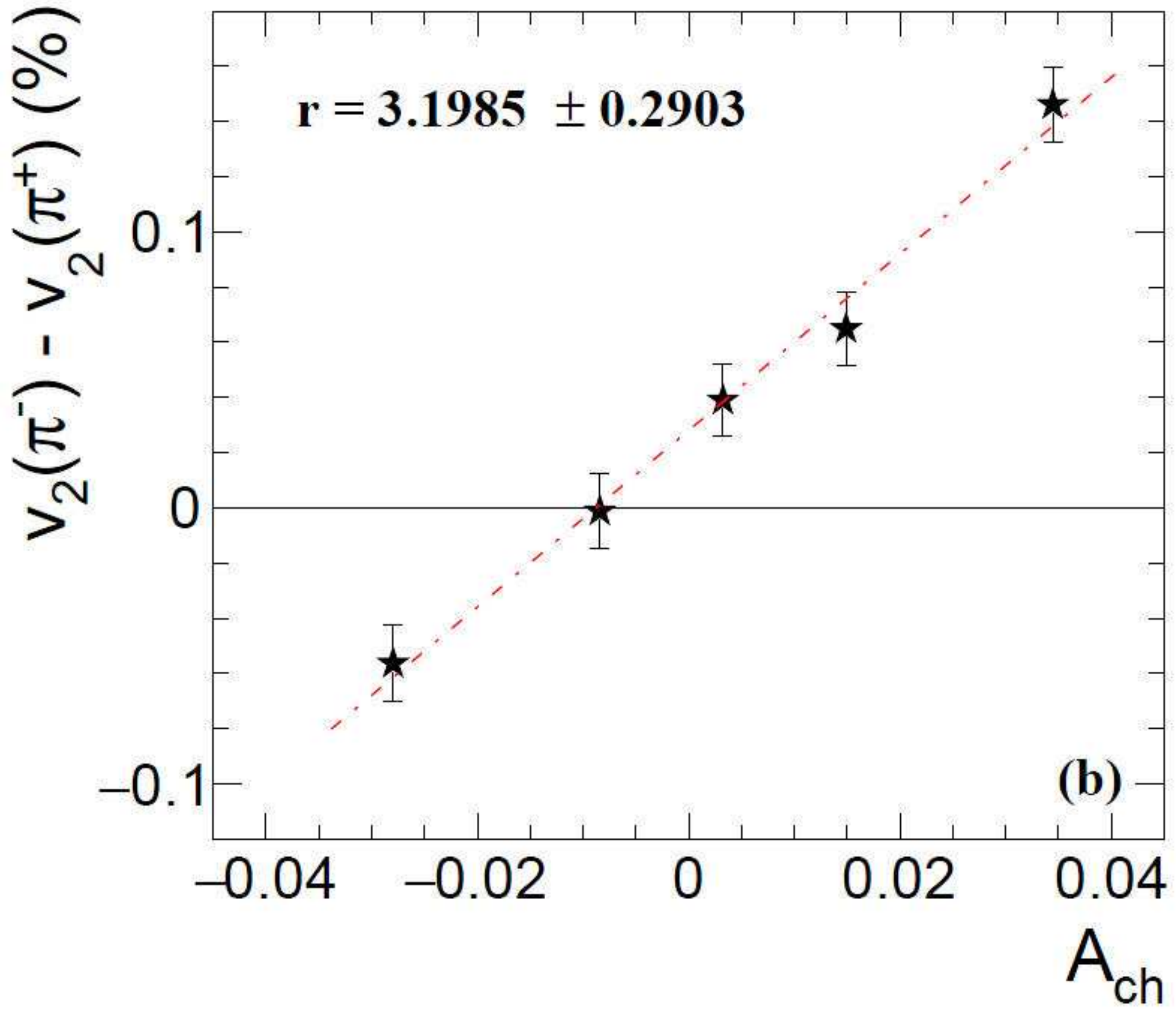}\;\;\;\;\;\;\;\;
\includegraphics[height=4cm]{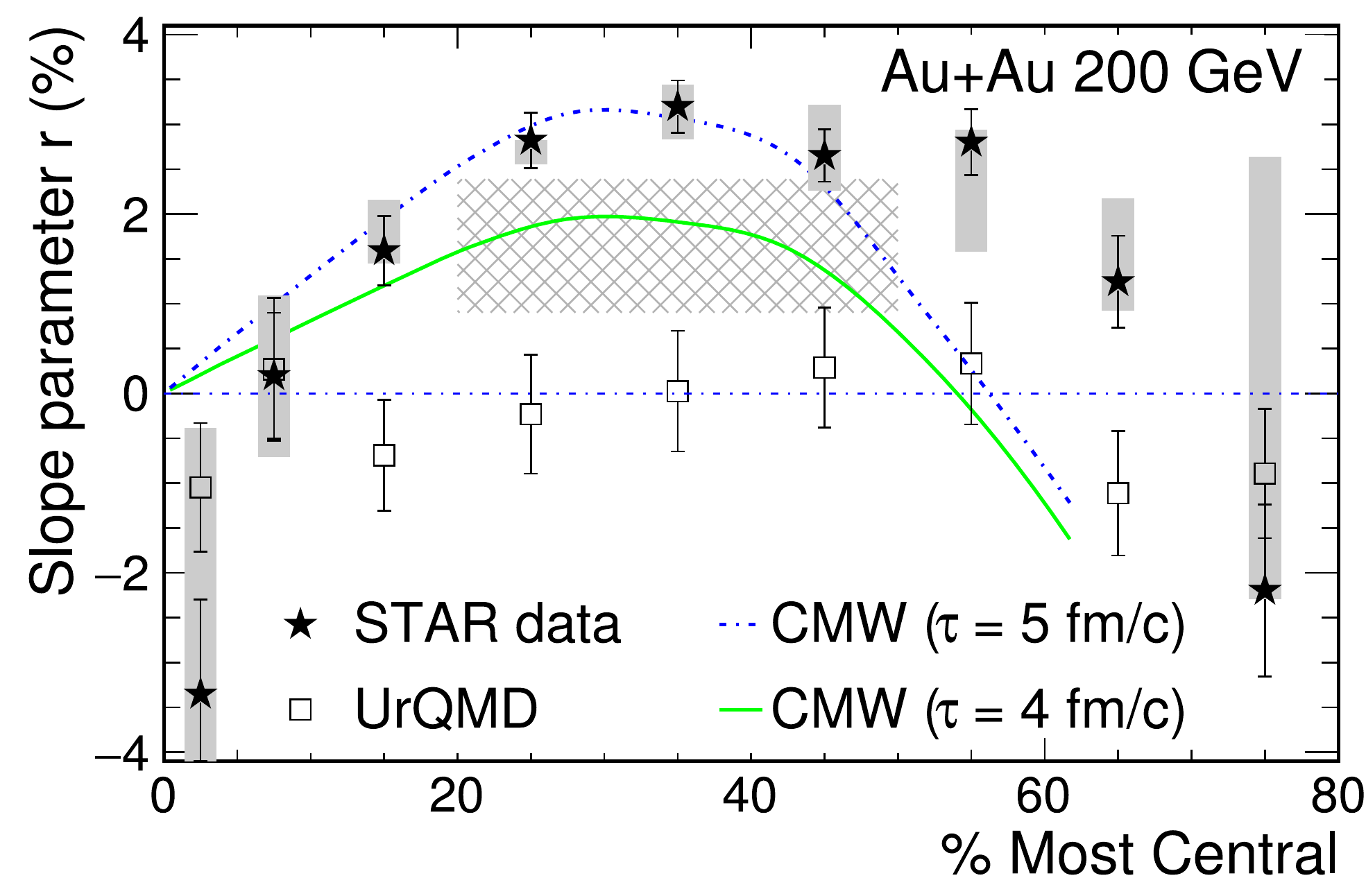}
\caption{(Left) The $v_2$ splitting between $\p^-$ and $\p^+$ as a function of charge asymmetry parameter $A_{\pm}$ measured by STAR Collaboration. (Right) The measured slop parameter $r$ as well as the UrQMD and CMW predictions as functions of the centrality. Figures are from Ref.~\cite{Adamczyk:2015eqo}.}
\label{figcmwexp}
\end{center}
\end{figure}

However, ambiguities exist in the interpretation of the experimental results as there are background effects which can also lead to $v_2$ splitting between $\p^+$ and $\p^-$. A variety of possible background effects were proposed, see Refs.~\cite{Deng:2012pc,Stephanov:2013tga,Dunlop:2011cf,Xu:2012gf,Song:2012cd,Bzdak:2013yla,Hatta:2015hca} and review articles, Refs.~\cite{Huang:2015oca,Kharzeev:2015znc}, for discussions. It should be noted that none of the above backgrounds can successfully explain all the features of the experimental data. It is quite plausible that various effects (including CMW) are coupled to produce the measured pattern. Before we can successfully subtract the background effects, we cannot make a conclusive claim. On the experimental side, it will be important to perform more detailed measurements such as the measurement of the $v_2$ splitting versus transverse momentum and rapidity, the measurement for other hadrons like kaons, the measurement of splitting in other harmonic flows like $v_3$, and the measurement of the cross-correlation of different observables. On the theoretical side,
an anomalous hydrodynamic simulation or a simulation based on kinetic theory for the CMW that incorporate both the dynamically evolving magnetic fields and the realistic initial charge distribution will be very useful.

Before we end this section, we note that, in addition to the magnetic-field induced anomalous transports, the electric field and vorticity can also induce certain anomalous transport phenomena and corresponding collective modes. For example, electric field can induce an axial current in the medium with both $\m_A$ and $\m_V$ nonzero which is called chiral electric separation effect (CESE) and, once coupled with the normal Ohm's current (the vector current induced by electric field), CESE leads to chiral electric wave and axial and vecor density waves~\cite{Huang:2013iia,Jiang:2014ura,Pu:2014cwa,Pu:2014fva}. The CESE may lead to observable effect in, e.g., Cu + Au collisions where persist electric fields from Au to Cu nuclei exist~\cite{Ma:2015isa}. The fluid vorticity can induce the so-called chiral vortical effects (CVEs) which are the vortical analogues of CME ~\cite{Kharzeev:2007tn,Erdmenger:2008rm,Banerjee:2008th,Son:2009tf} and the corresponding collective mode, chiral vortical wave (CVW)~\cite{Jiang:2015cva}. The CVE and CVW may have important consequences in noncentral heavy-ion collisions where nonzero fluid vorticity can be generated as a result of angular momentum conservation~\cite{Csernai:2013bqa,Becattini:2015ska,Jiang:2016woz,Deng:2016gyh}. Phenomenologically, the CVE can induce baryon number separation with respect to the reaction plane~\cite{Kharzeev:2010gr} and CVW may cause a $v_2$ splitting between $\L$ and $\bar{\L}$~\cite{Jiang:2015cva}. Recently, the STAR Collaboration has reported the first measurement of CVE in Au + Au collisions at $\sqrt{s}=200$ GeV~\cite{Zhao:2014aja} and the data is consistent with the expectation of CVE. But, similar with the case of CME and CMW detections, the data contains contributions from background effects like the transverse momentum conservation and the local baryon number conservation, and we need more efforts to understand the data.

\section{Heavy-quark dynamics in magnetic fields}

\label{sec:HQ}

In the heavy ion collisions, the dominant process for the heavy-quark production
is the initial hard scatterings among the partons from the colliding nuclei.
This is because the thermal excitation is suppressed due to the large value of
the heavy-quark mass compared to the temperature of the quark-gluon plasma (QGP).
Therefore, the heavy quarks will serve as a probe of the dynamics in the magnetic field
persisting in the early time of the collision events.
In this section, we discuss effects of the magnetic field on the quarkonium spectra in Sec.~\ref{sec:QQbar}
and the heavy-quark transport in a hot medium and magnetic field in Sec.~\ref{sec:HQtrans}.

The quarkonium spectra in magnetic fields have been investigated by using
the Cornell potential model \cite{Alford:2013jva, Bonati:2015dka, Suzuki:2016kcs, Yoshida:2016xgm}
and the QCD sum rules \cite{Machado:2013yaa, Cho:2014exa, Cho:2014loa}
with the help of the lattice QCD simulations for the heavy quark potential \cite{Bonati:2014ksa, Bonati:2016kxj}
and the quark and gluon condensates \cite{Bali:2011qj,D'Elia:2011zu,Bali:2012zg,Bruckmann:2013oba, Bali:2013esa}.
The phenomenological consequences for the quarkonium production in the heavy-ion collisions
have been discussed in Ref.~\cite{Marasinghe:2011bt, Yang:2011cz, Machado:2013rta, Alford:2013jva, Liu:2014ixa, Guo:2015nsa},
which could be measured by the dilepton channel.

Most of the heavy quarks created in the hard scatterings will evolve in a hot medium
and be observed as open heavy flavor mesons in the final state (see Ref.~\cite{Andronic:2015wma} for a review).
Therefore, it is important to understand the transport property of the heavy quarks
in a hot medium and magnetic field.
In Sec.~\ref{sec:HQtrans}, we discuss the diagrammatic computation of the heavy-quark diffusion constant.
The study on the heavy-quark dynamics in the magnetized medium has been
just initiated \cite{Fukushima:2015wck, Das:2016cwd}, and is awaiting for further developments.

\subsection{Quarkonia in magnetic fields}

\label{sec:QQbar}

\subsubsection{Spin mixing and level repulsion}

We shall begin with the simplest description of effects of magnetic fields on quarkonium spectra.
In a weak magnetic field, 
the dominant effects of the magnetic fields can be described by the hadronic degrees of freedom.
This contrasts to the case of strong magnetic fields where
modifications of nonperturbative vacuum structure
and of the internal structure of hadrons are important
(see Refs.~\cite{Fukushima:2012kc, Chernodub:2011mc, Hidaka:2012mz, Liu:2014uwa, Hattori:2015aki, Andreichikov:2012xe,
Andreichikov:2013zba, Bali:2015vua, Luschevskaya:2014lga, Bali:2011qj, Taya:2014nha}
and references therein for the ongoing discussions). 
As long as the magnitude of the magnetic field is much smaller than 
the heavy quark mass, these effects will be subdominant contributions. 

In the hadronic degrees of freedom,
effects of magnetic fields appear in the mixing of spin eigenstates
\cite{Machado:2013rta,Alford:2013jva,Cho:2014exa,Cho:2014loa, Bonati:2015dka,Suzuki:2016kcs,Yoshida:2016xgm}.
This mixing is caused by the breaking of the spatial rotation symmetry in magnetic fields:
there remains only the azimuthal rotational symmetry with respect to the direction of the external magnetic field.
This indicates that only the spin state along the magnetic field
can persist as a good quantum number of the mesons, and thus that
there is a mixing between the spin single and triplet states,
$(S_{\rm total}, S_z) = (0,0) $ and $(1,0)$,
which are, for example, $ \eta_c$ and the longitudinal component of $ J/\psi$.

%

We examine general mixing patterns among the pseudoscalar $P$,
vector $V^\mu$, scalar $S$ and axial-vector $A^\mu$ quarkonia.
Possible interaction vertices among those fields are informed from
the Lorentz invariance and the parity and charge-conjugation symmetries.
The vertices relevant for interactions among {\it static} charmonia are found to be \cite{Cho:2014exa,Cho:2014loa}
\begin{eqnarray}
\Lag_{\gam \pv} &=&
\frac{ g_{\pv} }{ m_0 } e \tilde{F}_{\mu \nu} (\partial^{\mu} P) V^{\nu}
\label{eq:L_pv} \ ,
\\
\Lag_{\gam \va}
&=& i g_{\va} e \tilde{F}_{\mu \nu} V^{\mu} A^{\nu}
\label{eq:L_va} \ ,
\\
\Lag_{\gam \sa}
&=& \frac{ g_{\sa} }{ m_1 } e \tilde{F}_{\mu \nu} (\partial^{\mu} S) A^{\nu}
\label{eq:L_sa}
\ ,
\end{eqnarray}
with $m_0=(m_\ps+m_\V)/2$, $m_1=(m_S+m_\A)/2$,
and dimensionless effective coupling constants $g_\pv$, $g_\va$, and $g_\sa$.
The vertices proportional to $ F^{\mu\nu} \partial_\mu $ do not play any role
for static charmonia, because they pick up the vanishing spatial derivatives
and the temporal component of the vector or axial-vector fields.
The nonvanishing vertices are responsible for, e.g., radiative decay modes of quarkonia
such as $\Jp \rightarrow \eta_c + \gamma$ \cite{Brambilla:2012be,Pineda:2013lta}.
It is also worth mentioning that the three-point vertex (\ref{eq:L_pv})
induces the conversion of $ \eta_c$ to $J/\psi $ in a magnetic field.
The contribution to the $ J/\psi$ production from this conversion process
was examined in Ref~\cite{Yang:2011cz}.

\begin{figure}[!tb]
		\begin{center}
			\includegraphics[width=0.65\hsize]{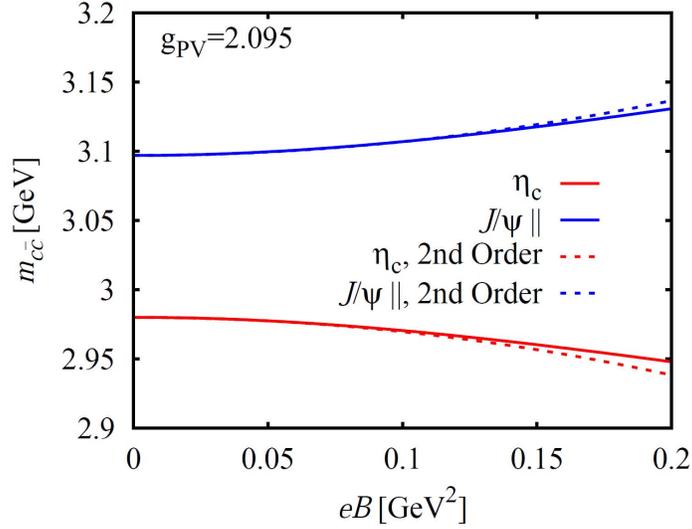}
\caption{Level repulsion in the mass spectra of the $ \eta_c$ and longitudinal $J/\psi $.
(Reproduced from Refs.~\cite{Cho:2014loa}) }
\label{fig:Qshift}
		\end{center}
\end{figure}
From the above vertices, one can identify two pairs of mixing partners:
the mixing between $P$ and $V$, and between $S$ and $A$,
are possible when they are at rest in external magnetic fields.
Since the $ P$ and $V $ are not mixed with the $ S$ and $ A$,
we shall focus on the mixing between $ S$ and $V $.
As mentioned in the beginning of this section,
a mixing effect arises only in the longitudinal component of the vector field $V_\parallel $,
and the other two transverse components persist as the energy eigenstates.
This can be easily confirmed by identifying the non-vanishing components in Eq.~(\ref{eq:L_pv}).
With a vanishing spatial momentum $q^\mu = (\omega, 0 , 0, 0)$ and the vertex (\ref{eq:L_pv}),
the coupled equations of motion are found to be
\beq
\left(
\begin{array}{cc}
- \omega^{2} +  m_{\ps}^{2} & - i \frac{g_{\pv}}{m_{0}} \omega e B \\
i \frac{g_{\pv}}{m_{0}} \omega e B & -\omega^{2} +  m_{\V}^{2} \\
\end{array}
\right)
\left(
\begin{array}{c}
P \\
V_{\parallel} \\
\end{array}
\right)
\label{Matrix_eq_etac_Jpsi}
=0.
\eeq
Following from the equations of motion (\ref{Matrix_eq_etac_Jpsi}),
we obtain the physical mass eigenvalues as 
\begin{equation}
\bar m_{V,P}^{2}=\frac{1}{2} \bigg( M_+^2+\frac{\gamma^2}{m_{0}^{2}}
\pm \sqrt{M_-^4+\frac{2\gamma^2 M_+^2}{m_0^2}+ \frac{\gamma^4}{m_0^4}} \bigg),
\label{eq:EFT}
\end{equation}
where $M_+^2=m_\ps^2+m_\V^2, M_-^2=m_\V^2-m_\ps^2$ and $\gamma=g_{\pv} e B$.
The upper and lower signs are for the vector and pseudoscalar states, respectively.
Expanding the right-hand side of Eq.~\eqref{eq:EFT} up to the second order in $\gamma$ and the leading order in
$\frac{1}{2}(m_\V-m_\ps)/m_{0} $, 
we find
\beq
\bar m_{P, V}^{2} (B) &=&
m_{V, P}^{2} \pm \frac{\gamma^2}{M_-^2}
,
\label{eq:Jpsi_2nd}
\eeq
with eigenvectors given by
\beq
&&
| P \rangle_{\scriptscriptstyle B} \ \, =  \bigg(1-\frac{1}{2} \frac{\gamma^2}{ M_-^4}  \bigg)
| P \rangle - i \frac{\gamma}{M_-^2}
| V \rangle   , \nonumber \\
&&
| V \rangle_{\scriptscriptstyle B} \, = -i \frac{\gamma}{M_-^2}
| P \rangle +  \bigg(1-\frac{1}{2} \frac{\gamma^2}{M_-^4} \bigg)
| V \rangle
\, .
\label{eq:wave}
\eeq
The mass eigenvalue (\ref{eq:Jpsi_2nd}) clearly shows the level repulsion
in the presence of the mixing effect.

In case of the mixing between $ \eta_c$ and $J/\psi $,
the coupling constant $g_\pv $ can be read off from
the measured radiative decay width, $ J/\psi \to \eta_c + \gam$ \cite{Cho:2014exa, Cho:2014loa}.
From the vertex (\ref{eq:L_pv}), the invariant amplitude of this decay process is written as
\beq
\M_\pv
&=& \langle \gamma {\rm P} | \mathcal{L}_{\scriptscriptstyle \gamma {\rm P V} } | {\rm V}  \rangle \nonumber \\
&=& - \frac{e g_\pv}{m_0} \epsilon_{\mu \nu \alpha \beta}
k_{\gamma}^{\mu} \epsilon_{\gamma}^{\nu} p_{\V}^{\alpha} \epsilon^{\beta}_{\V}
\, \ ,
\eeq
where $ k_{\gamma}^{\mu} $ and $\epsilon_{\gamma}^{\nu}$
($ p_{\V}^{\alpha} $ and $\epsilon^{\beta}_{\V}$) denote
the polarization and momentum of the photon ($ J/\psi$).
Summing the polarizations of the photon and averaging those of the vector meson,
we find
\beq
\Gamma [ {\rm V} \to \gamma {\rm P} ]
&=& \frac{1}{12} \frac{ e^2 g_{\pv}^{2} \tilde p^{3} }{  \pi m_0^{2} }
\label{eq:coupling_decay}
\,
\eeq
We have also integrated over the phase-space volume in the two-body final state.
The center-of-mass momentum in the final state is given by $\tilde p = (m_\V^2 - m_\ps^2) / (2 m_\V)$.
Using the measured radiative decay width,
$\Gamma_{\rm exp}[ J/\psi \to \gamma \, \eta_{c} ] = 1.579$ keV,
the coupling strength is estimated to be $g_\pv = 2.095 $.
A consistent value of the coupling strength is also obtained from computation in the quark degrees
of freedom with the help of the heavy quark expansion
and diagrammatic methods in an external magnetic field (see Appendix.~\ref{sec:coupling_FS}).

Figure~\ref{fig:Qshift} shows the mass shifts in the presence of the mixing effect.
We find the level repulsion between the $\eta_c $ and longitudinal $J/\psi$ mass spectra.
While the mass of $\eta_{c}$ decreases as $eB$ increases,
the mass of the longitudinal $J/\psi$ (denoted by $J/\psi_\parallel$) increases.
This result clearly provides the simple picture of the effect of weak magnetic fields.

The level repulsion occurs most strongly between a pair of adjacent states.
Therefore, the other excited states give rise to only subdominant contributions
to the mass shifts of $\eta_{c}$ and the longitudinal $ J/\psi$ in a weak magnetic field.
However, as the mass of the longitudinal $ J/\psi$ keeps increasing with an increasing magnetic field,
it will approaches the other excited states.
In such a strong magnetic field, many states will get involved in the mixing effects,
which will result in an intriguing mixing patterns of the spectra 
as discussed in the next section.

Moreover, the mixing effects induced by the minimal coupling (\ref{eq:L_pv}) to a magnetic field
will not be valid in the strong-field regime.
Since the strong magnetic field probes the inner structure of hadrons and modifies the QCD vacuum,
we will need to look into the dynamics of the fundamental degrees of freedom.
To investigate these effects, one can invoke on the Cornell potential models
and QCD sum rules with the help of lattice QCD simulations.
We will proceed to exploring these frameworks on top of the simplest
framework for the mixing effect discussed in this section.


\subsubsection{Cornell potential model in magnetic fields}

To study the nonrelativistic bound states of the heavy quark and antiquark
on the basis of the fundamental degrees of freedom,
one can use the Cornell potential model \cite{Eichten:1978tg,Eichten:1979ms}
with the help of the lattice QCD simulations.
The basic framework for the $ 1S$ states of $ \eta_c$, $J/\psi $, $\eta_b $, and $\Upsilon $,
was first elaborated by Alford and Strickland \cite{Alford:2013jva}.
The authors included the interaction with the background magnetic field through
the magnetic moments of constituents and the kinetic terms,
as well as the mutual interactions by the linear and Coulomb potentials in vacuum.
In this framework, the interactions through the magnetic moments give rise to
the mixing between the spin singlet and triplet states discussed in the previous section.
Based on this model, a systematic spectroscopy for excited states was recently performed by Suzuki and Yoshida
owing to an improvement of computational method \cite{Suzuki:2016kcs,Yoshida:2016xgm}.
Effects of strong magnetic fields on nonperturbative QCD vacuum were investigated by Bonati et al.
who computed the modification of the quark-antiquark potential by lattice QCD simulation \cite{Bonati:2014ksa,Bonati:2016kxj}
and plugged it into the bound-state problem \cite{Bonati:2015dka}.
These studies suggest an opportunity to explore the nature of QCD confinement
in the presence of a strong magnetic field.
In the following, we briefly summarize the framework for the potential model in a magnetic field
and the numerical results for the mass spectra.

The Hamiltonian for the two-body bound state in a magnetic field is given by
\cite{Machado:2013rta,Alford:2013jva,Suzuki:2016kcs,Yoshida:2016xgm,Bonati:2015dka}
\begin{eqnarray}
\Ham &=& 2m_Q + \sum_{i=1,2}  \Ham_0^{(i)}  + \Ham_\cor + \Ham_{\bm \mu}
\label{eq:H}
\end{eqnarray}
where
\begin{eqnarray}
\Ham_0^{(i)} &=&  \frac{1}{2m_Q} \left( \, \bp^{(i)} - q^{(i)} \bA(\br^{(i)}) \, \right)^2
\\
\Ham_\cor &=& 
\sigma_s r - \frac{a}{r} + b(r) \left( {\bm \sigma}^{(1)} \cdot  {\bm \sigma}^{(2)} \right)
\\
\Ham_{\bm \mu} &=& - \left( \bmu^{(1)} +  \bmu^{(2)} \right) \cdot \bB
\end{eqnarray}
with the gauge potential $ \bA(\br)$ for the external magnetic field $\bB $,
the Pauli matrix ${\bm \sigma} $, and the heavy-quark mass $ m_Q$.
The indices $ i=1,2$ denote the quark and antiquark, respectively, and thus $ q^{(1)} = - q^{(2)} (\equiv q)$ for quarkonia.
The mutual interaction is described by the Cornell potential $ \Ham_\cor $ which includes
the linear and Coulomb potentials as well as the spin-spin interaction.
For the nonrelativistic quark and antiquark, the magnetic moments read
\begin{eqnarray}
\Ham_{\bm \mu} &=& - \mu_Q \frac{g}{2} (\sigma^{(1)}_z - \sigma^{(2)}_z) B
\end{eqnarray}
where $ \mu_Q = q/ (2m_Q)$ and $ g \simeq 2$.
We took the magnetic field in $ +z$ direction without loss of generality.
The eigenstates of the Pauli matrices are spanned by the direct product
$ \vert \pm \pm \rangle \equiv \vert \pm \rangle \otimes \vert \pm \rangle$
where the left and right states in the product stand for the eigenstates of
$\sigma^{(1)}_z$ and $ \sigma^{(2)}_z$, respectively.
In the total-spin basis, the spin singlet state is given by
$ \vert 0, 0 \rangle = ( \vert + - \rangle - \vert - +  \rangle)/\sqrt{2}$,
while the triplet states by $ \vert 1, \pm 1 \rangle = \vert \pm \pm \rangle$
and $ \vert 1, 0 \rangle = ( \vert + - \rangle + \vert - +  \rangle)/\sqrt{2}$.
Therefore, the operation of $ \Ham_{\bm \mu} $ on these states results in
\begin{eqnarray}
&&
\Ham_{\bm \mu} \vert 1, +1 \rangle = \Ham_{\bm \mu} \vert 1, -1 \rangle = 0
\, ,
\\
&&
\Ham_{\bm \mu}
\left(\!\!
\begin{array}{ll}
\vert 0, 0 \rangle
\\
\vert 1, 0 \rangle
\end{array}
\!\!\right)
=
\left(
\begin{array}{ll}
0 & 2 \mu_Q B
\\
2 \mu_Q B & 0
\end{array}
\right)
\!\!
\left(\!\!
\begin{array}{ll}
\vert 0, 0 \rangle
\\
\vert 1, 0 \rangle
\end{array}
\!\!\right)
\, .
\end{eqnarray}
Consistent with the discussion in the previous section,
we find that, while the transverse modes $ \vert 1, \pm 1 \rangle$ are
eigenstates of the two-body Hamiltonian, there is a mixing between
$\vert 0, 0 \rangle $ and $\vert 1, 0 \rangle $ states
which need to be diagonalized to the energy eigenstates.

\begin{figure}[!htb]
		\begin{center}
			\includegraphics[width=0.6\columnwidth]{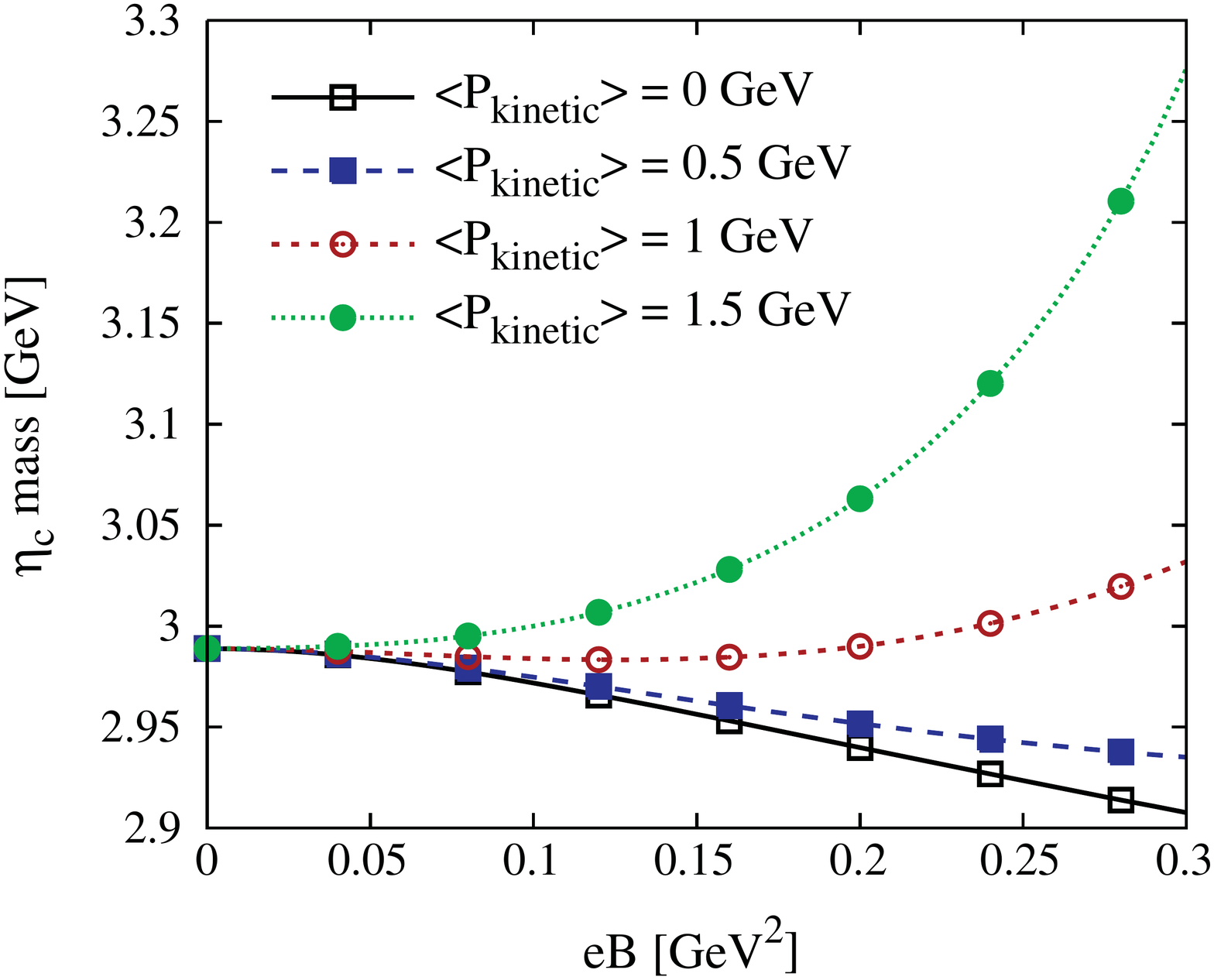}
			\includegraphics[width=0.6\columnwidth]{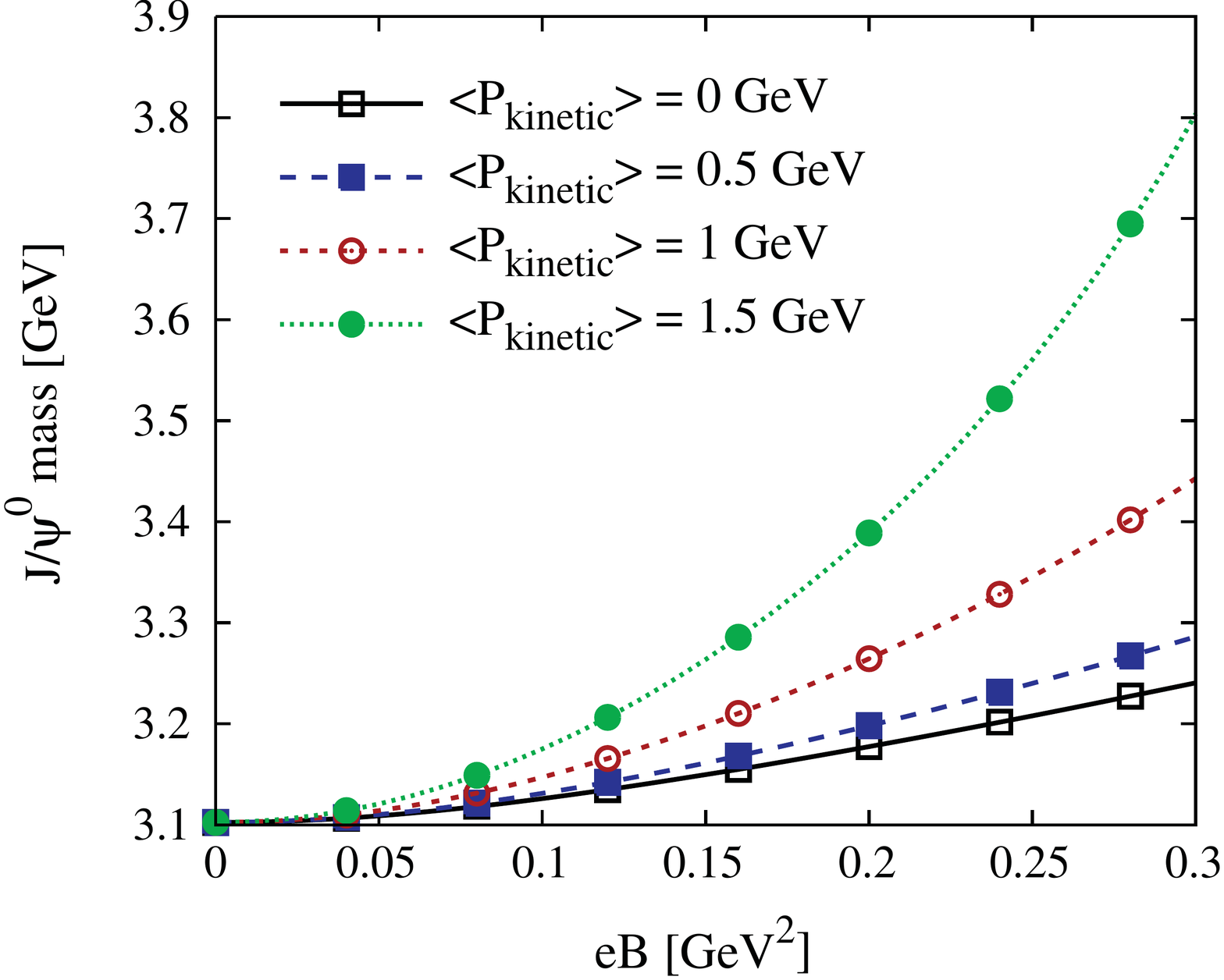}
			\includegraphics[width=0.6\columnwidth]{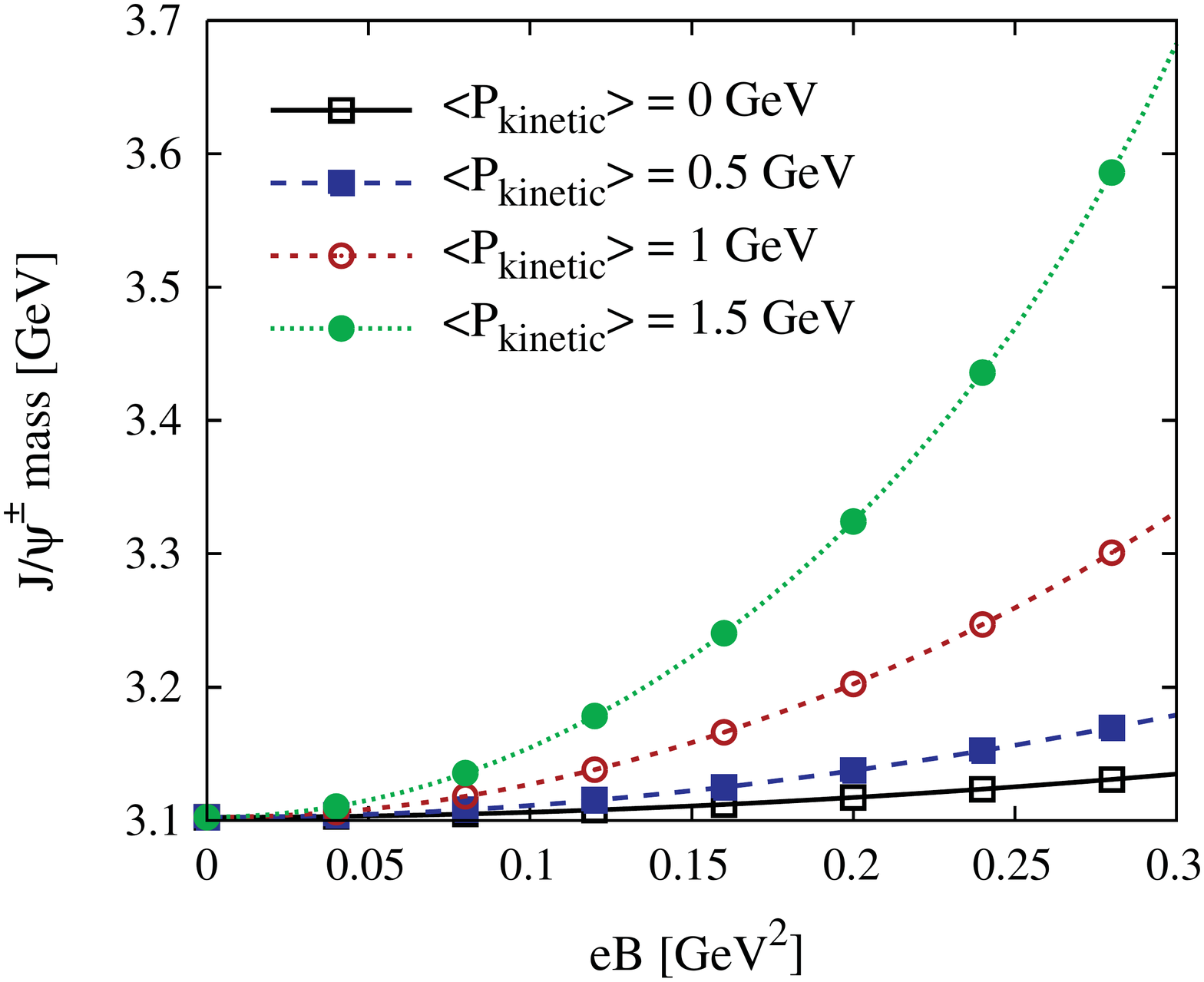}
		\end{center}
\caption{Ground-state charmonium spectra from the Cornell potential model. (Reproduced from Ref.~\cite{Alford:2013jva}.)}
\label{fig:AS_sch}
\end{figure}

A specific issue of the dynamics in magnetic fields is
the non-conservation of the transverse momentum.
The gauge potential for an external magnetic field will depend on
the transverse coordinate as $A^\mu_\ext = (0,A_\perp^i(\bx_\perp),0) $,
and apparently breaks the translational invariance in the transverse plane.
Whereas the momentum along the magnetic field is conserved,
the transverse momentum is not a conserved quantity, and is moreover gauge dependent.
For example, one of the transverse components is not conserved in the Landau gauge,
while neither of them are conserved in the symmetric gauge.
Nevertheless, it is clear that a constant magnetic field has a translational invariance
in the transverse plane, so that there must be a conserved momentum
which works as the generator of the translation.
This momentum is called the pseudomomentum,
and its generalization to the two-body system is defined as
(see Refs.~\cite{Alford:2013jva,Andreichikov:2013zba} and references therein)
\begin{eqnarray}
{\bm K} &=& \sum_{i=1,2} \left(\bp^{(i)} + q^{(i)} \bA^{(i)} \right)
\, ,
\end{eqnarray}
where the labels denote the quark and antiquark as before.
This momentum is related to the center-of-mass kinetic momentum as
$\langle {\bm P}_{\rm kinetic} \rangle = {\bm K} - q \bB \times \langle \br \rangle $
with $q \equiv q^{(1)} = - q^{(2)}  $ and the relative coordinate $\br = \br_1 - \br_2 $.
The same issue of the momentum nonconservation is seen
in diagrammatic calculations in the presence of a magnetic field.
In such calculations, the gauge- and translational-symmetry breaking parts
are encoded in the so-called Schwinger phase appearing from
the gauge-dependent wave function of fermions in a magnetic field \cite{Schwinger:1951nm}.
Effects of the Schwinger phase on the Bethe-Salpeter equation for light mesons
were discussed in Ref.~\cite{Hattori:2015aki}.

The Schr\"odinger equation for the two-body system (\ref{eq:H})
can be solved numerically for a fixed value of the conserved momentum $ {\bm K}$.
The parameters for the Cornell potential are so tuned to reproduce the vacuum spectrum
in Ref.~\cite{Alford:2013jva,Suzuki:2016kcs,Yoshida:2016xgm},
and are imported from the lattice simulation in a finite magnetic field in Ref.~\cite{Bonati:2015dka}.
In those papers, the numerical codes were carefully tested to reproduce
the analytic solution for a harmonic potential.

Figure~\ref{fig:AS_sch} shows the charmonium spectra from Ref.~\cite{Alford:2013jva} with the potential in vacuum.
At the vanishing momentum $ \langle {\bm P}_{\rm kinetic} \rangle = 0$ GeV,
one sees that the masses of the $ \eta_c$ and longitudinal $ J/\psi$
decreases and increases, respectively, as expected from the mixing effect.
On the other hand, the modification in the transverse $ J/\psi$ is weak without the mixing effect.
The masses of the moving charmonia, which were obtained by subtracting
the center-of-mass kinetic energy, tend to increase with an increasing its momentum.
The bottomoium spectra shown in Ref.~\cite{Alford:2013jva} have qualitatively the same behavior
as in charmonium spectra, with suppression in magnitude
due to the smaller magnetic moments of the bottom quark and antiquark.

\begin{figure}[!t]
 \begin{center}
   \includegraphics[clip,width=0.85\columnwidth]{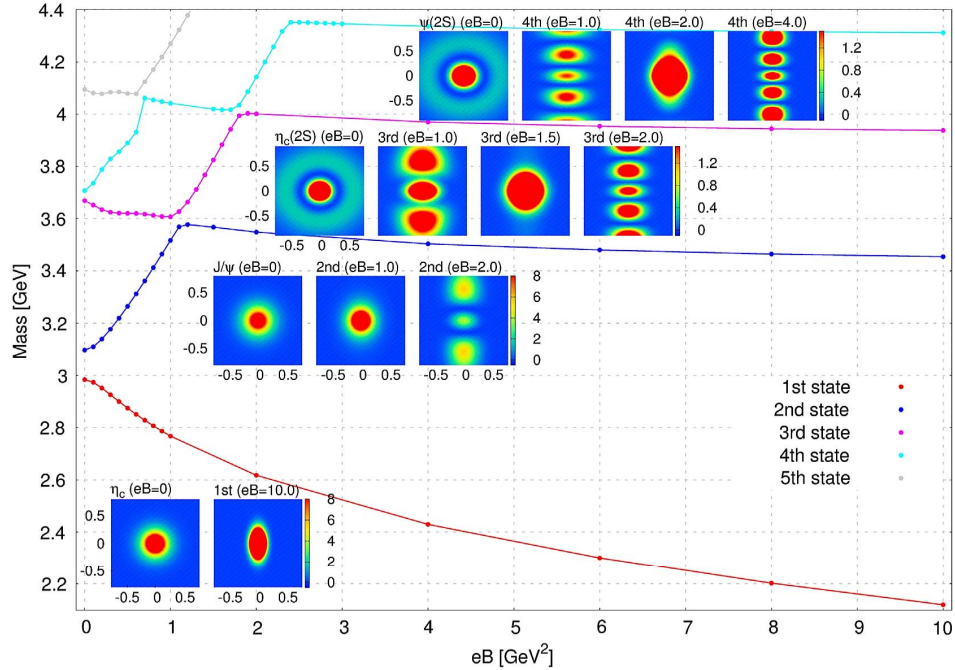}
    \end{center}
    \vspace{-1.5cm}
     \caption{Charmonium spectroscopy by the Cornell potential model. 
Small windows exhibit the spatial extensions of the wave functions 
in the transverse and longitudinal directions with respect to the $ B$ 
which are shown as the horizontal and vertical axes, respectively. 
(Reproduced from Ref.~\cite{Suzuki:2016kcs}.)}
 \label{fig:SY}
\end{figure}

This study was extended to the excited states of quarkonia,
which was made possible by using the cylindrical Gaussian expansion method proposed in Ref.~\cite{Suzuki:2016kcs}.
Since this method does not resort to the imaginary time evolution $\tau \to \infty $
to extract the spectra \cite{Alford:2013jva}, it would be more suitable for investigating the excited states.
The result of the systematic spectroscopy is shown in Fig.~\ref{fig:SY}.
In a weak magnetic field, the level repulsions between the nearest mixing partners
give rise to the dominant modification of the mass spectra. 
As the magnetic field becomes stronger, there appear the level-crossing points
where the wave functions are strongly mixed.
The density plots in the small windows show the transition of the wave functions.
One can see the squeezing of the wave function along the magnetic field in the strong-field regime
as expected from the shrink of cyclotron orbits.
Including the contributions of the excited states, the $ \eta_c$ state is,
as it is the lowest-lying state, pushed down additively by all of the mixing partners,
and its mass monotonically decreases.
This systematic computation was very recently extended to the spectra of bottomonia and open heavy flavors \cite{Yoshida:2016xgm}.

\begin{figure}[!h]
\begin{minipage}[t]{0.49\hsize}
		\begin{center}
			  \includegraphics[clip,width=\columnwidth]{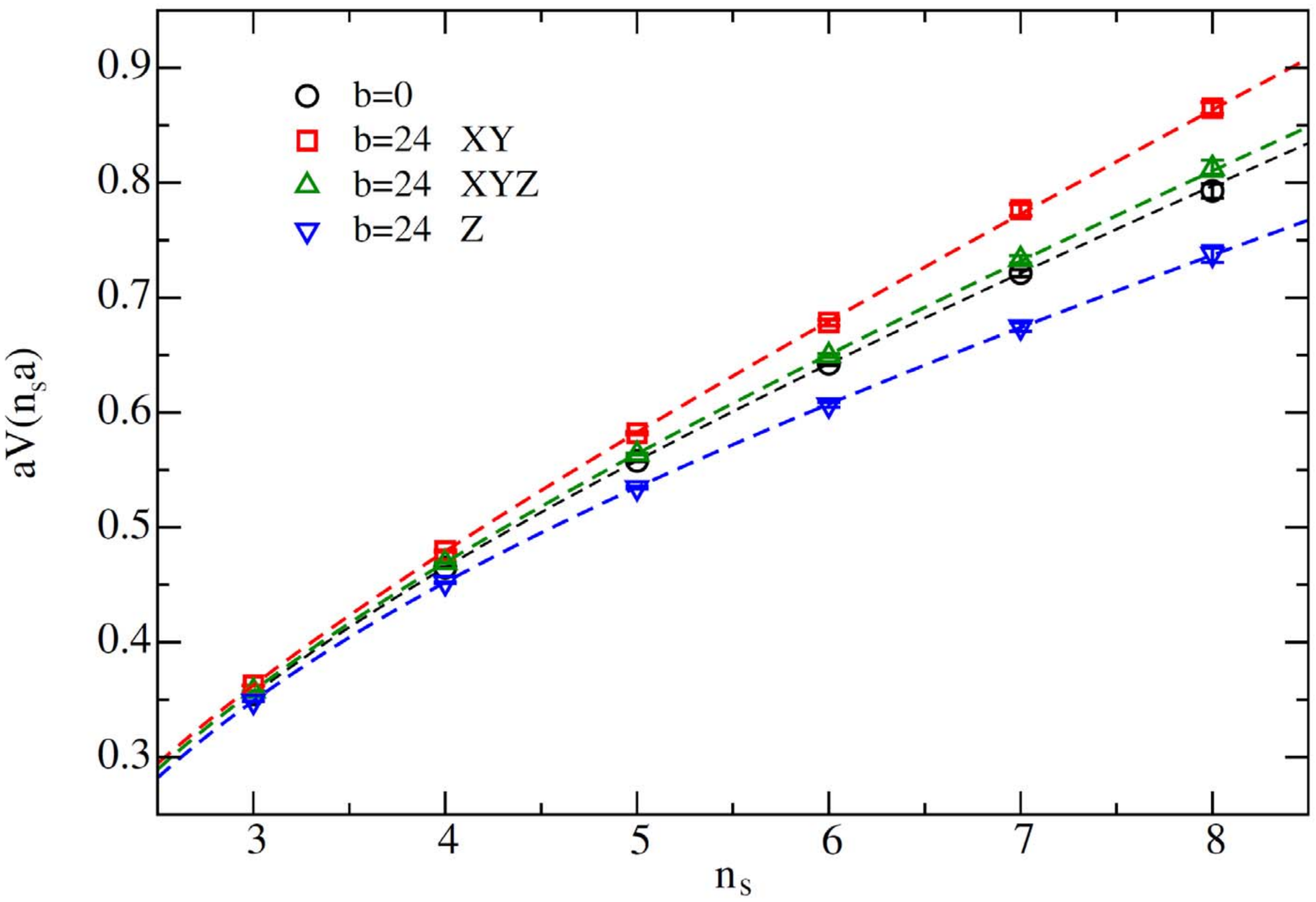}
		\end{center}
\label{fig:poles}
\end{minipage}
\begin{minipage}[t]{0.49\hsize}
		\begin{center}
			\includegraphics[clip,width=0.9\hsize]{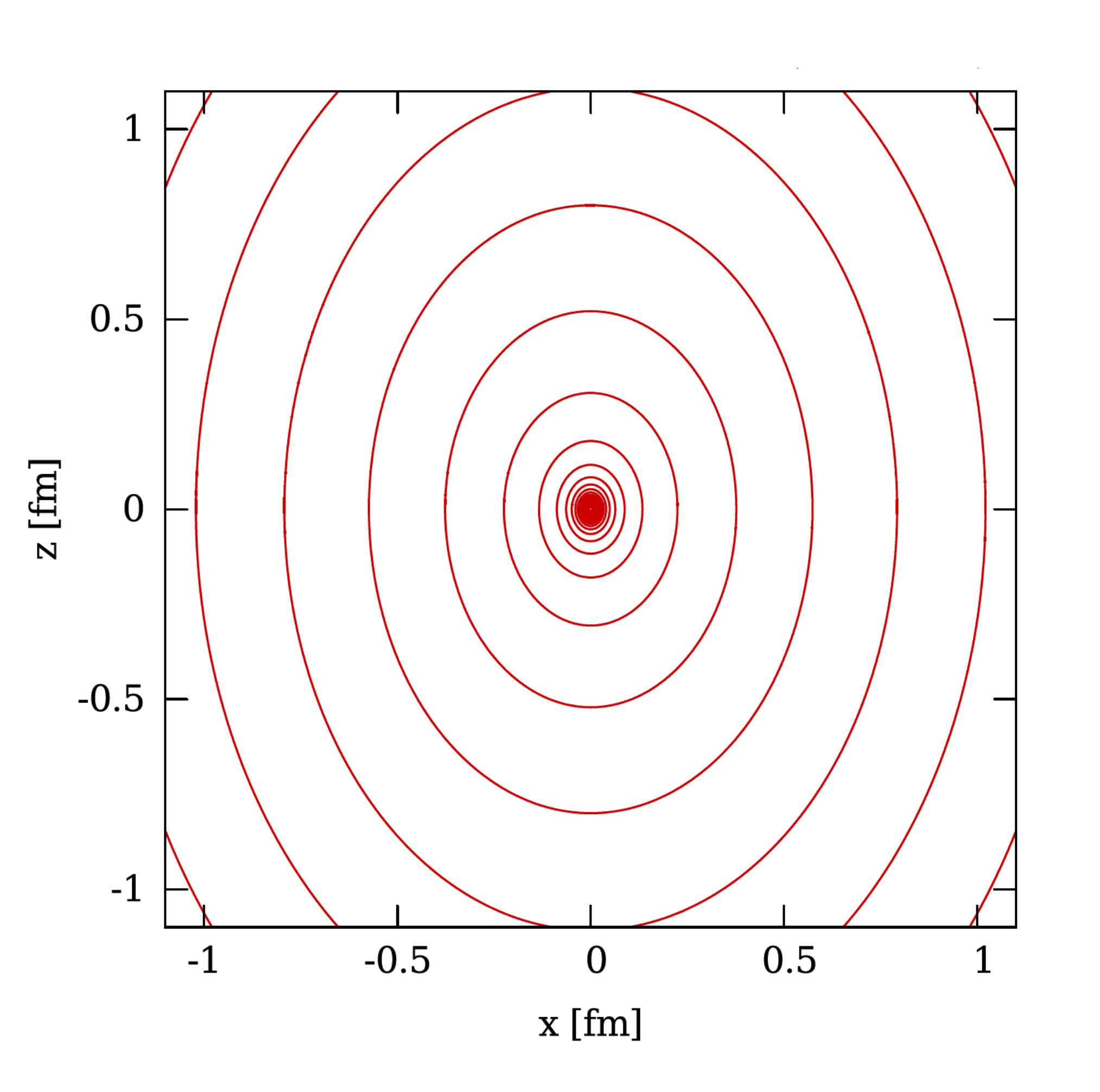}
		\end{center}
 \end{minipage}
    \vspace{-.5cm}
\caption{The $q \bar q $ potential from the lattice QCD simulation as a function of
the $ q \bar q$ separation (left) and of the azimuthal angle (contour plot on right). 
The separation is specified by the lattice coordinate $ n_s$ and spacing $ a$, 
and the magnetic field is applied in the $ z$-direction. 
The potential without a magnetic field ($ b=0$) is compared with 
the anisotropic potentials in the transverse ($ XY$) and the longitudinal $ (Z)$ directions, 
as well as the average of them ($ XYZ$), at a finite magnetic field 
with $b=24$ that is the discretized magnetic flux on the periodic lattice. 
(Reproduced from Refs.~\cite{Bonati:2014ksa,Bonati:2015dka}.)}
 \label{fig:potential_aniso}
\end{figure}

\begin{figure}[!t]
 \begin{center}
   \includegraphics[clip,width=0.65\hsize]{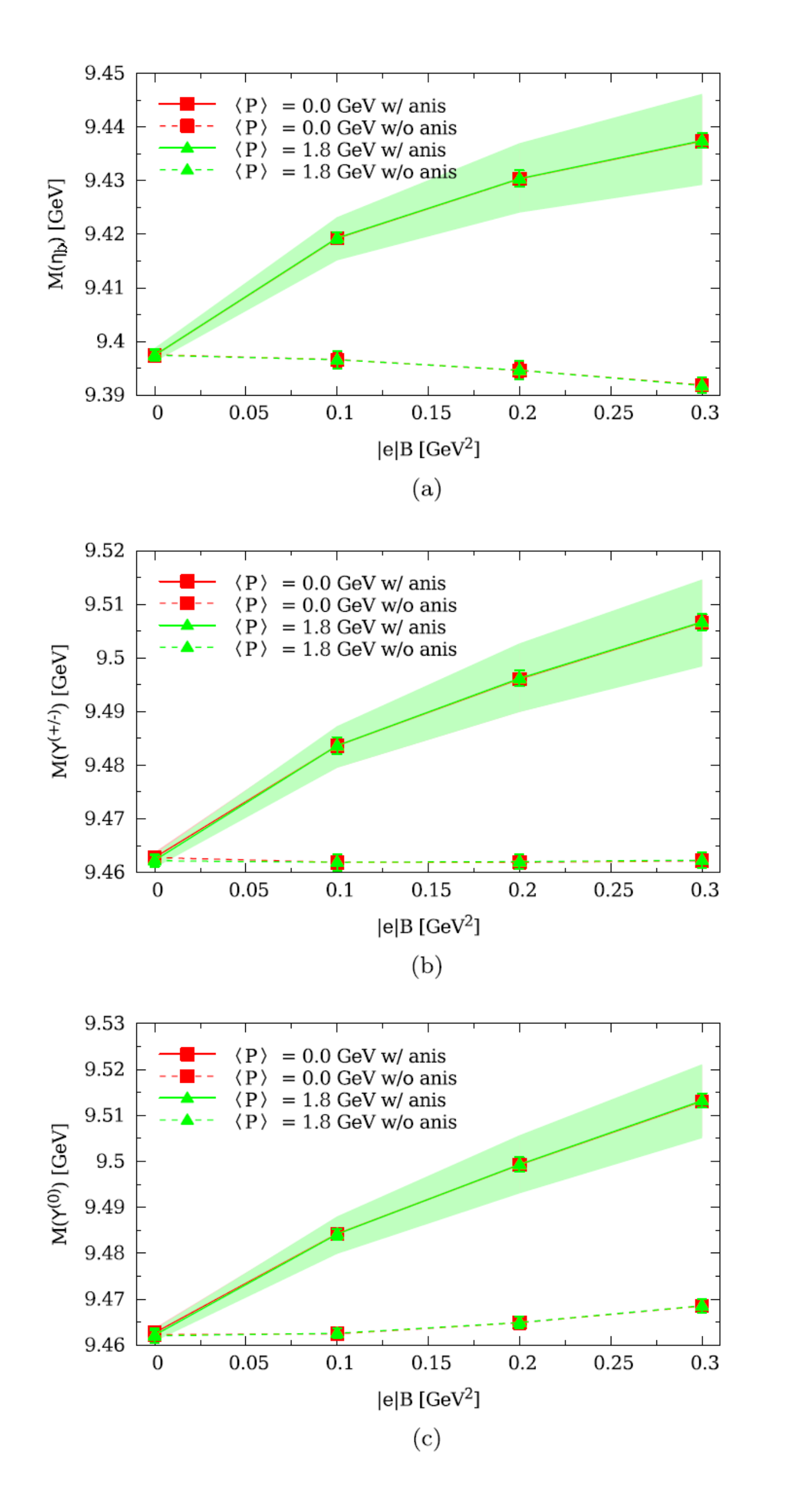}
    \end{center}
    \vspace{-.5cm}
     \caption{Bottomonium spectra with and without including the anisotropic potential from the lattice QCD.
     (Reproduced from Ref.~\cite{Bonati:2015dka}.)}
 \label{fig:bottom_aniso}
\end{figure}

\clearpage

So far, we have not considered the modification of the Cornell potential by the magnetic field.
Effects of a magnetic field on the potential was invastigated in Ref.~\cite{Bonati:2014ksa} by lattice QCD simulation.
The authors extracted the strengths of the potential in two cases where
the separation between the quark and antiquark is in parallel and transverse to the magnetic field.
The left panel in Fig.~\ref{fig:potential_aniso} shows the results from the lattice simulation.
Compared to the potential in vacuum, the strength of the potential is enhanced and suppressed
in the transverse $ (XY)$ and longitudinal $ (Z)$ directions, respectively.
This behavior may be understood from the fact that the screening due to the vacuum polarization of
light-quark pairs is coupled to only the temporal and parallel ($ z$) components of the gluon field.
The full azimuthal-angle and temperature dependences
were recently studied by the lattice QCD simulation \cite{Bonati:2016kxj}, and
those effects were also discussed by the strong couping method \cite{Rougemont:2014efa, Dudal:2014jfa, Sadofyev:2015hxa}.

The right panel in Fig.~\ref{fig:potential_aniso} shows the contour plot of the potential
which was obtained by interpolating the values in the transverse $ (XY)$ and longitudinal $ (Z)$ directions
with respect to the azimuthal-angle dependence.
In Ref.~\cite{Bonati:2015dka}, the quarkonium spectra were investigated by including
this anisotropic potential as well as the mixing effects.
Figure~\ref{fig:bottom_aniso} shows the bottomonium spectra with and without
the inclusion of the anisotropic potential.
The authors argued that effects of the anisotropic potential
can be seen in the bottomonium spectra more strongly than in the charmonium spectra, because the magnetic moment
of the bottom quark is suppressed by the larger value of the quark mass and the smaller electric charge.
Without the anisotropic potential, the spectra barely change due to the suppression of the mixing effects,
while the effects of the anisotropic potential act to increase the bottomonium masses in all three of the channels.
The effect of the anisotropic potential is also much stronger than the momentum dependence of the spectra.

On the basis of the potential model, several phenomenological consequences have been also discussed.
Prior to moving to the next section, we briefly summarize these studies.

\begin{itemize}

\item In the presence of the mixing effect, the wave function of
the mass eigenstate is the mixture of the two components,
so that the mass eigenstate shares the decay modes with both of the mixing partners.
This means, for example, that the physical $ \eta_c$ and $ \eta_b$ are allowed to decay into the dilepton,
while the dilepton decay channels of the physical longitudinal $ J/\psi$ and $\Upsilon $ are suppressed \cite{Alford:2013jva}.
As mentioned in Ref.~\cite{Alford:2013jva}, a similar modification of the positronium decay mode is known.

\item The Lorentz forces acting on a moving quark and antiquark pair
are oriented in the opposite direction. Therefore, the force will act to break up the bound state,
implying a possible dissociation mechanism in a strong magnetic field \cite{Marasinghe:2011bt, Alford:2013jva}.
This effect may be also understood as an effect of an electric field appearing
in the rest frame of a moving quarkonium which is connected to the original magnetic field by the Lorentz boost.
The electric field will stretch the separation between the quark and antiquark pair,
and induces a tunneling through the binding potential \cite{Marasinghe:2011bt}.

\item Mass shifts of $ D$ mesons were also examined with use of the potential model \cite{Machado:2013rta}.
In turn, this mass shifts induce the shift of the $ D \bar D$ threshold above the quarkonium states.
Based on the color evaporation model,
modification of the charmonium production rate was discussed \cite{Machado:2013rta}.

\item A time dependence of the magnetic field was introduced in the potential model \cite{Guo:2015nsa}.
Based on the time-dependent model, the authors investigated
the anisotropic production rate of high-$ p_T$ charmonia.
The feed-down effect from the excited states was also examined.

\end{itemize}


\subsubsection{QCD sum rules in magnetic fields}
\label{sec:QCDSR}

Effects of the magnetic field 
on the meson spectra were studied also by using QCD sum rule (QCDSR)
for open heavy flavor mesons \cite{Machado:2013yaa,Gubler:2015qok}
and quarkonia \cite{Cho:2014exa,Cho:2014loa}.
While we will here focus on the QCD sum rule for charmonia, extension to bottomonia is straightforward
and extension to other systems will be discussed in the last of this section.

In the QCD sum rules \cite{Shifman:1978bx,Shifman:1978by,Reinders:1984sr,Shifman:1998rb},
the spectra of bound states are extracted from
the time-ordered current correlator defined by
\begin{eqnarray}
\Pi^J(q) = i \int \!\! d^4x \, e^{iqx} \langle 0 \vert T[ J(x) J(0) ] \vert 0 \rangle
\label{eq:JJ}
\ ,
\end{eqnarray}
where the superscript $J$ specifies a channel.
The dispersion relation relates the hadronic spectral function in the time-like region ($q^2 = - Q^2 \geq 0$)
to the correlator in the deep Euclidean region ($Q^2 \to \infty$) as
\begin{eqnarray}
\Pi^\scJ( Q^2 )  = \frac{1}{\pi} \int_0^\infty
\frac{ \, {\rm Im} \, \Pi^\scJ (s) \, }{ s + Q^2 } \, ds
\ + \ ({\rm subtractions})
\label{eq:disp0}
.
\end{eqnarray}

One may evaluate the correlator on the left-hand side
on the basis of the Operator Product Expansion (OPE) in the deep Euclidean region.
Here, we shall focus on the charmonia created by the currents:
\begin{eqnarray}
J_P = i \bar c \gamma^5 c , \ J^\mu_V = \bar c \gamma^\mu c
.
\end{eqnarray}
Up to the dimension-4 operators, the OPE can be written down as
\begin{eqnarray}
\Pi &=& C_0 + C_{G^2} \langle G^a_{\mu\nu} G^{a \, \mu\nu} \rangle
+ C_{B^2}^{\mu\nu\alpha\beta} \langle F^\ext_{\mu\nu} F^\ext_{\alpha\beta} \rangle
\label{eq:ope}
\, .
\end{eqnarray}
The Wilson coefficients, $ C_0 $, $ C_{G^2} $, and $C_{B^2} $,
are responsible for the short-distance processes, so that one may
perturbatively calculate the diagrams for the hard quark loops with
insertions of the soft external gluon/magnetic fields (see Refs.~\cite{Cho:2014exa,Cho:2014loa} for explicit expressions).
This can be done by using standard methods in the Fock-Schwinger gauge
which is also as known as the fixed-point gauge \cite{Reinders:1984sr,Novikov:1983gd}.

On the other side of the dispersion relation (\ref{eq:disp0}),
we assume an Ansatz for the functional form of the spectral function
$\rho(s) = \pi^{-1} {\rm Im} \, \Pi^\scJ (s) $ to extract the lowest-lying pole.
In conventional analyses, 
the lowest-lying pole is simply separated from the continuum as
\begin{eqnarray}
\rho_{\rm vac}(s) =
f_0 \delta (s -m ^2) +  \frac{1}{\pi} {\rm Im} \, \Pi^\scJ_{\rm pert} (s) \theta(s - s_{\rm th}).
\label{eq:pole}
\end{eqnarray}
The parameters $s_{\rm th}$ and $f_0$ are the threshold of the continuum and the coupling strength
between the current and the lowest-lying pole, respectively.

\begin{figure}[t!]
 \begin{center}
   \includegraphics[clip,width=0.5\hsize]{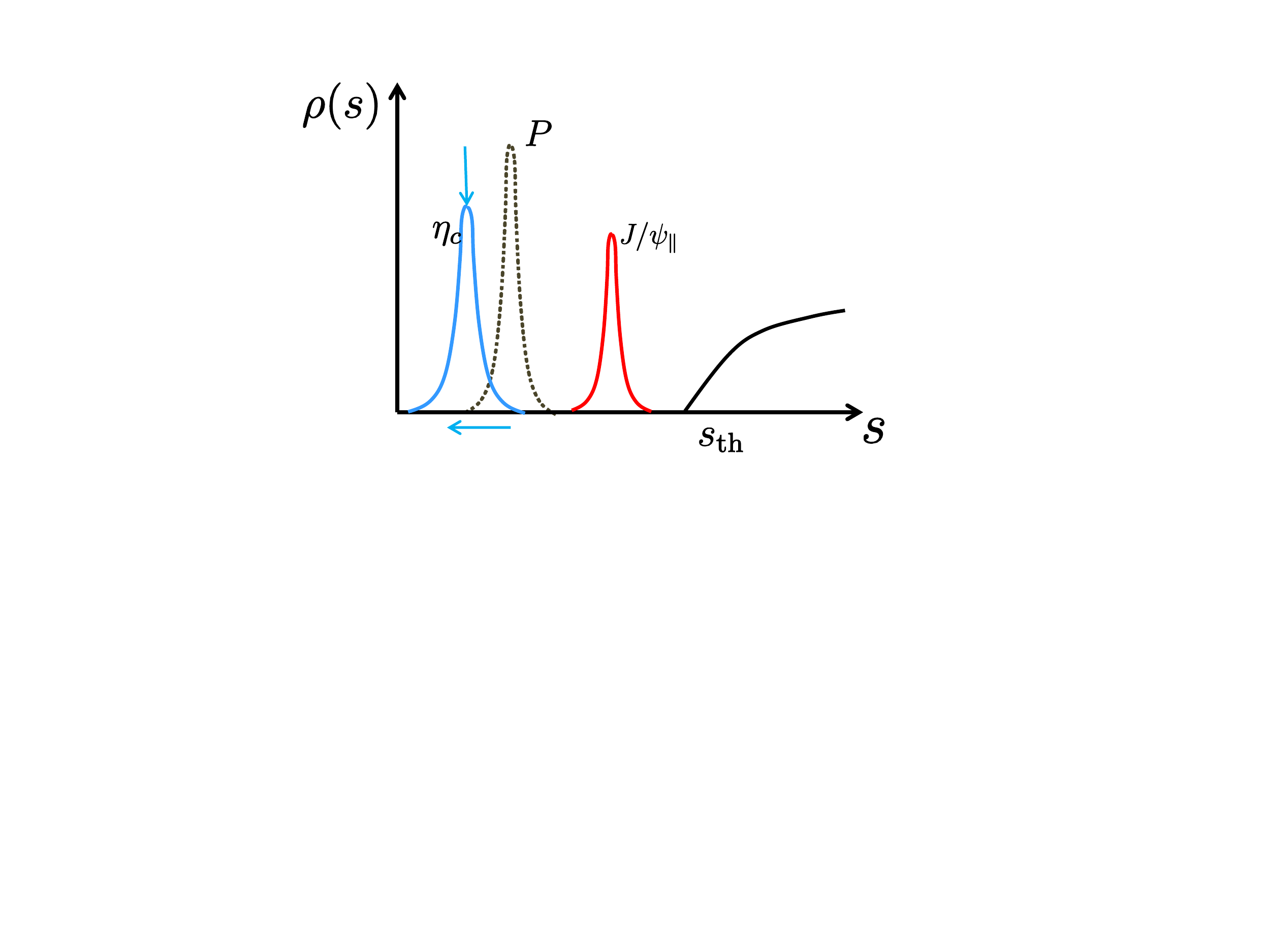}
    \end{center}
    \vspace{-.5cm}
     \caption{Sketch of the spectral function in the pseudoscalar channel. The longitudinal $ J/\psi_\parallel$ pole emerges,
     which induces the repulsive shifts of the poles and the reduction of the residue of the $ \eta_c$ pole.
     Similar happens in the longitudinal vector channel.}
 \label{fig:spectralF}
\end{figure}

We, however, need to carefully reexamine the Ansatz (\ref{eq:pole}) in the presence of a magnetic field,
because effects of a magnetic field should be consistently taken into account
on both sides of the dispersion relation (\ref{eq:disp0})
\cite{Machado:2013yaa,Gubler:2015qok,Cho:2014exa,Cho:2014loa}. 
The effect in the hadronic degrees of freedom is nothing but the mixing effect discussed in the previous sections.
In the language of the spectral function, the mixing effect can be seen as the emergence of a new pole for the mixing partner
which is the longitudinal $J/\psi$ in the pseudoscalar channel and the $ \eta_c$ in the vector channel.
A schematic picture in Fig.~\ref{fig:spectralF} shows the emergence of the longitudinal $J/\psi$
in the pseudoscalar channel, which induces the repulsive shifts of the poles
and the reduction of the residue of $\eta_c$.
It is important to include these effects in order to
properly maintain the normalization of the spectral function.

In the pseudoscalar channel, the second-order term in $ eB$ can be obtained by first converting the current
to the psedoscalar meson with the strength $f_0$ and then using the results
for the mixing effect shown in Eqs.~(\ref{eq:Jpsi_2nd}) and (\ref{eq:wave}).
Therefore, the correlator can be written in the hadronic degrees of freedom as \cite{Cho:2014exa}
\begin{eqnarray}
\Pi_{\rm 2nd}^P (q^2) =
f_0  \left[ \
\frac{\vert \langle P \vert \etac \rangle_{\scriptscriptstyle B} \vert^2}{q^2-m_\etac^2}
+ \frac{\vert \langle P \vert \Jp \rangle_{\scriptscriptstyle B} \vert^2}{q^2-m_\Jp^2}
\ \right]
\label{eq:2nd}
.
\end{eqnarray}
Inserting Eqs.~(\ref{eq:Jpsi_2nd}) and (\ref{eq:wave}) into Eq.~(\ref{eq:2nd})
and expanding up to the second order in $eB$, we find
\begin{eqnarray}
&\hspace{-0.25cm}
\Pi^P_{\rm 2nd}(q^2)
=
f_0 \frac{\gamma^2}{M_-^4 }
\left[\frac{1}{q^2 - m_{\V}^2} - \frac{1}{q^2 - m_{\ps}^2}
- \frac{ M_-^2 }{ (q^2-m_{\ps}^2)^2 }  \right]
\label{eq:exp}
.
\end{eqnarray}
Interpretation of the terms in Eq.~\eqref{eq:exp} are as follows.
The first term corresponds to production of the on-shell longitudinal $J/\psi$
from the pseudoscalar current via an off-shell $\eta_c$.
The second term with a negative sign is needed to preserve the normalization,
because the coupling of $\eta_c$ to the current must be reduced to
balance the emergence of the coupling to $J/\psi$.
This is confirmed in Eq.~(\ref{eq:2nd}), where these two terms come from
overlaps between the properly normalized unperturbated and perturbated states
obtained as $\vert ( P \vert \etac )_{\scriptscriptstyle B} \vert^2 \sim 1 - (\gamma/M_-^2)^2$
and $\vert ( P \vert \Jp )_{\scriptscriptstyle B} \vert^2 \sim (\gamma /M_-^2)^2$.
The third term has a double pole on the $\eta_c$ mass with a
factor $M_-^2$, corresponding to a virtual transition to the longitudinal $J/\psi$
state between on-shell $\eta_c$ states, which is nothing but the
origin of the mass shift due to the mixing effect. In
Eq.~(\ref{eq:2nd}), this term comes from an expansion with respect
to the mass correction shown in Eq.~(\ref{eq:Jpsi_2nd}). Clearly, if one
includes this mixing term in the phenomenological spectral function,
its effect is subtracted out from the total mass shift obtained from the QCDSR,
and thus can be separated from the residual effects of $B$-fields, not described in the hadronic level.
A similar calculation can be applied to the longitudinal vector channel.
On the phenomenological side,
those terms should be included as well as the vacuum contributions (\ref{eq:pole})
in both the pseudoscalar and longitudinal vector channels.

\begin{figure}
		\begin{center}
			\includegraphics[width=0.4\hsize]{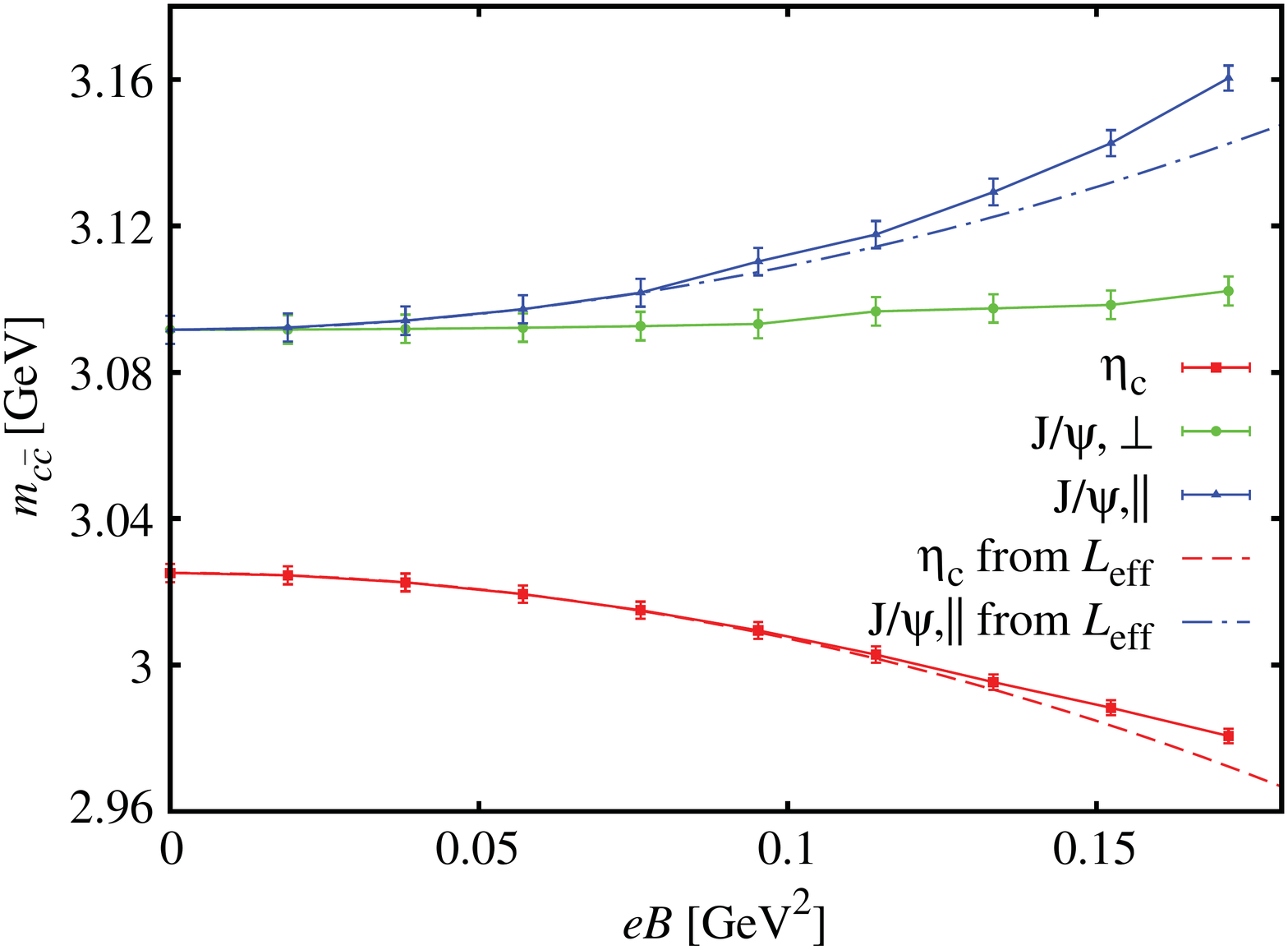}
		\end{center}
        \vspace{-0.5cm}
\caption{The $\eta_c $ and $J/\psi $ mass spectra from the QCD sum rule
together with the results from the hadronic effective model shown in Fig.~\ref{fig:Qshift}.
(Reproduced from Ref.~\cite{Cho:2014exa}).
}
\label{fig:QCDSR}
\end{figure}

To make the dispersion integral (\ref{eq:disp0})
dominated by the pole contribution and to improve the convergence of the OPE,
we use the Borel transform defined by
\begin{eqnarray}
\B[f(Q^2)] \equiv \!\!
\lim_{ \substack{Q^2, n \rightarrow \infty \\ Q^2/n = M^2 } }
\frac{(Q^2)^{n+1}}{n!} \left( - \frac{d \ }{dQ^2 } \right)^n f(Q^2)
\label{eq:Borel}
.
\end{eqnarray}
By applying the Borel transform to the both sides of Eq.~(\ref{eq:disp0}),
the contribution of the delta function for the lowest-lying pole
is picked up to result in a mass formula:
\begin{eqnarray}
m^2 =  - \frac{\partial \ }{ \partial(1/M^2)} \ln[ \M_{\rm OPE} - \M_{\rm cont} - \M_{\rm mix}]
\label{eq:mass}
\, ,
\end{eqnarray}
where $ \M_{\rm OPE} $, $ \M_{\rm cont} $, and $ \M_{\rm mix}$
are the Borel transform of the OPE (\ref{eq:ope}), the perturbative continuum,
and those induced by the magnetic field \eqref{eq:exp}, respectively.
More details and explicit forms are given in Refs.~\cite{Cho:2014exa,Cho:2014loa}.

Figure~\ref{fig:QCDSR} shows the mass spectra from the QCDSR
together with those from the hadronic effective model shown in Fig.~\ref{fig:Qshift}.
First, the results from these two methods are consistent with each other,
indicating the dominance of the mixing effect in the shown regime.
They are also consistent with the results from the potential model discussed
in the previous section \cite{Alford:2013jva,Suzuki:2016kcs,Yoshida:2016xgm,Bonati:2015dka}.
Also, the modification in the transverse components of $ J/\psi$ (green line) is much smaller
than the others due to the absence of the mixing partner.
On top of these observation, one finds deviations between the results from the two methods.
Although they are small deviations in the current regime of the magnetic field,
they suggest the precursor of other effects which are not explained by the mixing effect.
It would be natural to expect the emergence of such effects in stronger magnetic fields.
Possible origin of these deviations were discussed in Ref.~\cite{Cho:2014loa}.

One caveat is that the modification of the gluon condensate in a magnetic field
has not been included in the evaluation of the OPE (\ref{eq:ope}).
Since the lattice QCD simulation has provided the values of the gluon condensate
in finite magnetic field and temperature \cite{Bali:2013esa}, it will be interesting to investigate
effects of these modifications by using the QCDSR.

Also, applications of the QCDSR to light mesons and open heavy flavor mesons
are an interesting topic, because these mesons contain a light quark component
which reflects the modification of the quark condensates.
In Refs.~\cite{Machado:2013yaa,Gubler:2015qok},
the OPE was performed by taking into account not only the modification of $\langle \bar q q \rangle $
but also the emergence of the tensor-type condensate $\langle \bar q \sigma^{\mu\nu} q \rangle $
which have been calculated by the lattice QCD simulations \cite{Bali:2011qj,D'Elia:2011zu,Bali:2012zg,Bruckmann:2013oba}.
Furthermore, the mass spectra of the open heavy flavor mesons would serve as
a signature of the magnetically induced QCD Kondo effect \cite{Ozaki:2015sya}
which was proposed on the basis of a close analogy to the QCD Kondo effect in the dense QCD \cite{Hattori:2015hka,Yasui:2016svc}.
The readers are referred to the literature for the interesting analogy
between the (1+1)-dimensional low-energy dynamics
at high densisy \cite{Hong:1998tn,Hong:1999ru,Schafer:2003jn}
and in strong magnetic fields \cite{Gusynin:1994xp, Hong:1996pv, Hong:1997uw, Kojo:2012js}.


 \subsection{Transport of heavy quarks in a hot medium and magnetic field}

\label{sec:HQtrans}

\begin{figure*}[t!]
 \begin{center}
   \includegraphics[clip,width=0.9\columnwidth]{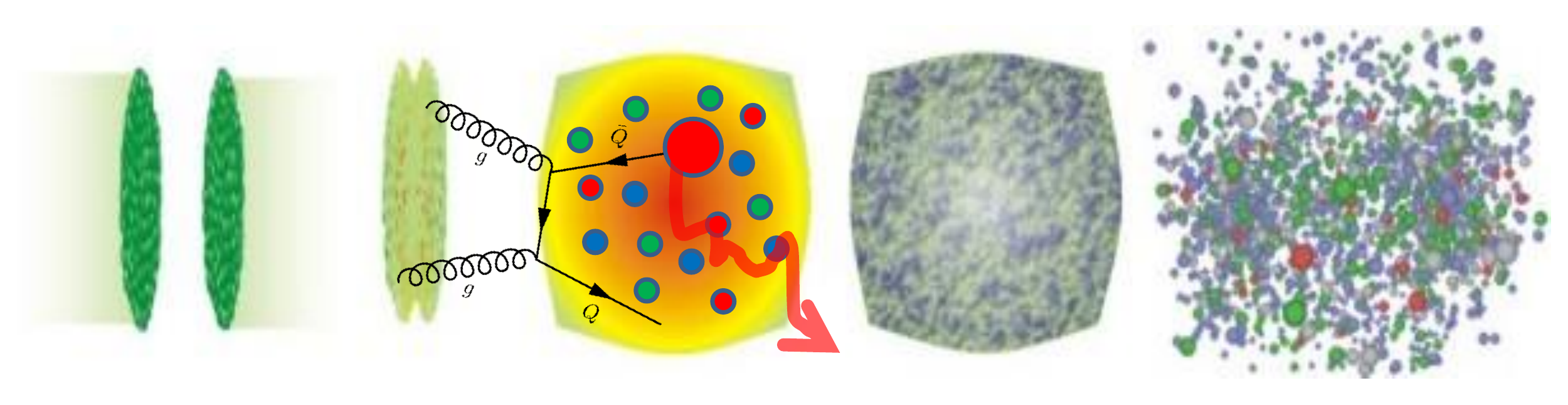}
    \end{center}
    \vspace{-.5cm}
     \caption{Brownian motion of heavy quarks created by the initial hard processes.}
 \label{fig:Brownian}
\end{figure*}

As mentioned in the beginning of this section,
the heavy quarks are dominantly created by the initial hard scatterings
among the partons from the colliding nuclei (see Fig.~\ref{fig:Brownian}).
Therefore, the heavy quarks typically have the power law $ p_T$ spectrum according to the perturbative QCD.
This spectrum is however gradually modified by the multiple interactions with
the thermal quarks and gluons in the QGP, and the spectrum approaches a thermal distribution
which is realized at a large enough time.
However, the QGP has a finite lifetime, and to what extent the thermalization
is achieved depends on the transport property of the QGP.
Therefore, the heavy quarks will probe the transport property of the QGP.

In the RHIC and LHC experiments, the nuclear modification factor $ R_{AA}$
and the anisotropic spectrum $ v_2$ of the open heavy flavors have been measured
(see a review~\cite{Andronic:2015wma} and references therein).
Both of them are thought to be sensitive to the heavy-quark thermalization,
and are closely related to each other \cite{Moore:2004tg}.
However, a consistent theoretical modeling, which simultaneously
reproduces $ R_{AA}$ and $v_2 $, is still under investigation.
Effects of the magnetic field could be new ingredients to resolve the discrepancy between
the theoretical and experimental results.
In this section, we summarize a recent study on the heavy quark dynamics
in a hot medium and magnetic field on the basis of Langevin equation
and the perturbative computation of the heavy-quark diffusion constant.


The dynamics of the low-energy heavy particles is modelled
as a Brownian motion caused by the random kicks by the thermal particles,
which is described by the Langevin equations
\cite{Moore:2004tg}:
\begin{eqnarray}
\label{eq:Langevin}
\frac {d p_z}{ dt}
 = -\eta_\parallel p_z + \xi_z \,,
 \qquad
\frac{d\bm p_\perp}{ dt} =
   -\eta_\perp \, \bm p_\perp + \bm\xi_\perp \,.
\end{eqnarray}
Since the external magnetic field provides a preferred spatial direction,
we have a set of two equations for the heavy quark motions,
parallel and perpendicular to the magnetic field that is oriented in the $z$-direction.
The random forces are assumed to be white noises,
\begin{eqnarray}
 \langle \xi_{z}(t) \xi_z(t')\rangle=\kappa_\parallel \delta(t-t')\,,\qquad
 \langle \xi_{\perp}^i(t)\xi_\perp^j(t')\rangle=\kappa_\perp
  \delta^{ij}\delta(t-t') \quad (i,j=x,y)
\end{eqnarray}
and these coefficients, $\kappa_\parallel$ and $\kappa_\perp$, are
related to the drag coefficients, $\eta_\parallel$ and $\eta_\perp$,
through the fluctuation-dissipation theorem as
\begin{eqnarray}
\label{eq:fdtheorem}
  \eta_{\parallel} = 2 M_{Q}T \kappa_{\parallel} \, , \qquad
  \eta_{\perp}= 2 M_{Q}T \kappa_{\perp}
\, .
\end{eqnarray}
Here, we would like to mention a recently numerical study \cite{Das:2016cwd} where the authors
investigated the Lorentz force exerting on the heavy quark
with the inclusion of the Lorentz force in the Langevin equation (\ref{eq:Langevin}).

\begin{figure}[t]
		\begin{center}
			\includegraphics[width=0.75\hsize]{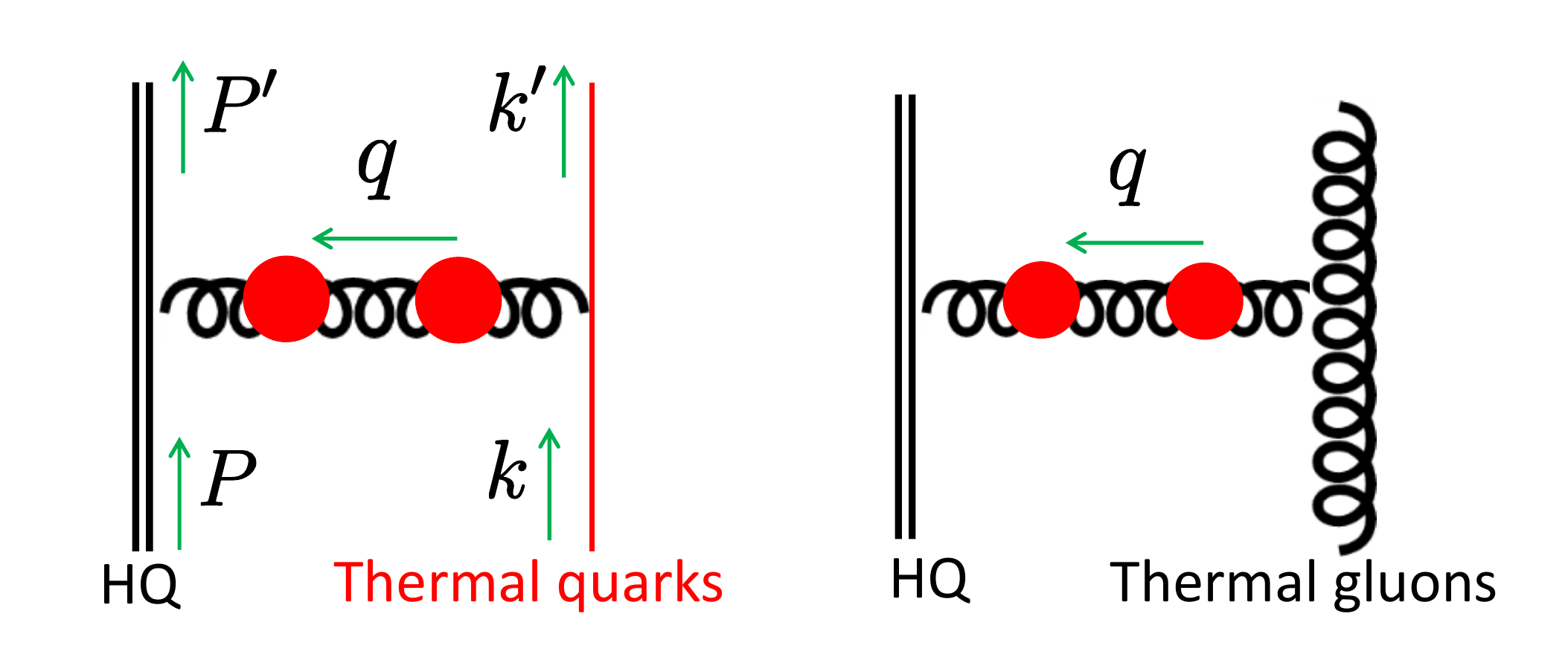}
		\end{center}
        \vspace{-0.5cm}
\caption{Coulomb scattering amplitudes contributing to the heavy-quark (HQ) momentum diffusion.
A magnetic field acts on the quark loop in the polarization and the thermal-quark scatterers.}
\label{fig:scatterings}
\end{figure}

In addition to effects of the Lorentz force exerting on the heavy quark as a {\it direct} effect of the magnetic field,
the diffusion and drag coefficients will be changed because
the light quarks in the hot medium are strongly affected by the magnetic field \cite{Fukushima:2015wck,Finazzo:2016mhm}.
At the leading order in $ g_s$, the anisotropic momentum diffusion coefficients, $\kappa_\parallel$ and
$\kappa_\perp$, can be defined by
 \begin{eqnarray}
 \kappa_{\parallel} = \int {d^3\bm q}\,
  \frac{d\Gamma(\bm q)}{d^3\bm q}\, q_{z}^2 \,,
 \qquad
 \kappa_{\perp} = {1\over 2}\int {d^3\bm q}\,
\frac{d\Gamma(\bm q)}{ d^3\bm q}\, \bm q_{\perp}^2
  \, .
\label{kappa}
\end{eqnarray}
where $ q$ is the amount of the momentum transfer from the thermal particles to the heavy quark,
and the static limit ($ q^0 \to 0$) is assumed in the above definitions.
The momentum transfer rate $   {d\Gamma(\bm q)\over d^3\bm q}$ is
computed from the gluon-exchange diagrams shown in Fig.~\ref{fig:scatterings}.
Effects of a magnetic field appear in two places (highlighted by red).
While the gluons are not directly coupled to the magnetic field,
(i) the Debye screening mass and (ii) the dispersion relation of the thermal-quark scatterers
will be changed as discussed below.
The Debye mass is necessary for cutting off the infrared divergence in the forward scattering.

In a strong magnetic field, the fermion wave function is strongly squeezed along the magnetic field,
corresponding to the small radius of the cyclotron orbit.
Indeed, from the Landau level quantization and the Zeeman effect,
the dispersion relation of fermions in the lowest Landau level (LLL)
becomes the (1+1) dimensional, i.e.,
$ \epsilon^2 = m^2 + p_z^2$ for a massive fermion and
$ \epsilon = \pm p_z$ for massless fermions with right- and left-handed chiralities.
To be specific, we focus on the strong-field regime such that
the transition from the LLL to the hLL states are suppressed according to a hierarchy $ T^2 \ll eB $.

(i) First, the Debye screening mass is obtained form the gluon self-energy
computed by the standard method in finite temperature field theory in the (1+1) dimension.
The gluon self-energy is completely factorized into the transverse and longitudinal parts as
\begin{eqnarray}
\Pi^{\mu\nu} (q)  = q_f \frac{ eB}{2\pi} e^{ - \frac{\vert\bq_\perp\vert^2}{2\vert q_f eB \vert} }
\Pi_{1+1}^{\mu\nu} (q_\parallel)
\, ,
\end{eqnarray}
where the factor of $ q_f \frac{ eB}{2\pi}$ comes from the degeneracy factor
in the transverse phase space, and the Gaussian is the wave function of the LLL state.
The longitudinal part is nothing but the (1+1)-dimensional polarization tensor
which is known as the Schwinger model:
\begin{eqnarray}
\Pi_{1+1}^{\mu\nu} (q_\parallel) = {\rm tr}[t^at^a]
\frac{g_s^2}{\pi}  f(q^0,q_z)
(q_\parallel^2 g_\parallel^{\mu\nu} - q^\mu_\parallel q^\nu_\parallel)
\, .
\end{eqnarray}
Here, the longitudinal momentum is defined by $ q_\parallel^\mu = (q^0,0,0,q_z)$.
There is only one gauge-invariant tensor structure in the (1+1) dimension,
so that the temperature correction is, if any, contained in a function $f(q^0,q_z) $.
However, an important observation is that
there is no temperature or density correction in the massless case \cite{Dolan:1973qd,Baier:1991gg,Fukushima:2015wck},
and $ f(q^0,q_z) = 1 $.
Accordingly, in massive case, a temperature correction is proportional
to the quark mass, and is suppressed by $m_q/T $ (see Ref.~\cite{Fukushima:2015wck} for explicit expressions).
Therefore, in both cases, the Debye mass in strong magnetic fields ($eB \gg T^2 $)
is given by the vacuum part of the gluon self-energy \cite{Fukushima:2011nu,Hattori:2012je,Hattori:2012ny}.
Namely, we find $ m_D^2 \sim q_f \frac{eB}{2\pi} \cdot \frac{g_s^2}{\pi} $ 
which is much larger than the usual thermal mass squared $ \sim (g_s T)^2$.

(ii) The change in the dispersion relation of fermion scatterers is
important for the kinematics of the Coulomb scattering process \cite{Sadofyev:2015tmb,Fukushima:2015wck}.
We shall consider a massless fermion scatterer in the lowest Landau level (see Fig.~\ref{fig:scatterings}).
For the massless quarks to be on-shell both in the initial and final states,
the energy transfer has to be $q^0 = \pm (k_z^\prime - k_z) $, and thus $ q^0 = \pm q_z$.
Note that the chirality does not change at any perturbative vertex,
so that the signs appear only as the overall ones.
When taking the static limit $q^0 \to 0 $ in the definition of the diffusion constant,
one can immediately conclude that the longitudinal-momentum transfer
is kinematically prohibited in the massless case.
On the other hand, this constraint does not apply to the transverse momentum transfer,
so that the transverse momentum transfer is allowed.
This finite contribution to the diffusion constant can be obtained either by the direct computation
of the matrix elements in Fig.~\ref{fig:scatterings} or by using the cutting rule.
At the leading order in $ \alpha_s$, the transverse momentum diffusion constant
in the massless limit is obtained as \cite{Fukushima:2015wck}
\begin{eqnarray}
 \kappa_\perp^{\rm LO} \sim \alpha_s^2 T
  \biggl({eB\over 2\pi}\biggr) \ln \alpha_s^{-1}
\label{eq:LO}
\, .
\end{eqnarray}
Therefore, strong magnetic fields give rise to an anisotropy of the momentum diffusion constant.

Nonvanishing contributions to the longitudinal momentum transfer come from
either the finite quark-mass correction or the contribution of the gluon scatterers.
In Ref.~\cite{Fukushima:2015wck}, the contribution of the mass correction was obtained as
\begin{eqnarray}
\kappa_\parallel^{\rm LO,\,massive} \sim
\alpha_s    m_q^2 \sqrt{\alpha_s eB}
    \label{eq:LO_mass}
    \, ,
\end{eqnarray}
while the gluon contribution is obtained by substituting the Debye mass $ m_D^2 \sim \alpha_s eB$
for the conventional one $ \sim (g_sT)^2$ in Refs.~\cite{Moore:2004tg,CaronHuot:2007gq} as
\begin{eqnarray}
 \kappa^{\rm LO,\,gluon}_\parallel
\sim \alpha_s^2 T^3 \ln \alpha_s^{-1}
\label{eq:LO_gluon}
  \, .
\end{eqnarray}
We have assumed a hierarchy $ \alpha_s eB \ll T^2 \ll eB$.
Complete expressions including the prefactors are found in Ref.~\cite{Fukushima:2015wck}.
It is instructive to compare the mass correction \eqref{eq:LO_mass} with the gluon contribution \eqref{eq:LO_gluon}.
The ratio is written as
\begin{eqnarray}
\frac{\kappa_\parallel^{\rm LO,\,massive} }{\kappa_\parallel^{\rm LO,\,gluon}} \sim
\frac{\alpha_s (\alpha_s eB)^{1/2} m_q^2 }{ \alpha_s^2 T^3}
 = \biggl( \frac{m_q^2 }{ \alpha_s eB}\biggr)
 \biggl( \frac{\alpha_s eB }{ T^2}\biggr)^{1/2}
 \biggl( \frac{eB}{T^2}\biggr)
 \,.
\end{eqnarray}
While the first two factors are small in our working regime,
the last factor can be large.  Therefore, the massive contribution
$\kappa_\parallel^{\rm LO,\, massive}$ could be in principle as
comparably large as $\kappa_\parallel^{\rm LO,\, gluon}$, and this
happens when $eB\sim \alpha_s(T^6/m_q^4)$.  However, to be consistent
with our assumed regime, $\alpha_s\,eB\ll T^2$, we have a constraint
of $\alpha_s\ll m_q^2/T^2$, which is not quite likely true in the heavy ion collisions.
Hence,  the longitudinal momentum diffusion constant
is dominated by the gluon contribution $\kappa_\parallel^{\rm LO,\,gluon}$.

Now that we obtained the leading contributions in Eqs.~(\ref{eq:LO}) and (\ref{eq:LO_gluon}),
the anisotropy in the momentum diffusion constant and the drag force is estimated to be
\begin{eqnarray}
\frac{\kappa_\parallel^{\rm LO,\,gluon}}{\kappa_\perp^{\rm LO}}
=
\frac{\eta_\parallel^{\rm LO,\,gluon}}{\eta_\perp^{\rm LO}}
\sim \frac{T^2}{eB} \ll 1
\, .
\end{eqnarray}
This large anisotropy is induced by the enhancement of the density of states
in the quark contribution $ \kappa_\perp^{\rm LO}$, which is proportional to $ eB$.
On the other hand, a magnetic field does not change the phase space volume of the thermal gluons
in the gluon contribution $ \kappa_\parallel^{\rm LO,\,gluon}$, but changes only the Debye screening mass.

\begin{figure*}
\begin{minipage}[t]{0.49\hsize}
		\begin{center}
			\includegraphics[width=0.8\hsize]{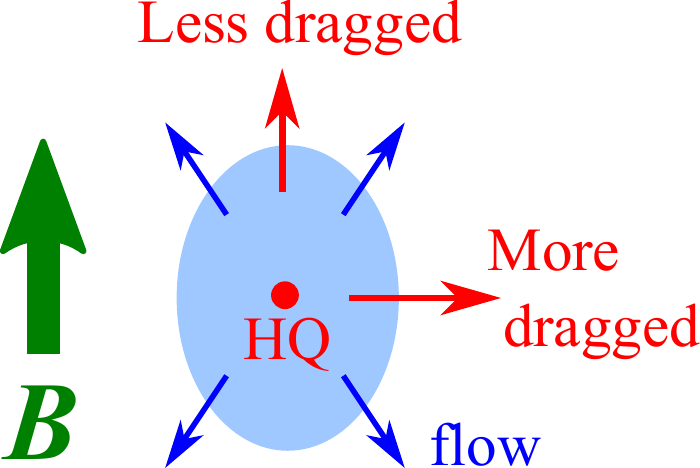}
		\end{center}
 \end{minipage}
\begin{minipage}[t]{0.49\hsize}
		\begin{center}
			\includegraphics[width=\hsize]{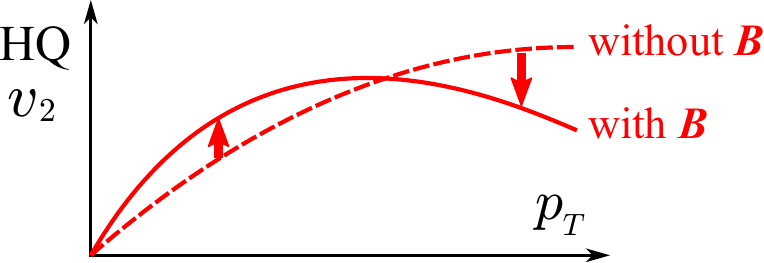}
		\end{center}
\end{minipage}
\caption{Schematic pictures of the anisotropic drag force in a magnetic field (left) and
the resultant anisotropic heavy-quark spectrum in the final state (right).
(Reproduced from Ref.~\cite{Fukushima:2015wck})}
\label{fig:drag}
\end{figure*}

Based on the above findings, a toy model for the dynamical modeling
was discussed in Ref.~\cite{Fukushima:2015wck}
by using the Fokker-Planck equation \cite{Moore:2004tg}.
Since the friction exerting on the heavy quarks is stronger
in the direction transverse to the magnetic field (see Fig.~\ref{fig:drag}),
the heavy-quark momentum in the final state will acquire an anisotropy.
In the small momentum region, the heavy quarks will be dragged by the flow more strongly in the transverse direction,
while they will feel a larger resistance in the transverse direction
as the momentum increases.
Therefore, there is a turnover in the anisotropy
of the heavy quark spectrum (see the right panel in Fig.~\ref{fig:drag}).

It will be interesting to implement the anisotropic diffusion and drag coefficient
in the numerical simulations. As the magnetic field in the heavy-ion collisions is
time dependent, it is also necessary to investigate the weak-field limit as
the first step to consider the time dependence.
These issues are left as open questions.
Finally, the drag force in the presence of the anomalous currents
was also investigated in Ref.~\cite{Rajagopal:2015roa, Stephanov:2015roa, Sadofyev:2015tmb}.

\section {Summary}\label{sec:summ}
In summary, we have discussed the recent progresses in understanding the novel quantum phenomena in heavy-ion collisions. We first briefly reviewed some special features of the magnetic fields generated in heavy-ion collisions from both numerical and analytical approaches in \sect{sec:fields}. The magnetic fields in RHIC Au + Au and in LHC Pb + Pb collisions can be comparable to or even larger than the QCD confinement scale, $\L^2_{\rm QCD}$, and thus are able to significantly influence the physics of strongly interacting matter.

The interplay between the magnetic fields and the non-trivial topological configuration of QCD leads to a number of anomalous transports, for example, the chiral magnetic effect (CME), chiral separation effect (CSE), chiral magnetic wave (CMW). These effects, once experimentally confirmed, provide ``soft probes" of the topological sector of QCD in heavy-ion collisions. We discussed the basic mechanisms that lead to the occurrence of CME, CSE, and CMW in \sect{sec:cme} and \sect{sec:cmw}. The experimental search of CME and CMW in heavy-ion collisions has been performed for years and the data already shows very encouraging features that are consistent with the expectations of CME and CMW. However, background effects which are not related to CME and CMW exist and make the interpretation of the experimental measurements ambiguous. More efforts from both experimental and theoretical sides are certainly needed before we can make any conclusive claims. We discussed the current experimental status of the search of CME and CMW in \sect{sec:corrcme} and \sect{sec:expcmw}.

In the latter half of this short review, we discussed effects of
the magnetic fields on the heavy quark sector which are expected
to be observed by hard probes in the heavy-ion collisions.
In Sec.~\ref{sec:QQbar}, we first discussed the quarkonium spectra in magnetic fields,
which has close connection to the recent lattice QCD simulations
on the properties of the QCD vacuum, especially the confinement potential.
Extension to the finite temperature and/or density is left for the future study.
Finally, in Sec.~\ref{sec:HQtrans}, we discussed a recent study on the heavy-quark transport
in a hot medium and magnetic field on the basis of the Langevin equation.
The momentum diffusion constant was computed by using the joint resummation
for Hard Thermal Loop and strong magnetic field.
While the computation has been performed for the lowest Landau level in the strong-field limit so far,
extension to other regimes, e.g., weak field, and inclusion of the higher Landau levels are
relevant for the phenomenology in the heavy-ion physics.

The topics covered by the present review are still fast developing, a number of important issues should be further investigated in future. Here we briefly discuss a few of them. 1) Although there are already thorough numerical simulations of the magnetic fields in the initial stage of the heavy-ion collisions, the precise time evolution of the magnetic fields is still unknown, see discussions in \sect{sec:fields}. The knowledge of the precise time evolution of the magnetic fields is very important for quantitative understanding of a number of magnetic-field induced phenomena, like the CME, CMW, etc. Thus detailed simulations of the time dependence of the magnetic field by using, for example, the parton cascade models or by using magnetohydrodynamics are very desirable. 2) The experimental measurements of the CME and CMW contain significant background contributions, see \sect{sec:anoma}; to extract the CME or CMW signals we have to understand these background contributions at the quantitative level which still remains an unsolved problem and will be one of the most important directions to pursue in this field. 3) Heavy flavors will serve 
as an alternative probe to explore the dynamics in the strong magnetic field 
and to constrain the spacetime profile of the magnetic field. 
Therefore, it will be important to develop the numerical modelling 
of the heavy-flavor dynamics with the inclusion of the effects of the magnetic field. 
It is also interesting to investigate other transport coefficients. 
For example, the electrical conductivity will be an important input for the magnetohydrodynamics, 
which in turn will be relevant for the estimate of the magnetic-field profile. 
The techniques and ingredients developed in the computation of the heavy-quark diffusion coefficients 
have been applied in a few works for the jet quenching parameter \cite{Li:2016bbh} 
and the electrical conductivity \cite{Hattori:2016lqx,Hattori:2016cnt} in the strong-field limit. 
The other regime, regarding the strength of the magnetic field, still needs to be explored 
for the phenomenology of the heavy-ion collisions.

\vspace{1cm}

\emph{Acknowledgments}---
K.H. thanks Kei Suzuki for useful comments on the manuscript.
K.H. and X.G.H are supported by Shanghai Natural Science Foundation with Grant No. 14ZR1403000, 1000 Young Talents Program of China,
and  by NSFC with Grant No. 11535012. K.H. is also supported by China Postdoctoral Science Foundation
under Grant No. 2016M590312 and is grateful to support from RIKEN-BNL Research Center.


\appendix

\section{Mixing strengths from the Bethe-Salpeter amplitudes}
\label{sec:coupling_FS}

It would be instructive to see the computation of the coupling strength at the three-point vertex
among two quarkonia and an external magnetic field by using the Bethe-Salpeter amplitudes obtained
in the ladder approximation and the heavy-quark limit \cite{Oh:2001rm,Song:2005yd}.
This computation involves a typical technique for the perturbation theory
in the presence of a magnetic field.

For the static charmonium carrying a momentum $q =  ( 2m - \eb, 0,0,0)$ with $\eb$ being the binding energy,
the Bethe-Salpeter amplitudes for $\eta_c$ and $J/\psi$ are, respectively, given by
\begin{eqnarray}
\Gamma^5 (p,p-q) &=& \left(\epsilon_0 + \frac{\bp^2}{m} \right) \!
\sqrt{\frac{m_\cc}{N_c}}  \, \psi_\swave (\bp) \, P_+ \gamma^5 P_-
\label{eq:G5}
,
\\
\Gamma^\mu (p,p-q) &=& \left(\epsilon_0 + \frac{\bp^2}{m} \right) \!
\sqrt{\frac{m_\cc}{N_c}} \, \psi_\swave (\bp) \, P_+ \gamma^\mu P_-
,
\label{eq:Gvec}
\end{eqnarray}
where we have the projection operators $P_\pm = \frac{1}{2} ( 1 \pm \gamma^0 ) $
and the ground-state wave function of the S-wave bound state $\psi_\swave (\bp) $.
The mass $m_\cc$ is those of $\eta_c$ and $J/\psi$, which are degenerated in the heavy-quark limit.
The number of the color is $N_c=3$.

\begin{figure}[b]
		\begin{center}
			\includegraphics[width=\hsize]{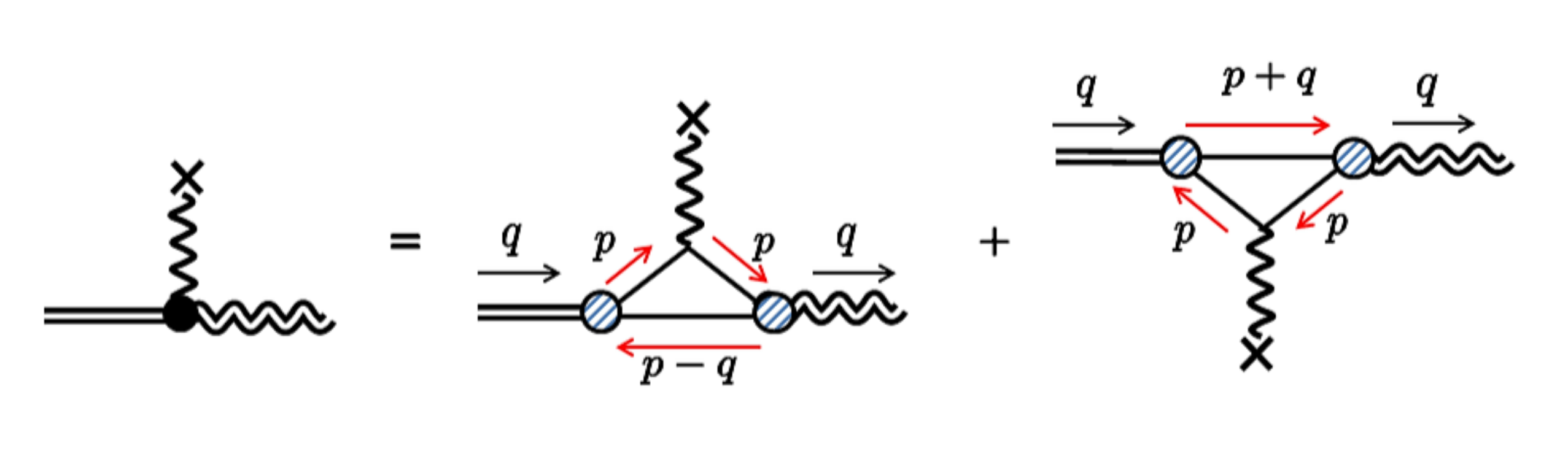}
		\end{center}
        \vspace{-0.5cm}
\caption{An effective coupling strength from triangle diagrams.
Shaded vertices show form factors given by the Bethe-Salpeter amplitudes.
(Reproduced from Ref.~\cite{Cho:2014loa}.)
}
\label{fig:triangles}
\end{figure}

We show a calculation of a coupling strength in the mixing between
$\eta_c$ and the longitudinal $J/\psi$ from triangle diagrams (Fig.~\ref{fig:triangles}).
Interactions between quarks and external magnetic fields are taken into account
by employing the Fock-Schwinger gauge.
In this gauge, the quark propagators with one and two insertions of constant external fields are expressed as \cite{Reinders:1984sr}
\begin{eqnarray}
S_1 (p) &=& - \frac{i}{4} Q_\EM F_{\alpha \beta} \frac{1}{(p^2-m^2+ i \varepsilon)^2}
\label{eq:S1}
\\
&& \hspace{.5cm} \times
\left\{ \sigma^{\alpha\beta} (\slashed p + m) + ( \slashed p + m) \sigma^{\alpha\beta} \right\}
\nonumber
\ ,
\\
S_2 (p) &=&  - \frac{1}{4} Q_\EM^2 F_{\alpha \beta} F_{\mu\nu} \frac{1}{(p^2-m^2+ i \varepsilon)^5}
\label{eq:S2}
\\
&& \hspace{.5cm} \times
(\slashed p +m) \left\{ f^{\alpha\beta \mu\nu} + f^{\alpha\mu\beta\nu} + f^{\alpha\mu\nu\beta} \right\} (\slashed p +m)
\nonumber
\ ,
\end{eqnarray}
where $Q_\EM$ denotes an electromagnetic charge of a quark
and the gamma matrix structures are given by
\begin{eqnarray}
\sigma^{\alpha\beta} \ &=& \frac{i}{2} [\gamma^\alpha, \gamma^\beta]
\ ,
\\
f^{\alpha\beta \mu\nu} &=&
\gamma^\alpha ( \slashed p + m) \gamma^\beta ( \slashed p + m) \gamma^\mu ( \slashed p + m) \gamma^\nu
\ .
\end{eqnarray}


The coupling strength can be read off from the sum of the two diagrams
$ i \M^\mu  = i \M_a^\mu + i \M_b^\mu  $, where the each diagram is written down as
\begin{eqnarray}
&&
i \M_a^\mu  = - \!\! \int \! \frac{d^4p}{(2\pi)^4}
\tr \left[ \, \Gamma_5^\dagger (p-q,p)  S_1(p) \right.
\nonumber
\\
&& \hspace{3.3cm} \left. \times
\Gamma^\mu (p,p-q)  S_0 (p-q)  \, \right]
\label{eq:M1}
\ ,
\\
&&
i \M_b^\mu  = - \!\! \int \! \frac{d^4p}{(2\pi)^4}
\tr \left[ \,  \Gamma_5^\dagger (p+q,p) S_0 (p+q) \right.
\nonumber
\\
&& \hspace{3.3cm} \left. \times
\Gamma^\mu (p,p+q) S_1(p) \, \right]
\label{eq:M2}
\ .
\end{eqnarray}
In the leading order of the heavy-quark expansion,
the $ p^0$-integral is easily performed, and we find
\begin{eqnarray}
i \M_a^\mu  &=& i \M_b^\mu
\nonumber
\\
&=& 2 Q_\EM  \tilde F^{0\mu}  \int \frac{ d^3\bp}{(2\pi)^3} \vert \psi_\swave (\bp) \vert^2
\ .
\end{eqnarray}
From the normalization of the wave function,
\begin{eqnarray}
\int \frac{ d^3\bp}{(2\pi)^3} \vert \psi_\swave (\bp) \vert^2 = 1
,
\end{eqnarray}
the amplitude is independent of the wave functions, and
the sum of two triangle diagrams is obtained as
\begin{eqnarray}
i \M^\mu
&=& 4 Q_\EM  \tilde F^{0\mu}
\ .
\end{eqnarray}
By contracting with the polarization vector for
the longitudinal (transverse) vector state $\epsilon^\mu = (0,0,0,1)$ ($\tilde \epsilon^\mu = (0, n_\perp, 0)$),
the amplitude vanishes for the transverse modes as
\begin{eqnarray}
i \M_\mu  \tilde \epsilon^\mu &=& 0
\, ,
\end{eqnarray}
while the longitudinal mode has a nonvanishing amplitude
\begin{eqnarray}
i \M_\mu  \epsilon^\mu &=& 4 Q_\EM  \tilde F^{0\mu} \epsilon_\mu
= 4 Q_\EM  B
\label{eq:pv}
\, .
\end{eqnarray}
Therefore, the coupling strength in Eq.~(\ref{eq:L_pv}) is found to be
\begin{eqnarray}
g_\pv = 4 Q_\EM
\label{eq:coupling}
\, .
\end{eqnarray}
The coupling strength depends only on the electric charge,
and is given by $g_{\pv} = 8/3 \simeq 2.66 $ ($g_{\pv} = 4/3 \simeq 1.33 $)
for the transition between $\eta_c$ and $J/\psi$ ($\eta_b$ and $\Upsilon$).
This is consistent with the value obtained by fitting the measured radiative decay width [see below Eq.~(\ref{eq:coupling_decay})],
but is slightly overestimated.
The radiative decay widths in $J/\psi \rightarrow \eta_c + \gamma$ and $\Upsilon \rightarrow \eta_b + \gamma$
computed with the coupling strength (\ref{eq:coupling}) agree with
the leading-order results by the potential Non-Relativistic QCD (pNRQCD) \cite{Brambilla:2012be,Pineda:2013lta}.
The overestimate can be improved with the inclusion of the subleading terms \cite{Brambilla:2012be,Pineda:2013lta}.
Extension to the open heavy flavors was done in Ref.~\cite{Gubler:2015qok}.

\bibliography{ref}

\end{document}